\documentclass[11pt,a4paper]{article}
\pdfoutput=1

\usepackage[colorlinks=true, linkcolor=black!50!blue, urlcolor=blue, citecolor=blue, anchorcolor=blue]{hyperref}
\usepackage[font=small,labelfont=bf,margin=0mm,labelsep=period,tableposition=top]{caption}
\usepackage[a4paper,top=1.5cm,bottom=2cm,left=1.5cm,right=1.5cm,bindingoffset=0mm]{geometry}

\usepackage{placeins,cite}
\usepackage{graphicx}
\usepackage{float}
\usepackage{afterpage}
\usepackage{epsfig}
\usepackage{amssymb}
\usepackage{amsmath}
\usepackage{bm}
\usepackage{multirow}
\usepackage{url}
\usepackage{xcolor}
\usepackage{ulem}
\usepackage{url}
\usepackage{booktabs,multirow}
\usepackage{textcomp}
\graphicspath{{../}{./figures/}}

\makeatletter
\def\thickhline{%
             \noalign{\ifnum0 =`}\fi\hrule \@height \thickarrayrulewidth \futurelet
             \reserved@a\@xthickhline}
\def\@xthickhline{\ifx\reserved@a\thickhline
                \vskip\doublerulesep
                \vskip -\thickarrayrulewidth
                \fi
                \ifnum0 =`{\fi}}
\makeatother
\newlength{\thickarrayrulewidth}
\usepackage{tikz}
\usetikzlibrary{positioning}
\usetikzlibrary{shapes.geometric, arrows}
\usetikzlibrary{arrows.meta}
\usepackage{varwidth}
\usepackage{xcolor}
\definecolor{mtplotlib1}{HTML}{1f77b4}
\definecolor{mtplotlib2}{HTML}{ff7f0e}
\definecolor{mtplotlib3}{HTML}{2ca02c}
\definecolor{mtplotlib4}{HTML}{d62728}

\tikzset{%
  >={Latex[width=2mm,length=2mm]},
            base/.style = {rectangle, rounded corners, draw=black,
                           minimum width=4cm, minimum height=1cm,
                           text centered}, 
            mystyle/.style={rectangle, rounded corners, draw=black,
            minimum width=12cm, minimum height=1cm,
            text centered}, 
    col0/.style = {base, fill=white!30},
    col1/.style = {base, fill=mtplotlib1!30},
    col11/.style = {mystyle, fill=mtplotlib1!30},
    col2/.style = {base, fill=mtplotlib2!30},
    col3/.style = {base, fill=mtplotlib3!30},
    col4/.style = {base, minimum width=2.5cm, fill=mtplotlib4!15,}
}

\newcommand{\be}{\begin{equation}}
\newcommand{\ee}{\end{equation}}
\newcommand{\bea}{\begin{eqnarray}}
\newcommand{\eea}{\end{eqnarray}}
\newcommand{\bi}{\begin{itemize}}
\newcommand{\ei}{\end{itemize}}
\newcommand{\ben}{\begin{enumerate}}
\newcommand{\een}{\end{enumerate}}
\newcommand{\la}{\left\langle}
\newcommand{\ra}{\right\rangle}
\newcommand{\lc}{\left[}
\newcommand{\rc}{\right]}
\newcommand{\lp}{\left(}
\newcommand{\rp}{\right)}

\def\frac#1#2{{{#1}\over {#2}}}
\def\gsim{\mathrel{\rlap{\lower4pt\hbox{\hskip1pt$\sim$}}
    \raise1pt\hbox{$>$}}}       
\def\lsim{\mathrel{\rlap{\lower4pt\hbox{\hskip1pt$\sim$}}
    \raise1pt\hbox{$<$}}}

\newcommand{\draft}[1]{}

\def\beq{\begin{equation}}
\def\eeq{\end{equation}}

\numberwithin{equation}{section}
\numberwithin{figure}{section}
\numberwithin{table}{section}

\usepackage{tabularx}
\newcolumntype{C}[1]{>{\centering\arraybackslash}p{#1}}

\definecolor{darkblue}{rgb}{0.0,0,0.5}
\definecolor{darkgreen}{rgb}{0.0,0.3,0.0}
\definecolor{redish}{rgb}{0.675,0,0.2}
\definecolor{red}{rgb}{0.8,0,0}
\definecolor{green}{rgb}{0,0.6,0}
\definecolor{bluish}{rgb}{0.2,0.2,0.675}
\definecolor{mygrey}{rgb}{0.6,0.6,0.6}

\usepackage{tikz} 
\usetikzlibrary{shapes,arrows,positioning,automata,backgrounds,calc,er,patterns,arrows.meta}
\usepackage{tikz-feynman}
\tikzfeynmanset{compat=1.0.0}
\usepackage{varwidth}
\usepackage{xcolor}
\definecolor{mtplotlib1}{HTML}{1f77b4}
\definecolor{mtplotlib2}{HTML}{ff7f0e}
\definecolor{mtplotlib3}{HTML}{2ca02c}
\definecolor{mtplotlib4}{HTML}{d62728}

\tikzset{%
  >={Latex[width=2mm,length=2mm]},
            base/.style = {rectangle, rounded corners, draw=black,
                           minimum width=4cm, minimum height=1cm,
                           text centered}, 
            mystyle/.style={rectangle, rounded corners, draw=black,
            minimum width=12cm, minimum height=1cm,
            text centered}, 
    col0/.style = {base, fill=white!30},
    col1/.style = {base, fill=mtplotlib1!30},
    col11/.style = {mystyle, fill=mtplotlib1!30},
    col2/.style = {base, fill=mtplotlib2!30},
    col3/.style = {base, fill=mtplotlib3!30},
    col4/.style = {base, minimum width=2.5cm, fill=mtplotlib4!15,}
}

\usepackage{tabularx}
\usepackage{subcaption}
\newcolumntype{C}[1]{>{\centering\arraybackslash}p{#1}}

\begin{document}
\newgeometry{top=1.5cm,bottom=1.5cm,left=1.5cm,right=1.5cm,bindingoffset=0mm}

\vspace{-2.0cm}
\begin{flushright}
Nikhef-2023-009 \\ CERN-TH-2023-165\\
\end{flushright}
\vspace{0.6cm}

\begin{center}
  {\Large \bf The LHC as a Neutrino-Ion Collider}\\
  \vspace{1.1cm}
  {\small
    Juan M. Cruz-Martinez$^{1}$, Max Fieg$^{2}$, Tommaso Giani$^{3,4}$, Peter Krack$^{3,4}$, Toni M\"akel\"a$^{5}$,  \\[0.1cm]
    Tanjona Rabemananjara$^{3,4}$, and Juan Rojo$^{3,4}$
  }\\
  
\vspace{0.7cm}

{\it \small
    ~$^1$CERN, Theoretical Physics Department, CH-1211 Geneva 23, Switzerland \\[0.1cm]
    ~$^2$ Department of Physics and Astronomy, University of California, Irvine, CA 92697 USA  \\[0.1cm]
    ~$^3$Department of Physics and Astronomy, Vrije Universiteit, NL-1081 HV Amsterdam\\[0.1cm]
    ~$^4$Nikhef Theory Group, Science Park 105, 1098 XG Amsterdam, The Netherlands\\[0.1cm]
    ~$^5$National Centre for Nuclear Research, Pasteura 7, Warsaw, PL-02-093, Poland \\[0.1cm]
 }

\vspace{1.0cm}

{\bf \large Abstract}

\end{center}

Proton-proton collisions at the LHC generate a high-intensity collimated beam of neutrinos
in the forward (beam) direction, characterised by energies of up to several TeV.
The recent  observation of LHC neutrinos by FASER$\nu$ and SND@LHC
signals that this hitherto ignored particle beam is now available for scientific inquiry.
Here we quantify the impact that neutrino deep-inelastic scattering (DIS)
measurements at the LHC would have
on the parton distributions (PDFs) of protons and heavy nuclei.
We generate projections for DIS structure functions
for FASER$\nu$ and SND@LHC at Run III,
as well as for the FASER$\nu$2, AdvSND, and FLArE experiments
to be  hosted at the proposed
Forward Physics Facility (FPF) operating concurrently with the High-Luminosity LHC (HL-LHC).
We determine that up to one million electron- and muon-neutrino DIS interactions
within detector acceptance can be expected by the end of the HL-LHC,
covering a kinematic region in $x$ and $Q^2$ overlapping with that
of the Electron-Ion Collider.
Including these DIS projections into
global (n)PDF analyses, specifically PDF4LHC21, NNPDF4.0,
and EPPS21,  reveals a significant reduction of PDF uncertainties, in particular
for strangeness and the up and down valence PDFs.
We show that LHC neutrino data enables improved theoretical
predictions for core processes at the HL-LHC, such as Higgs and weak gauge
boson production.
Our analysis demonstrates that exploiting the LHC neutrino beam effectively
provides CERN with a ``Neutrino-Ion Collider''
without  requiring modifications in its accelerator infrastructure.

\clearpage

\tableofcontents
\section{Introduction and motivation}
\label{sec:introduction}

Proton-proton collisions at the LHC produce a high-intensity collimated flux of neutrinos.
These neutrinos are characterised by the largest energies ever achieved in laboratory experiments,
reaching up to several TeV~\cite{Kling:2021gos}.
Due to the lack of dedicated instrumentation in the far-forward region,
until recently these neutrinos avoided detection.
The recent observation of LHC neutrinos~\cite{FASER:2023zcr,SNDLHC:2023pun,CERN-FASER-CONF-2023-002} by the
  FASER~\cite{FASER:2019dxq,FASER:2022hcn} and SND@LHC~\cite{SHiP:2020sos,SNDLHC:2022ihg} far-forward experiments
demonstrates that this hitherto discarded beam can now be deployed for physics studies.
Beyond the ongoing Run III, a dedicated suite of upgraded far-forward
neutrino experiments would be hosted by the proposed
Forward Physics Facility (FPF)~\cite{Anchordoqui:2021ghd,Feng:2022inv} operating
concurrently with the High-Luminosity LHC (HL-LHC)~\cite{Azzi:2019yne,Cepeda:2019klc}.
Current and future LHC neutrino experiments enable unprecedented scientific opportunities
for particle and astroparticle physics both within the Standard Model and beyond it,
as summarised in~\cite{Anchordoqui:2021ghd,Feng:2022inv}
and references therein.

Measurements of neutrino structure functions~\cite{Conrad:1997ne,Mangano:2001mj,Candido:2023utz} in deep-inelastic scattering
(DIS) are sensitive probes of the parton distributions (PDFs)
of nucleons and nuclei~\cite{Ethier:2020way,Gao:2017yyd,Kovarik:2019xvh},  in particular
concerning (anti)quark  flavour separation and
strangeness~\cite{NuTeV:2007uwm,CCFR:1994ikl,Faura:2020oom,Alekhin:2014sya}.
Constraints arising from charged-current neutrino scattering provide information on
complementary flavour combinations as compared to neutral-current charged-lepton DIS.
Several  experiments have measured neutrino
DIS structure functions over a wide range of energies, and neutrino data
from CHORUS~\cite{CHORUS:2005cpn}, NuTeV~\cite{NuTeV:2001dfo},
CCFR~\cite{Yang:2000ju}, NOMAD~\cite{NOMAD:2013hbk}, CDHS~\cite{Berge:1989hr},
and other experiments is routinely  included in global
determinations of proton~\cite{NNPDF:2021njg,Hou:2019efy,Bailey:2020ooq} and nuclear
PDFs~\cite{Eskola:2021nhw,AbdulKhalek:2022fyi,Muzakka:2022wey}.

As compared to  previous neutrino DIS experiments,  neutrino
scattering at the LHC involves energies of up to a factor 10  higher.
Furthermore, large event rates are expected, with
up to one million muon neutrinos interacting at the
FPF detectors~\cite{Anchordoqui:2021ghd,Feng:2022inv}.
Initial estimates~\cite{Feng:2022inv} indicate that an extension of the coverage of
available neutrino DIS data by an order of magnitude both at small-$x$
and large-$Q^2$ should be possible.
However, quantitative projections for the kinematic reach
and experimental accuracy expected at current
and future LHC neutrino experiments are not available.
The lack of these projections has prevented detailed studies assessing the impact
of LHC neutrino data in global analyses of proton and nuclear PDFs, comparable to
those performed for the HL-LHC~\cite{AbdulKhalek:2018rok,Azzi:2019yne}, the Electron-Ion Collider (EIC)~\cite{AbdulKhalek:2021gbh,Khalek:2021ulf,AbdulKhalek:2019mzd}, and the
Large Hadron-electron Collider (LHeC)~\cite{AbdulKhalek:2019mps,LHeC:2020van,LHeCStudyGroup:2012zhm}. 

Here we bridge this gap by quantifying
the expected impact of  LHC neutrino structure functions on proton and nuclear PDFs.
To this end, we produce simulations for  FASER$\nu$ and SND@LHC at Run III 
as well as for the proposed FPF experiments~\cite{Anchordoqui:2021ghd,Feng:2022inv,Batell:2021blf,Batell:2021aja}, FLArE, AdvSND, and FASER$\nu$2.
For each experiment, we determine the expected event yields in bins of $(x,Q^2,E_\nu)$
satisfying acceptance and selection cuts,
generate pseudo-data for  inclusive and charm 
structure functions, 
and estimate their dominant systematic uncertainties.
Subsequently, we study their impact on the proton and nuclear PDFs by means of both the Hessian profiling~\cite{Paukkunen:2014zia,  Schmidt:2018hvu, AbdulKhalek:2018rok, HERAFitterdevelopersTeam:2015cre}
of  PDF4LHC21~\cite{PDF4LHCWorkingGroup:2022cjn} (for protons) and EPPS21~\cite{Eskola:2021nhw}
(for tungsten nuclei)
within the {\sc\small xFitter}~\cite{Alekhin:2014irh, Bertone:2017tig, xFitter:2022zjb, xFitter:web} open-source QCD analysis framework,
as well as with the direct inclusion in the open-source NNPDF4.0 fitting code~\cite{NNPDF:2021uiq}.

Our analysis reveals that  LHC neutrino structure functions can provide  stringent constraints
on the light quark and antiquark PDFs, especially on the up and down
valence quarks and on strangeness, as compared to state-of-the-art global analyses.
We also  find that accounting for the main systematic uncertainties does not significantly
degrade the sensitivity achieved in the baseline fits considering only statistical errors.
We quantify the impact in our results of charm-tagged structure functions (large), study the relevance
of final-state lepton-charge identification capabilities (moderate), and compare the constraints
provided by different experiments (finding that the overall sensitivity is dominated by FASER$\nu$2).
We also study the implications of the resulting improvement in PDF precision
on core processes at the HL-LHC, finding a theory error reduction
of up to a factor two in the most optimistic scenario
for selected  Higgs and gauge boson production cross-sections. 

Our results demonstrate that the availability of far-forward neutrino detectors
at the LHC effectively
provides CERN with a charged-current counterpart of the EIC,
with similar kinematic reach and complementary sensitivity on hadronic
structure.
Therefore, LHC neutrino experiments realise, upon Lorentz-boosting, the analog of
a ``Neutrino-Ion Collider'' at CERN
without the need of new accelerator infrastructure or additional energy consumption.

The outline of this paper is as follows.
Sect.~\ref{sec:dis_pseudodata} describes the procedure
adopted to generate projections for neutrino DIS structure functions at the LHC
and the methodology used to include these into PDF fits.
The impact of such LHC structure function measurements on proton and nuclear
PDFs is quantified in Sect.~\ref{sec:protonPDFs}, with the
associated implications for precision phenomenology
at the HL-LHC assessed in Sect.~\ref{sec:pheno}.
We summarise in Sect.~\ref{sec:summary}, where we also consider possible
directions for follow-up research.
Additional results are collected in two appendices:
App.~\ref{app:comparisons_with_HLLHC} revisits
the phenomenological studies of
Sect.~\ref{sec:pheno} using the HL-LHC PDF impact projections
presented in~\cite{AbdulKhalek:2018rok}, while App.~\ref{app:nPDF_impact_appendix}
quantifies the stability of the nPDF impact projections with respect
to input variations.

\section{Deep-inelastic scattering with LHC neutrinos}
\label{sec:dis_pseudodata}

Here we describe the procedure adopted 
to generate projections for the kinematic coverage
and uncertainties associated to  measurements
of neutrino-nucleus scattering at the LHC far-forward experiments.
First, we summarise the theoretical description of differential
neutrino scattering in terms of DIS structure functions.
Then, we present an  overview of the operative and proposed
LHC far-forward neutrino detectors that
are considered in the present study and indicate their
acceptance and performance parameters.
By convoluting the expected electron and muon neutrino fluxes
with the acceptance and scattering rates of each
of these detectors,
we evaluate the event yields in bins of $x$, $Q^2$,
and $E_\nu$ and the associated statistical and systematic uncertainties.
Finally, we discuss the procedure adopted to generate pseudo-data
for LHC neutrino structure functions
and to quantify their impact
 into proton and nuclear PDF determinations.

 \subsection{Neutrino DIS revisited}
 \label{sec:nudis_revisited}

The double-differential cross-section for neutrino-nucleus charged-current scattering,
see~\cite{Candido:2023utz} and references therein,
can be expressed in terms of three
independent structure functions $F_2^{\nu A}$, $xF_3^{\nu A}$
and $F_L^{\nu A}$:
\begin{align}
	\label{eq:neutrino_DIS_xsec_FL}
	&\frac{d^2\sigma^{\nu A}(x,Q^2,y)}{dxdy} =  \frac{G_F^2s/4\pi}{\lp 1+Q^2/m_W^2\rp^2}\\ 
	& \lc Y_+F^{\nu A}_2(x,Q^2) - y^2F^{\nu A}_L(x,Q^2) +Y_- xF^{\nu A}_3(x,Q^2)\rc, \nonumber
\end{align}
where $Y_\pm = 1 \pm (1-y)^2$ and with a counterpart expression for anti-neutrino scattering,
\begin{align}
	\label{eq:antineutrino_DIS_xsec_FL}
	&\frac{d^2\sigma^{\bar{\nu} A}(x,Q^2,y)}{dxdy} =  \frac{G_F^2s/4\pi}{\lp 1+Q^2/m_W^2\rp^2} \\
	& \lc Y_+F^{\bar{\nu} A}_2(x,Q^2) - y^2F^{\bar{\nu} A}_L(x,Q^2) -Y_- xF^{\bar{\nu} A}_3(x,Q^2)\rc,
	\nonumber
\end{align}
with $s=2m_N E_\nu$ being the neutrino-nucleon centre-of-mass energy squared, $m_N$ the nucleon mass,
$E_\nu$ the incoming neutrino energy,
and $y=Q^2/(2x m_n E_{\nu})$ the inelasticity.
In the case of tau-neutrino scattering, tau-lepton mass effects
may be relevant and Eqns.~(\ref{eq:neutrino_DIS_xsec_FL}) and~(\ref{eq:antineutrino_DIS_xsec_FL}) receive additional contributions
from the $F_4$ and $F_5$ structure functions.
We neglect these effects, since here we focus on electron and muon neutrino scattering.

Structure functions depend on both $x$ and $Q^2$, while the differential
cross-section depends also on the neutrino energy $E_\nu$, or alternatively
on the inelasticity $y$.
Further, structure functions $F^{\nu A}_i(x,Q^2)$ and $F^{\bar{\nu} A}_i(x,Q^2)$ depend on the nuclear target $A$ entering neutrino scattering through the
nuclear modifications of the free-nucleon PDFs.
Eqns.~(\ref{eq:neutrino_DIS_xsec_FL}) and~(\ref{eq:antineutrino_DIS_xsec_FL}) are valid provided
the hadronic 
invariant mass $W$  is above the resonance production threshold,
\be
\label{eq:W2_invmass}
W^2 = \lp m_N^2 + Q^2 \frac{(1-x)}{x} \rp \gsim \lp 2\,{\rm GeV} \rp^2\, .
\ee
In addition, here we  restrict ourselves to the DIS region with perturbative momentum
transfers $Q^2 \gsim 2$ GeV$^2$, such that the structure functions can be decomposed as
\begin{align}
	\label{eq:sfs_pqcd}
	F^{\nu A}_i(x,Q^2) = \sum_{j=q,\bar{q},g} & \int_x^1 \frac{dz}{z}\, C_{i,j}^{\nu N}
	(z,\alpha_s(Q^2))f^{(A)}_j\lp \frac{x}{z},Q^2\rp \, , \nonumber \\
	& \text{ with } \quad i = 2,3,L \, ,
\end{align}
 expressed
 in terms of a convolution of partonic scattering cross-sections  $C_{i,j}^{\nu N}(x,\alpha_s)$ and
of process-independent PDFs $f^{(A)}_j\lp x,Q^2\rp$.
A similar expression holds for charm production~\cite{Faura:2020oom}, which requires
accounting also for charm mass effects~\cite{Gao:2017kkx}.
Here the theory pipeline adopted
to evaluate neutrino structure functions using Eq.~(\ref{eq:sfs_pqcd}) is provided by 
{\sc\small EKO}~\cite{Candido:2022tld}
and {\sc\small YADISM}~\cite{yadism,Candido:2023utz}
interfaced to {\sc\small PineAPPL}~\cite{Carrazza:2020gss}
for the generation of fast interpolation grids. 

Different neutrino structure functions provide complementary sensitivity
 to the partonic flavour decompositions of nucleons.
 To illustrate this feature, consider a leading order  calculation
 for a proton target with $n_f=4$ active quark flavours,
a diagonal CKM matrix, and no heavy quark mass effects.
 The resulting $F_2^{\nu p}$ and $xF_3^{\nu p}$ structure functions read
\begin{align}
	 F_2^{\nu p}(x,Q^2) &= 2x\lp f_{\bar{u}} + f_{d} + f_{s} + f_{\bar{c}} \rp(x,Q^2) \, , \nonumber  \\
	F_2^{\bar{\nu} p}(x,Q^2) &= 2x\lp f_u + f_{\bar{d}} + f_{\bar{s}} + f_c \rp(x,Q^2) \, , \label{eq:neutrinoSFs_proton} \\
	xF_3^{\nu p}(x,Q^2) &= 2x\lp -f_{\bar{u}} + f_d +f_s - f_{\bar{c}}\rp(x,Q^2)  \, , \nonumber\\
	xF_3^{\bar{\nu} p}(x,Q^2) &= 2x\lp f_u - f_{\bar{d}} -f_{\bar{s}} + f_{c}\rp(x,Q^2) \, . \nonumber
\end{align}
 The corresponding expressions for a neutron target are obtained from isospin symmetry
 \begin{align}
 	 F_2^{\nu n}(x,Q^2) &= 2x\lp f_{\bar{d}} + f_{u} + f_{s} + f_{\bar{c}} \rp(x,Q^2) \, , \nonumber  \\
 	F_2^{\bar{\nu} n}(x,Q^2) &= 2x\lp f_d + f_{\bar{u}} + f_{\bar{s}} + f_c \rp(x,Q^2) \, , \label{eq:antineutrinoSFs_neutron} \\
 	xF_3^{\nu n}(x,Q^2) &= 2x\lp -f_{\bar{d}} + f_u +f_s - f_{\bar{c}}\rp(x,Q^2)  \, , \nonumber\\
 	xF_3^{\bar{\nu} n}(x,Q^2) &= 2x\lp f_d - f_{\bar{u}} -f_{\bar{s}} + f_{c}\rp(x,Q^2) \, , \nonumber
 \end{align}
 while for an isoscalar, free-nucleon target denoted by $N$ one has
 \begin{align}
 	\noindent
 	&F_2^{\nu N}(x,Q^2)= 2x\lp f_{u^+} + f_{d^+} + 2f_s + 2f_{\bar{c}} \rp(x,Q^2), \nonumber  \\
 	&F_2^{\bar{\nu} N}(x,Q^2)= 2x\lp f_{u^+} + f_{d^+} + 2f_{\bar{s}} + 2f_c \rp(x,Q^2), \label{eq:neutrinoSFs_isoscalar} \\
 	&xF_3^{\nu N}(x,Q^2)= 2x\lp f_{u^-} + f_{d^-} +2f_s - 2f_{\bar{c}}\rp(x,Q^2), \nonumber\\
 	&xF_3^{\bar{\nu} N}(x,Q^2)= 2x\lp   f_{u^-} + f_{d^-}-2f_{\bar{s}} +2 f_{c}\rp(x,Q^2), \nonumber \qquad
 \end{align}
 in terms of the valence and sea PDF combinations defined by the following relations
\begin{align}
	 f_{q^+} (x,Q^2) &\equiv \lp f_{q}+f_{\bar{q}}\rp(x,Q^2) \, , \nonumber \\
	f_{q^-} (x,Q^2) &\equiv \lp f_{q}- f_{\bar{q}}\rp(x,Q^2) \, .
\end{align}
 We note that, even for isoscalar targets, separate measurements
 for neutrinos and antineutrinos will not be equivalent, since in general
 the strange and charm sea asymmetries $f_{s^-}$ and
 $f_{c^-}$ are not expected to vanish~\cite{Sufian:2018cpj,Sufian:2020coz}.

 In the projections presented here, when interpreting the LHC neutrino structure
 functions in terms of proton PDFs, we will assume an isoscalar free-nucleon target and neglect
 nuclear PDF modifications, along the lines of Eq.~(\ref{eq:neutrinoSFs_isoscalar}).
 Accounting for nuclear modifications in a global proton
 PDF fit is possible by means of the procedure developed
 in~\cite{Ball:2020xqw,Ball:2018twp} based on
 the theory covariance matrix approach~\cite{NNPDF:2019vjt,NNPDF:2019ubu}.
 On the other hand, when evaluating structure functions
 for a tungsten (W) target, we keep into account both
 nuclear corrections and that
 the target is not isoscalar when evaluating the physical observables.

 It is also illustrative to compare the PDF dependence of neutrino structure functions
 at LO with that of their counterparts for neutral-current
 scattering with a charged lepton projectile.
 Within the same assumptions, for energies below
 the $Z$-boson mass, $Q^2 \ll m_Z^2$, the corresponding
decomposition is
\begin{align}
	 F_2^{\ell p}(x,Q^2) &= x\lp \frac{4}{9}\lc f_{u^+} + f_{c^+}\rc
	+ \frac{1}{9}\lc f_{d^+} + f_{s^+}\rc\rp(x,Q^2) \, , \nonumber  \\
	F_2^{\ell n}(x,Q^2) &= x\lp \frac{4}{9}\lc f_{d^+} + f_{c^+}\rc
	+ \frac{1}{9}\lc f_{u^+} + f_{s^+}\rc\rp(x,Q^2) \, , \nonumber   \\
	F_2^{\ell N}(x,Q^2) &= x\lp \frac{5}{18}\lc f_{u^+} + f_{d^+}\rc
	+ \frac{1}{9} f_{s^+} + \frac{4}{9} f_{c^+} \rp(x,Q^2) \, , \label{eq:NC_chargedlepton} 
\end{align}
 with $xF_3$ being negligible in this region.
 Comparing Eqns. (\ref{eq:neutrinoSFs_proton})--(\ref{eq:neutrinoSFs_isoscalar})
 with Eq.~(\ref{eq:NC_chargedlepton}) showcases the complementarity between
 neutrino and charged-lepton DIS in terms of sensitivity
 to different flavour PDF combinations.
 This implies that the best sensitivity on the quark
  flavour separation in the nucleon
 would be provided by combining neutrino DIS from the
 LHC far-forward experiments with measurements on charged-lepton
 DIS at the EIC.

 \subsection{LHC far-forward neutrino experiments}
 \label{sec:neutrinoDetectors}

 The calculation of differential neutrino scattering event rates
 at the LHC far-forward detectors involves two main ingredients: the energy
 and flavour dependence of the incoming neutrino flux crossing
 the detector fiducial volume, on the one hand,
 and the scattering rates within the detector acceptance, on the other hand.
 Here we summarise the main features of the existing and future
 far-forward detectors considered, in particular concerning
 their acceptance and expected performance.
 We focus on  muon-neutrino scattering, which benefits from the highest rates and is less
 affected by theoretical uncertainties in the production mechanism, but
 provide also predictions for the subdominant electron-neutrino structure functions.
 
 The kinematics of a charged-current neutrino DIS event $(x,Q^2, E_\nu)$,
 or alternatively $(x,Q^2, y)$, are uniquely specified by the measurement of three independent
 final-state variables,
 such as $\lp E_\ell,\theta_\ell, W^2\rp$ or $\lp E_\ell,\theta_\ell, E_h \rp$,
 with $E_\ell$ and $\theta_\ell$ being the energy and polar angle of the outgoing
 charged lepton and $E_h$ the total energy of the hadronic final state.
 Most neutrino detectors can only access $E_h$, given that measuring the invariant mass $W^2$ requires
 fully reconstructing the final state.
 A measurement of the three kinematic variables $\lp E_\ell,\theta_\ell, E_h \rp$ then fixes the DIS kinematics as:
 \bea
 E_\nu &=& E_h + E_\ell \, , \nonumber \\
 Q^2 &=& 4 ( E_h + E_\ell) E_\ell \sin^2 \lp \theta_\ell/2\rp \, ,  \label{eq:dis_kinematic_mapping}\\
 x&=& \frac{4 ( E_h + E_\ell) E_\ell \sin^2 \lp \theta_\ell/2\rp}{2m_N E_h} \, .\nonumber
 \eea
 These relations also reflect how systematic uncertainties affecting the measurement
 of $\lp E_\ell,\theta_\ell, E_h \rp$ translate into uncertainties in the
 reconstructed values of  $(x,Q^2, E_\nu)$ modifying the expected
 binned event rates.

 \paragraph{Detector overview.}
 Table~\ref{tab:FPF_experiments} summarises,
 for each of the far-forward LHC neutrino experiments considered in this
 work,
 their pseudo-rapidity coverage, target material, whether
  they can identify the sign of the outgoing charged lepton,
  the acceptance for the charged lepton and hadronic final state,
  and the expected reconstruction performance.
  We consider separately acceptance and performance for electron-neutrinos and
  muon-neutrinos.
  In these projections we assume that FASER$\nu$ and SND@LHC acquire data
  for Run III ($\mathcal{L}=150$ fb$^{-1}$), while FASER$\nu$2,
  AdvSND, and FLArE take data
  for the complete HL-LHC period  ($\mathcal{L}=3$ ab$^{-1}$).
   In the case of FLArE, we consider projections for fiducial volumes corresponding to both 10 and 100 tonne detectors.
  In the following, we provide details about the information collected in
  Table~\ref{tab:FPF_experiments}.

\begin{table}[t]
  \centering
  \small
  \renewcommand{\arraystretch}{1.50}
\begin{tabularx}{\textwidth}{Xccccc}
\toprule
Detector &  Rapidity &  Target & Charge ID & Acceptance  & Performance \\
\midrule
\midrule
\multirow{3}{*}{FASER$\nu$}  &  \multirow{3}{*}{ $\eta_\nu \ge 8.5$}  &   \multirow{2}{*}{Tungsten}  & \multirow{3}{*}{muons}      &   $E_\ell,E_h \gsim 100$ GeV   &      $\delta E_\ell \sim 30\% $    \\
&   &   \multirow{2}{*}{(1.1 tonnes)}  &       &  $\tan \theta_\ell \lsim 0.025 $ (charge ID)   &
$\delta \theta_\ell \sim 0.06$ mrad        \\
&   &     &       &  reco $E_h$ \& charm ID   &      $\delta E_h \sim 30\%$     \\
\midrule
\multirow{2}{*}{SND@LHC}  & \multirow{2}{*}{ $7.2 \le \eta_\nu \le 8.4$}   &  Tungsten   &   \multirow{2}{*}{n/a}    &  $E_\ell,E_h \gsim 20 $ GeV     &    \multirow{2}{*}{n/a}    \\
  &    &  (0.83 tonnes)   &  &  $\theta_\mu \lsim 0.15, \theta_e \lsim 0.5$         &       \\
\midrule
\midrule
\multirow{3}{*}{FASER$\nu$2}  & \multirow{3}{*}{ $\eta_\nu \ge 8.5$}  & \multirow{2}{*}{Tungsten}    &   \multirow{3}{*}{muons}     &   $E_\ell,E_h \gsim 100$ GeV  &    $\delta E_\ell \sim 30\% $     \\
  &   &  \multirow{2}{*}{(20 tonnes)}   &       &  $\tan \theta_\ell \lsim 0.05$ (charge ID)  &   $\delta \theta_\ell \sim 0.06$ mrad      \\
  &   &     &       &  reco $E_h$ \& charm ID   &  $\delta E_h \sim 30\%$        \\
\midrule
\multirow{3}{*}{AdvSND-far}  &   \multirow{3}{*}{ $7.2 \le \eta_\nu \le 8.4$}  &
\multirow{2}{*}{Tungsten}   &   \multirow{3}{*}{muons}    &  $E_\ell,E_h \gsim 20 $ GeV  & \multirow{3}{*}{n/a}          \\
  &   &   \multirow{2}{*}{(5 tonnes)}  &        & $\theta_\mu \lsim 0.15, \theta_e \lsim 0.5$     &           \\
  &   &     &       &  reco $E_h$   &           \\
\midrule
\multirow{3}{*}{FLArE ({\bf *})}  & \multirow{3}{*}{$\eta_\nu \ge 7.5$} & \multirow{2}{*}{LAr}  & \multirow{3}{*}{muons}  &  $E_\ell,E_h \gsim 2$ GeV, $E_e \lsim 2$ TeV    &    $\delta E_e \sim 5\% $,  $\delta E_\mu \sim 30\% $ \\
&   &  \multirow{2}{*}{(10,~100~tonnes)}   &   & $\theta_\mu \lsim 0.025$, $\theta_e \lsim 0.5$ &    $\delta \theta_\ell \sim 15 $ mrad\\
 &   &     &  & reco $E_h$  &    $\delta E_h \sim 30\% $   \\
  \bottomrule
\end{tabularx}
\vspace{0.2cm}
\caption{\small For each of the far-forward LHC neutrino experiments considered,
   we indicate their neutrino pseudo-rapidity coverage, target material, whether
  they can identify the sign of the outgoing charged lepton,
  the acceptance for the charged lepton and hadronic final state,
  and the expected reconstruction performance.
  We consider separately acceptance and performance for electron and muon
  neutrinos.
  For FLArE, we assume that muons would be measured in the FASER2 spectrometer
  situated downstream in the FPF cavern.
  See the description of each experiment in the text for more details.
  For our projections we assume that FASER$\nu$ and SND@LHC acquire data
  for the Run III period ($\mathcal{L}=150$ fb$^{-1}$), while FASER$\nu$2, AdvSND, and FLArE take data
  for the complete HL-LHC period ($\mathcal{L}=3$ ab$^{-1}$).
  In the case of FLArE, we consider projections for fiducial volumes corresponding to both 10 and 100 tonne detectors.
  \label{tab:FPF_experiments}
}
\end{table}


\paragraph{FASER$\nu$.}
The ForwArd Search ExpeRiment (FASER) detector
and its companion FASER$\nu$~\cite{FASER:2019aik,FASER:2019dxq,FASER:2023zcr,FASER:2022hcn}
are located at the TI12 tunnel of the CERN accelerator complex.
Both detectors are aligned
with the collision axis line-of-sight (LOS)
and have been acquiring data since the beginning of Run III in 2022.
Neutrino scattering takes place in the FASER$\nu$ 
detector, composed by interleaved emulsion and tungsten plates and
adding up to a mass of 1.1 tonnes with a fiducial volume of $20~\rm{cm} \times 25~\rm{cm} \times 80~{\rm cm}$.
The FASER apparatus is immersed in a magnetic field,  providing charged-lepton
identification thanks to two 1 m-long dipole magnets with $B=0.57$ T
and another 1.5 m-long magnet in front of the spectrometer. 
Neutrino detection and identification can be carried out either using the emulsion
films, which have the key benefit of excellent position and angular resolution,
or instead using the electronic detector components of FASER, which enable the tagging
of the outgoing downstream energetic muons.
FASER$\nu$ is sensitive to neutrinos with pseudorapidity $\eta_\nu \ge 8.5$
and can also identify charm-tagged events.
To identify the charge of the lepton from a neutrino interaction, the lepton is required to pass through the FASER spectrometer.
This imposes an angular requirement on the outgoing charged lepton, indicated
in Table~\ref{tab:FPF_experiments}.
Also, to identify a DIS interaction the emulsion detector requires at least 5 charged tracks
to emerge from the interaction vertex.
We implement this last constraint
in our simulations by requiring a minimum hadronic energy when calculating the event yields, as the charged track multiplicity is expected to grow with $W$~\cite{Aachen-Bonn-CERN-Munich-Oxford:1981lfk,FASER:2019dxq}. 

\paragraph{SND@LHC.}
In the same manner as FASER, the SND@LHC experiment~\cite{SNDLHC:2022ihg}
is located in a service tunnel (TI18)
around 500 meters from the ATLAS interaction point and has been taking data
since the  beginning of Run III.
SND@LHC is installed off the LOS axis in order to cover the neutrino
pseudo-rapidity range of $7.2 \le \eta_\nu \le 8.4$.
With a total fiducial volume corresponding to a 830 kg detector with a length of 35 cm, it is composed by tungsten plates,
where neutrino scattering takes place, interleaved with nuclear emulsions and electronic tracker
components.
Downstream, the scattering volume is followed by a hadronic calorimeter and a muon tracking system.
The electromagnetic
 and hadronic energy deposits can be measured at the electronic detectors, with the emulsion
 components providing vertex reconstruction.
 The lack of magnetic field prevents the charge-sign identification of the outgoing charged leptons.

\paragraph{FASER$\nu$2.}
This is a proposed 20-tonne neutrino experiment located on the LOS
of the LHC neutrino beam to be installed in the FPF cavern.
It is based on the same technology as FASER$\nu$, and hence
relies on a emulsion-based detector optimised to identify heavy flavour particles, including
tau leptons and charm and beauty particles, arising from neutrino interactions.
It would be sensitive to neutrinos with pseudorapidity $\eta_\nu \ge 8.5$.
The FASER$\nu$2 detector is composed of 3300 emulsion layers interleaved with 2-mm-thick tungsten plates,
for a total volume of  tungsten of $40~\rm{cm} \times 40~\rm{cm} \times 6.6~{\rm m}$.
The combination of FASER$\nu$2  with the nearby FASER2 detector, equipped with a spectrometer, makes measurements of the outgoing muon charge possible.
Given that FASER$\nu$2 is based on the same detector technology
as its predecessor, the same performance in terms of reconstruction
of final-state kinematics can be assumed.

 \paragraph{AdvSND.}
 This proposed experiment~\cite{Feng:2022inv} consists actually on  two detectors, a far-detector to be installed
 at the FPF with a coverage in neutrino pseudorapidity of $7.2 \le \eta_\nu \le 8.4$
 (i.e., off-LOS, same as SND@LHC) and a near detector installed somewhere else in the LHC
 complex and covering the range $4 \le \eta_\nu \le 5$.
Here we focus on the former.
 It would be
 composed (from upstream to downstream) by a target region
 for  vertex reconstruction and electromagnetic energy measurement, followed  by a hadronic calorimeter, a  muon 
 identification system, and finally  a magnet enabling muon charge and momentum measurements.
 The target region of the detector, where the neutrino interactions take place, is made of thin sensitive layers of emulsion interleaved with tungsten plates, for a total mass of 5 tonnes and fiducial length of 50 cm.
 This detector configuration will be able to track muons with energy $E_\nu \gsim 20$ GeV
 within an acceptance of 100 mrad and provide information on the charge
 of the  outgoing muon thanks to its magnet.
 The total energy of the hadronic final state will be measured
 in the hadronic calorimeter.
 No information on the expected performance of the AdvSND-far detector
 is available and hence no estimate of the systematic errors
 is carried out.

 \paragraph{FLArE.}
 Building upon recent progress in liquid noble gas neutrino detectors over the last decade (ICARUS, MicroBooNE, SBND, DUNE), this experiment~\cite{Batell:2021blf,Feng:2022inv}
 would rely on a modularized liquid argon (LAr) time projection detector.
 The use of LAr as a target is beneficial for final-state particle identification, track angle, and kinetic energy measurements with sub-millimeter spatial resolution in all dimensions.
 The detector will be equipped with a magnetized hadron/muon calorimeter downstream of the liquid argon volume
 for muon charge and momentum measurements.
 While muon neutrinos with energy $E_\nu \gsim 2$ GeV would not be fully
 contained in the  FLArE detector,
for our projections
we assume that outgoing high-energy
muons would be measured in the FASER2 spectrometer
  situated downstream in the FPF cavern.
 
 With an expected fiducial (active) mass of 10 tonnes (30 tonnes) and a length of 7 m, FLArE will
 detect final-state electrons with energies $E_{e}\lsim 2~\rm{TeV}$ and
 scattering angles up to 0.5 mrad,  while final-state muons
 with $\theta_\mu \lsim 0.025$ will be recorded by FASER2.
 Here we present  projections for this baseline design of the experiment, FLArE10,
 as well as for a potential larger variant based on a fiducial mass of 100 tonnes,
 denoted FLArE100.
 While FLArE100 is not one of the currently proposed FPF experiments, we nevertheless
 include it for illustration purposes.
 Reconstruction of the total energy of the hadronic final state will be
 possible. 
 In terms of performance, the targets
 are $\delta E_\mu \sim5\%$ of electron energy resolution,
 $\delta \theta_e \sim 15$ mrad of electron angular  resolution,
 and $\delta E_h \sim 30$\% for the hadronic energy.
 Since the muon energy from a charged-current interaction would be measured by FASER2, the
 performance parameters for $E_\mu$ are taken to be the same.

\subsection{Differential scattering event rates}
\label{sec:pseudo-data_generation}

For each of the LHC far-forward neutrino detectors
described in Table~\ref{tab:FPF_experiments}, we generate
projections for the expected DIS structure function
measurements as follows.
We want to evaluate the number of reconstructed charged-current neutrino interaction
events taking place in the fiducial volume of the detector when divided into bins of Bjorken-$x$,
momentum transfer $Q^2$, and neutrino energy $E_\nu$, that is,
\begin{align}
	\label{eq:event_yields}
	N_{\nu_e,{\rm ev}}^{(i)}\lp \nu_e ;  E_\nu; x; Q^2\rp
	\, ,\quad i=1,\ldots, N_{\rm bin},
\end{align}
with
\allowdisplaybreaks
\begin{align}
	E_{{\rm min}}^{(i)} &\le E_\nu \le E_{{\rm max}}^{(i)}, \nonumber \\
	x_{{\rm min}}^{(i)} &\le x \le x_{{\rm max}}^{(i)} , \nonumber \\
	Q_{{\rm min}}^{2(i)} & \le Q^2 \le Q_{{\rm max}}^{2(i)}, \nonumber
\end{align}
\allowdisplaybreaks
for electron neutrinos and for each of the bins composing the measurement, and with similar
expressions applying for muon neutrinos and antineutrinos.
These event yields determine the statistical
precision associated to a measurement of the double-differential cross-sections
Eqns.~(\ref{eq:neutrino_DIS_xsec_FL}) and~(\ref{eq:antineutrino_DIS_xsec_FL}).
Subsequently, we account for the expected reconstruction performance of the detector
in order to estimate the systematic uncertainties associated to these event yields.
For simplicity, in this initial study we consider a single bin in energy
and choose log-spaced bins in $x$ and $Q^2$; the eventual optimisation
of the binning selection is left for future work.

In this calculation, we adopt the neutrino fluxes evaluated in~\cite{Kling:2021gos} and used
for the FPF simulations presented in~\cite{Feng:2022inv}.
The bin-by-bin integrated event yields in Eq.~(\ref{eq:event_yields}) are
obtained by convoluting the incoming neutrino fluxes, for a given acceptance
of the target detector, with the corresponding neutrino differential cross-sections
Eqns.~(\ref{eq:neutrino_DIS_xsec_FL}) and~(\ref{eq:antineutrino_DIS_xsec_FL}).
Binned event yields are evaluated using
\begin{align}
  \label{eq:event_yields_calculation}
   N_{\rm ev}^{(i)} =& n_T L_T\int_{Q^{2(i)}_{\rm min}}^{Q^{2(i)}_{\rm max}}\int_{x^{(i)}_{\rm min}}^{x^{(i)}_{\rm max}}\int_{E_{\rm min}^{(i)}}^{E_{\rm max}^{(i)}} \frac{dN_{\nu}(E_\nu)}{dE_{\nu}} \\ 
   &\left(\frac{d^2\sigma(x,Q^2,E_{\nu})}{dxdQ^2}\right) {\cal A}(x,Q^2,E_{\nu}) dQ^2 dx dE_{\nu} \, , \nonumber
\end{align}
with $n_T$ is the nucleon density of the target detector material, $L_T$ its
length, and ${\cal A}(x,Q^2,E_{\nu})$ is an global acceptance factor which takes the form of a
step function and accounts for the experimental acceptances
in $E_\ell$, $\theta_\ell$, and $E_h$
listed in Table~\ref{tab:FPF_experiments}.

The incoming neutrino fluxes of~\cite{Kling:2021gos} account for the geometry
and neutrino pseudo-rapidity $\eta_\nu$ coverage of the considered detector and
are encoded in Eq.~(\ref{eq:event_yields_calculation}) as $dN_{\nu}(E_\nu)/dE_{\nu}$.
This neutrino flux takes into account both the prompt component associated with neutrinos from charmed hadron decay as well as a displaced component from light hadron decays.
The prompt component used in this work was simulated at NLO using {\sc\small POWHEG}~\cite{Nason:2004rx,Frixione:2007vw,Alioli:2010xd} matched to {\sc\small Pythia8}~\cite{Sjostrand:2014zea, Bierlich:2022pfr} for the parton shower and hadronisation, and the displaced component was simulated with {\sc\small EPOS-LHC}~\cite{Pierog:2013ria}.
In comparison with the calculation from~\cite{Kling:2021gos}, in addition to the improved charmed hadron production presented in~\cite{Sominka:2023},
the neutrino flux has been updated to $\sqrt{s} = 13.6~{\rm TeV}$.

As pointed out in Ref.~\cite{Kling:2021gos} there are notable neutrino flux uncertainties, as various event generators do not agree on the forward parent hadron spectra. 
If the spread of various generators' predictions was taken as a means of flux uncertainty, corresponding to a $\lesssim 50\%$ uncertainty on the interacting muon neutrino spectrum, this would be a significant systematic if left unresolved. However, it is noteworthy that many existing predictions are yet to be tuned for the purposes of experiments such as those planned for the FPF. Nevertheless, there are projections of FPF measurements which would reduce this uncertainty to the sub-percent level already in the context of the contemporary predictions, based on parametrizing their expected correlations~\cite{Kling:2023tgr}, as well as efforts to describe the uncertainty in a data-driven way while improving the modelling of forward hadronization~\cite{Fieg:2023kld}.
However, it is important to note that forward neutrino experiments actually constrain the product of flux and cross-section, and one must be assumed to measure the other. In a full analysis,  the flux and cross-section would be constrained simultaneously in a joint measurement, utilizing their different kinematic dependences on $ x,Q^2,E_{\nu}$ and neutrino rapidity $\eta_{\nu}$.
In our study, we aim to understand the full impact of FPF data on the PDF fit, thus motivating this future joint measurement.
To this aim, we take the neutrino flux to be known and focus on the irreducible systematics associated with event reconstruction. 
With this assumption, we will show that Run 3 measurements will not be sufficient to impact PDF fits. Instead, Run 3 measurements could be used to calibrate incoming neutrino fluxes, effectively reducing the large uncertainties by the time FPF data is collected in the future. The expected reduction of FPF neutrino flux uncertainties further justifies our choice to take the neutrino flux as known.

The triple integral in  Eq.~(\ref{eq:event_yields_calculation}) is evaluated numerically by means
of Monte Carlo sampling, by generating  $N_{\rm mc}$
sampling points in the $\lp x,Q^2,E_{\nu}\rp$ space
with the constraint that
\be 
0 < y \lp = Q^2/2m_N E_{\nu }x\rp <1 \, ,
\ee
and where $N_{\rm mc}$ is chosen to be large enough such that residual Monte Carlo integration
uncertainties are negligible for all the bins considered.

Eq.~(\ref{eq:event_yields_calculation}) can be generalised to charm-production events, with
the only difference being that now the neutrino scattering cross-section is restricted
to those processes leading to final-state charm quarks.
Assuming that charm quarks can be directly tagged by the detector one has
\begin{align}
	\label{eq:event_yields_charm}
	&N_{\rm ev,c}^{(i)} = n_T L_T\int_{Q^{2(i)}_{\rm min}}^{Q^{2(i)}_{\rm max}}\int_{x^{(i)}_{\rm min}}^{x^{(i)}_{\rm max}}
	\int_{E_{\rm min}^{(i)}}^{E_{\rm max}^{(i)}} \frac{dN_{\nu}(E_\nu)}{dE_{\nu}} \\
	&\left(\frac{d^2\sigma^{\nu N \to \ell + c+X}(x,Q^2,E_{\nu})}{dxdQ^2}\right) {\cal A}(x,Q^2,E_{\nu}) dQ^2 dx dE_{\nu} \, , 
	\nonumber
\end{align}
with the charm production cross-sections discussed in~\cite{Faura:2020oom}
and references therein.
Here we neglect efficiency and acceptance effects associated to $D$-meson
tagging, which can only be properly estimated by means
of a full detector simulation.
The acceptance correction ${\cal A}(x,Q^2,E_{\nu})$ in Eq.~(\ref{eq:event_yields_charm})
applies only to the charged leptons and hence is the same as in
the inclusive case.

Detectors without charm-tagging capabilities can still be sensitive to charm production via
the semileptonic decays of the $D$-mesons, resulting in the characteristic
dimuon topology, where
\be
 N_{\rm ev,2\mu}^{(i)} \approx N_{\rm ev,c}^{(i)} \times \mathcal{B}\lp c \to D \to \mu + X\rp \, ,
 \ee
 with $\mathcal{B}$ a numerical factor that accounts for charm hadronisation and the
 resulting semileptonic decay to a muon.
 Given that $\mathcal{B}\sim 0.1$, being able to tag directly charm quarks increases the event yields
 of charm production events by a factor of 10 as compared to reconstructing the dimuon final state.

 Table~\ref{tab:integrated_rates} summarises the predicted integrated event yields for the
 detectors
considered, separated into electron neutrinos and antineutrinos
and muon neutrinos and antineutrinos.
We used the PDF4LHC21 set to compute the differential cross sections that enter in Eq.~\ref{eq:event_yields_calculation}.
As mentioned above, for FLArE we assume that muon neutrinos interacting
in its fiducial volume will be measured
by the FASER$\nu$2 spectrometer.
These event yields are computed from Eq.~(\ref{eq:event_yields_calculation}) with the
requirement that the momentum transfer and the final-state invariant mass are restricted
to the DIS region,
\be
\label{eq:DISconditions}
Q^2 \ge 2~{\rm GeV}^2\quad{\rm and}\quad  W^2 \ge 4~{\rm GeV}^2 \, .
\ee
The numbers in parenthesis indicate the event rates corresponding to charm
production, assuming heavy flavour tagging capabilities.
In the case of the FLArE detector, we display results for two proposed options with fiducial
masses of 10 and 100 tonnes.
As opposed to the number of interacting neutrinos presented
in~\cite{Feng:2022inv}, we now account for the
detector acceptances listed in  Table~\ref{tab:FPF_experiments}, which reduce
the event yields by up to a factor 2.

\begin{table}[t]
  \centering
  \small
  \renewcommand{\arraystretch}{1.70}
\begin{tabularx}{\textwidth}{X||c|c|c||c|c|c}
\toprule
Detector & $\quad$ $N_{\nu_e}$ $\quad$ &$\quad$ $N_{\bar{\nu}_e}$$\quad$   &   $N_{\nu_e} + N_{\bar{\nu}_e}$ &
$\quad$$N_{\nu_\mu}$ $\quad$ & $\quad$ $N_{\bar{\nu}_\mu}$ $\quad$  &   $N_{\nu_\mu} + N_{\bar{\nu}_\mu}$ \\
\midrule
\midrule
FASER$\nu$  & 400 (62)    & 210 (38)  & 610 (100)  &  1.3k (200)  &  500 (90)  &  1.8k (290) \\
SND@LHC  &  180 (22)  & 76 (11)    & 260 (32)   &  510 (59) & 190 (25)   &  700  (83)\\
\midrule
\midrule
FASER$\nu$2  & 116k (17k)   & 56k (9.9k)   & 170k (27k)  & 380k (53k)  & 133k (23k)    & 510k (76k)   \\
AdvSND-far  &  12k (1.5k)  & 5.5k (0.82k)   & 18k (2.3k)  & 40k (4.8k)  & 16k (2.2k)   & 56k (7k)   \\
FLArE10 & 44k (5.5k) & 20k (3.0k)   &  64k (8.5k) &  76k (10k)&   38k (5.0k) &   110k (15k) \\
FLArE100 &   290k (35k)         &    130k (19k)         &       420k (54k)       &   440k (60k)      &  232k (30k)    &  670k (90k)  \\
  \bottomrule
\end{tabularx}
\vspace{0.2cm}
\caption{\small Integrated event yields for the six detectors considered,
  separated into electron neutrinos and antineutrinos,
  muon neutrinos and antineutrinos, and their sum.
  These event yields are computed from Eq.~(\ref{eq:event_yields_calculation})
  imposing DIS kinematics, $Q^2 \ge 2$ GeV$^2$ and $W^2 \ge 4$ GeV$^2$.
 The numbers in parenthesis indicate the event rates corresponding to charm
 production, Eq.~(\ref{eq:event_yields_charm}).
 In the case of the FLArE detector, we display results for two proposed
 options with fiducual masses of 10 and 100 tonnes respectively.
  \label{tab:integrated_rates}
}
\end{table}


\begin{table}[t]
  \centering
  \small
  \renewcommand{\arraystretch}{1.70}
  \begin{tabularx}{0.99\textwidth}{p{0.26\textwidth}|c|c|c}
    \toprule
    \multirow{2}{*}{Detector} & before cuts & after DIS and acceptance cuts  & acceptance efficiency \\
    & $N_{\nu_e} + N_{\bar{\nu}_e},~N_{\nu_\mu} + N_{\bar{\nu}_\mu}$ &  $N_{\nu_e} + N_{\bar{\nu}_e},~N_{\nu_\mu} + N_{\bar{\nu}_\mu}$ &
    $N_{\nu_e} + N_{\bar{\nu}_e},~N_{\nu_\mu} + N_{\bar{\nu}_\mu}$\\
    \midrule
    \midrule
    FASER$\nu$ & 1.2k, 4.1k  & 610, 1.8k  & 51\%, 44\% \\
    SND@LHC & 280, 860 & 260, 700 & 92\%, 81\% \\
    \midrule
    \midrule
    FASER$\nu$2 & 270k, 980k   & 170k, 510k  & 63\%, 52\%\\
    AdvSND-far & 19k, 66k &  18k, 56k & 95\%, 85\%\\
    FLArE10 & 65k, 202k  & 64k, 110k & 98\%, 55\%\\
    FLArE100 & 427k, 1.3M   & 420k, 670k &98\%, 52\%  \\
    \bottomrule
  \end{tabularx}
  \vspace{0.2cm}
  \caption{The number of electron and muon neutrinos interacting within the detector volume,
    compared with the results after applying the DIS requirements  ($W^2 > 4\,{\rm GeV}^2$ and $Q^2 > 2\,{\rm GeV}^2$)
    and the experimental acceptances from Table~\ref{tab:FPF_experiments}.
    The DIS requirement removes only  $\lesssim 1\%$ of the events, a consequence of the high energy of
    LHC neutrinos.
    The last column displays the acceptance efficiency, defined as the ratio between pre- and post-acceptance
    integrated event yields.
    While the specific efficiencies depend on the experiment, up to 50\% of the neutrinos
  interacting in the detector volume may fall outside detector acceptance. }
\label{tab:acceptance}
\end{table}


Several observations can be derived from Table~\ref{tab:integrated_rates}.
First, one appreciates the large increase in statistics from
the Run III experiments to the FPF ones, with for example a factor of
$\sim$ 250 increase in the muon neutrino yield between FASER$\nu$ and
FASER$\nu$2.
Second, the muon-neutrino scattering yield dominates over electron neutrino scattering by a factor
between 2 and 3, though the precise value of this ratio is affected by the large theory uncertainties
affecting forward electron neutrino production.
Third, charm production  represents around 15\% of the inclusive
yields, with both FASER$\nu$2 and FLArE100 resulting in around
80k recorded charm-production events.
Note that this result assumes $D$-meson tagging capabilities,
and if only dimuon events can be recorded
the yields would be reduced by a factor of 10.
Fourth, FASER$\nu$2 and FLArE100 lead to the largest absolute
yields of the FPF experiments, with a total of around 680k and 1.1M (non-tau) neutrino DIS events respectively, with 74k and 170k
events expected instead for AdvSND-far and for FLArE10.

\begin{figure*}[t]
    \centering
    \includegraphics[width=0.495\textwidth]{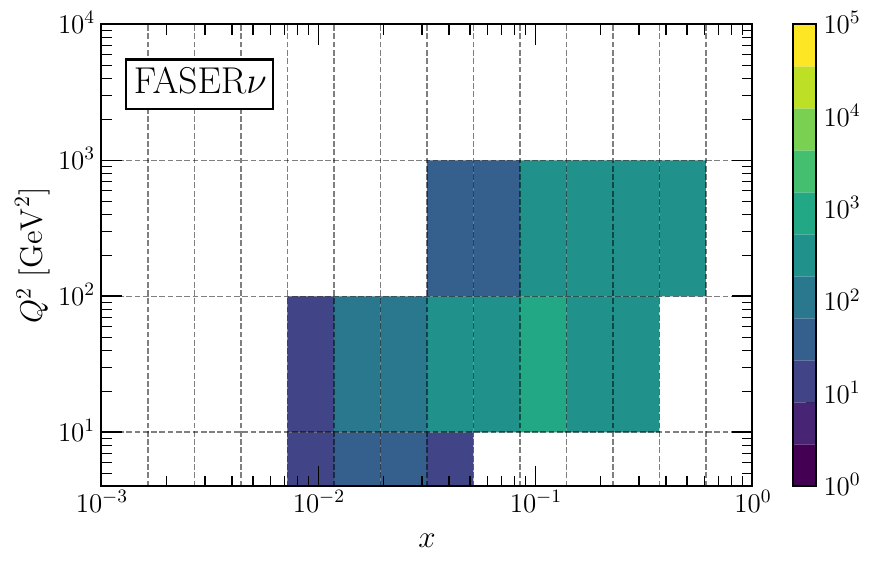}
    \includegraphics[width=0.495\textwidth]{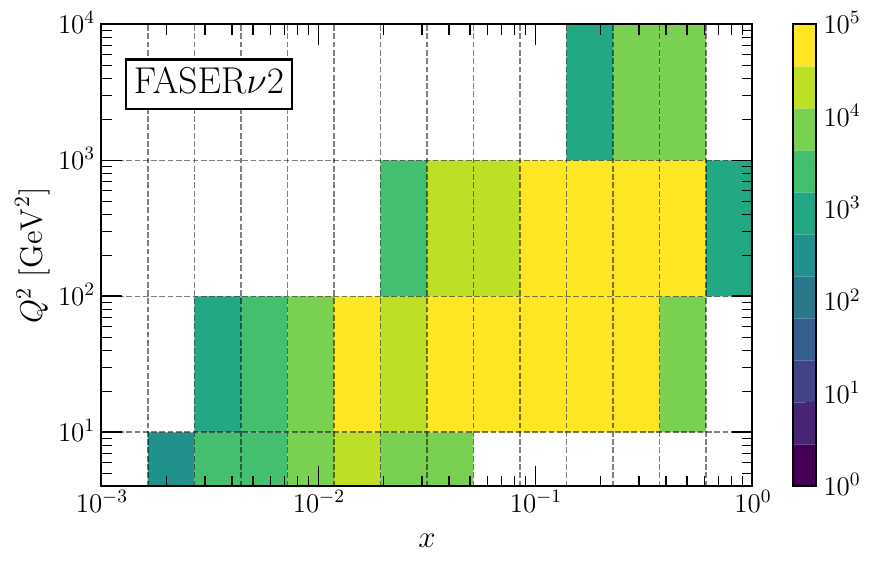}
    \includegraphics[width=0.495\textwidth]{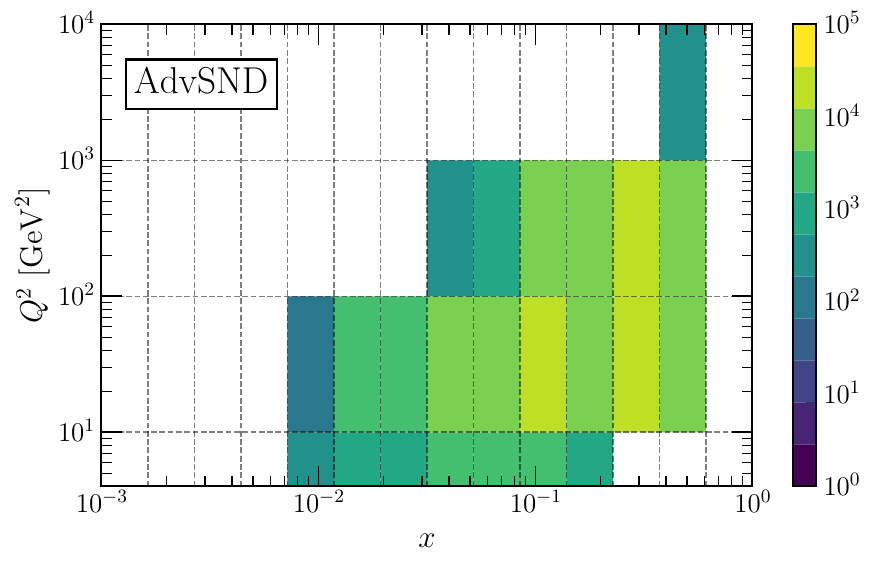}
    \includegraphics[width=0.495\textwidth]{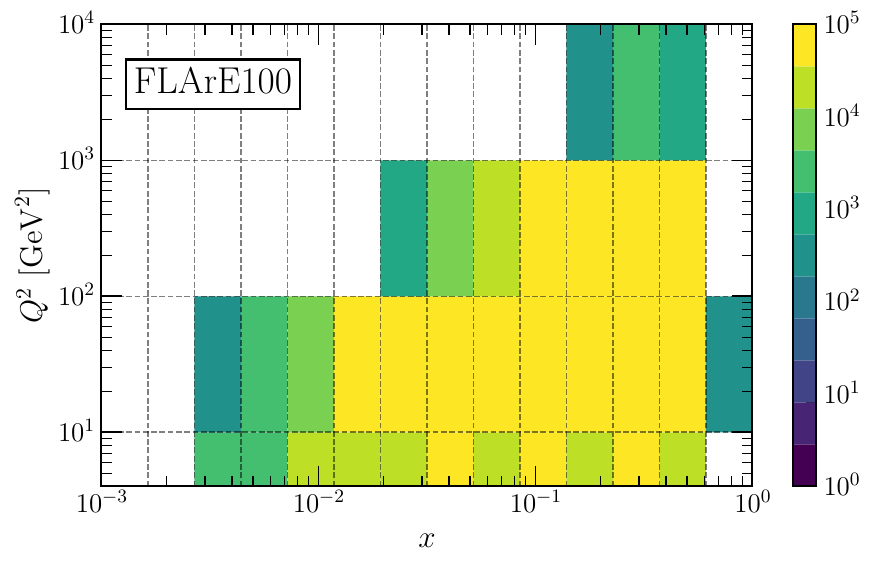}
    \caption{
    	\small The event yields per bin $N_{\rm ev}^{(i)}$,  Eq.~(\ref{eq:event_yields}), for 
    	muon-neutrino scattering at the FASER$\nu$, FASER$\nu$2, AdvSND, and FLArE100  experiments.
   		Selected events are restricted to the DIS region Eq.~(\ref{eq:DISconditions})
   		and only bins with $\ge 100$ events are retained except for FASER$\nu$ in which
   		bins with $\ge 10$ events are kept.
		Adding up the bins in each of the panels results into the numbers listed in
		Table~\ref{tab:integrated_rates}.}
    \label{fig:fasernu2_muon}
\end{figure*}

 Fig.~\ref{fig:fasernu2_muon} displays the differential
 event yields per bin,  Eq.~(\ref{eq:event_yields}),
 for muon neutrinos detected at the
 FASER$\nu$ (Run III) and the AdvSND, FASER$\nu$2, and FLArE100  (FPF) experiments,
 restricted
 to the DIS region defined by Eq.~(\ref{eq:DISconditions})
 and where only bins with $\ge 100$ events are retained, except for FASER$\nu$ in which bins with $\ge 10$ events are shown.
 Adding up the bins in each of the panels results into the inclusive yields listed in
 Table~\ref{tab:integrated_rates}.
 The clear improvement in going from the current FASER$\nu$ experiment
 to the FPF ones is visible both in terms
 of the number
 of events per bin as well as for the kinematic coverage.
 The FPF experiments benefit from large
 event rates for most of the region in $\lp x,Q^2\rp$ covered,
 leading to typical statistical uncertainties at the 1\% level or smaller,
 while for FASER$\nu$ the statistical uncertainties are larger due
 to the reduced event rates.
  
\begin{figure*}[t]
    \centering
    \includegraphics[width = 0.80\textwidth]{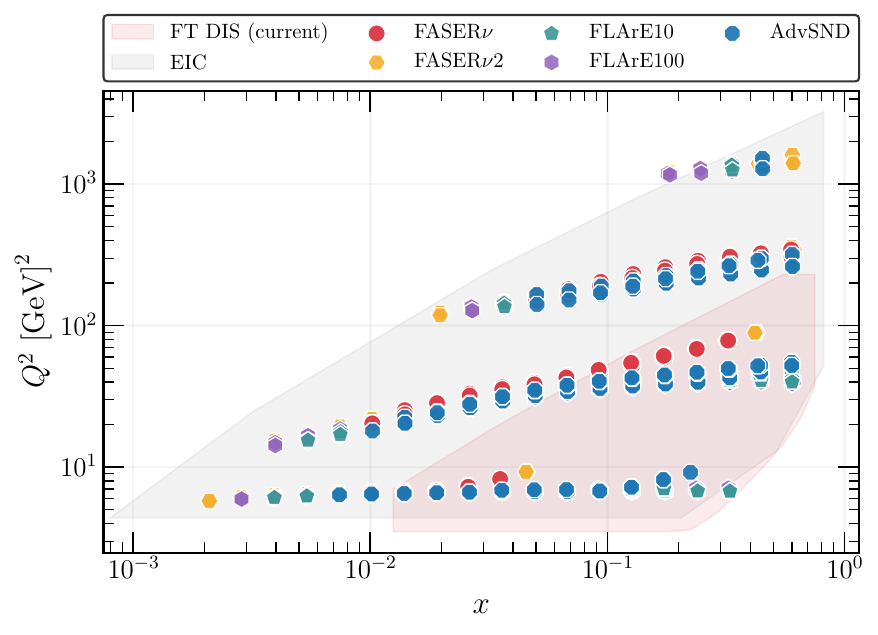}
    \caption{
      The kinematic coverage in the $(x,Q^2)$ plane of muon-neutrino scattering
      at the FASER$\nu$, FASER$\nu$2, FLArE10(100), and AdvSND experiments, see also Fig.~\ref{fig:fasernu2_muon},
      compared to that of electron-ion collisions at the EIC
      as well as to the coverage of existing neutrino
      fixed-target DIS measurements.
    }
    \label{fig:Kin_nNNPDF30_EIC_FPF}
\end{figure*}

From Fig.~\ref{fig:fasernu2_muon} one observes
how the kinematic coverage of the FPF far-forward experiments reaches down to
$x_{\rm min}\sim 3\times 10^{-3}$ at small-$x$ and up to $Q^2_{\rm max}\sim 10^4$ GeV$^2$
at large-$Q^2$, representing an extension
of around one order of magnitude in both directions as compared to available
DIS neutrino data.
To illustrate this,
Fig.~\ref{fig:Kin_nNNPDF30_EIC_FPF} compares
the kinematic coverage of FASER$\nu$, FASER$\nu$2, FLArE, and AdvSND, same as in
Fig.~\ref{fig:fasernu2_muon}, with that of electron-ion collisions
at the upcoming EIC~\cite{Khalek:2021ulf,AbdulKhalek:2021gbh} at the highest
centre-of-mass energies planned, as well as to available fixed-target
neutral- and charged-current DIS measurements.
The LHC neutrino experiments cover an $x$ region relevant
at hadron colliders for Higgs boson analyses,
precision electroweak measurements  such as the $W$-boson
mass~\cite{Amoroso:2023pey}, and new physics measurements sensitive
to the large-$x$ PDFs~\cite{Ball:2022qtp}.
FASER$\nu$2 and FLArE100 mostly overlap with the EIC coverage, providing a complementary handle
on the quark flavour decomposition in protons and heavy nuclei as compared
to the one provided by the EIC measurements.

The inclusive yields listed in Table~\ref{tab:integrated_rates}
differ from the total number of neutrinos interacting within the
detector volume~\cite{Feng:2022inv,Kling:2021gos} due to both the
DIS requirements  ($W^2 > 4\,{\rm GeV}^2$ and $Q^2 > 2\,{\rm GeV}^2$)
and the detector fiducial acceptances summarised in Table~\ref{tab:FPF_experiments}.
The former is found to be negligible, with DIS cuts removing
only  $\lesssim 1\%$ of the events, a consequence of the high energy of
LHC neutrinos.
The latter is quantified in Table~\ref{tab:acceptance}, displaying
the total number of electron and muon neutrinos before and after applying
acceptance cuts.
The acceptance efficiency, defined as the ratio between pre- and post-acceptance
integrated event yields, depends on the detector and can be up to 50\%.
For instance, FASER$\nu$2 loses 63\% and 52\% of the interacting electron and muon
neutrinos respectively due to the acceptance requirements.
From Table~\ref{tab:FPF_experiments}, is it also worth noting how FLArE has virtually
perfect acceptance for electron neutrinos.
One also observes that SND@LHC and AdvSND exhibit relatively
large acceptance efficiencies, which however are not sufficient to compensate for
the lower number of initial interacting neutrinos due to their off-axis location
as compared to FASER$\nu$(2) and FLArE.

\subsection{Statistical and systematic uncertainties}
\label{subsec:uncertainties}

The event yields displayed in Fig.~\ref{fig:fasernu2_muon} determine the associated
statistical uncertainty in each bin,
\be
\label{eq:statistical_uncertainties}
\delta^{\rm (stat)}  N_{\rm ev}^{(i)} = \sqrt{N_{\rm ev}^{(i)}} \, ,
\ee
such that the fractional statistical uncertainty per bin is $\delta^{\rm (stat)}_i=1/\sqrt{N_{\rm ev}^{(i)}}$.
Since we discard bins with less than 100 events for FPF experiments (FASER$\nu$2, FLArE, AdvSND) and
10 events for FASER$\nu$ and SND@LHC, the fractional statistical uncertainty
ranges between $\lsim 1\%$ and $\sim 30\%$, depending on the values of
$x$ and $Q^2$ associated to each bin.

\begin{figure*}[h]
    \centering
    \includegraphics[width = \textwidth]{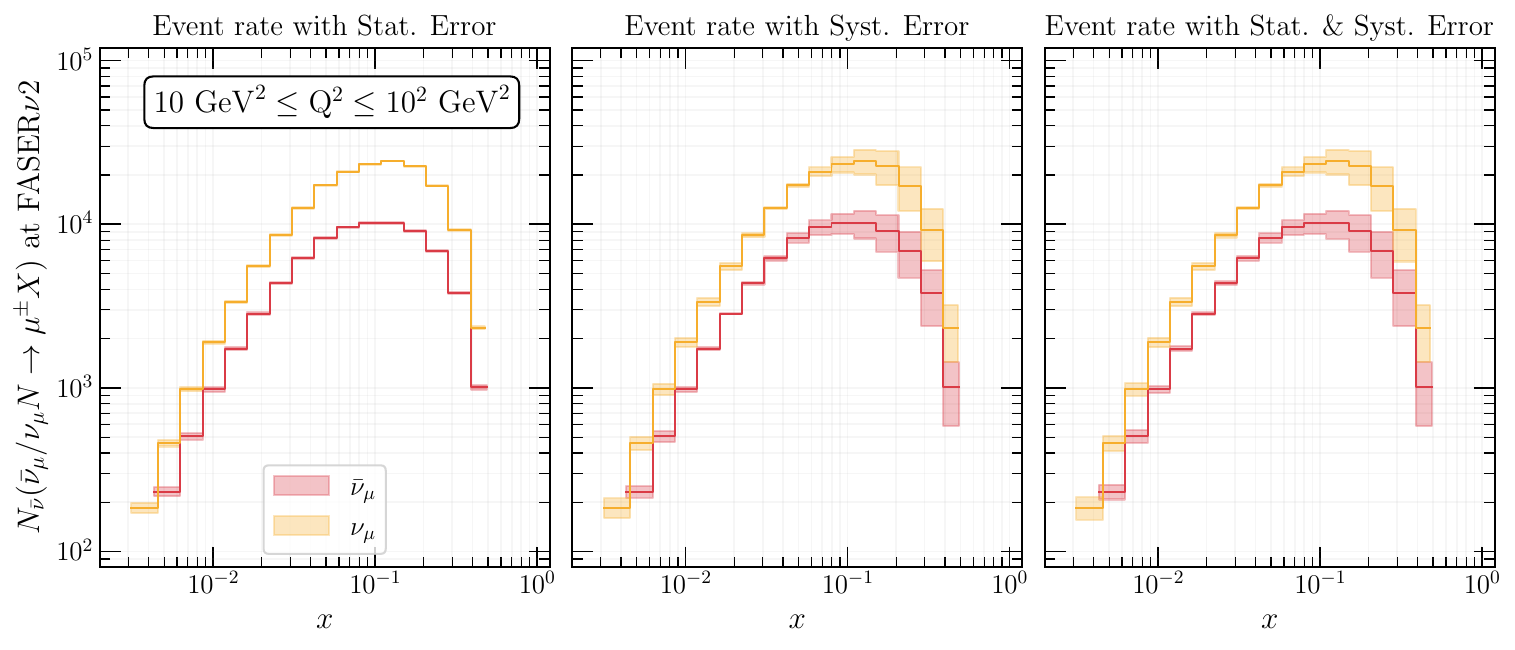}
    \caption{Same as Fig.~\ref{fig:fasernu2_muon} for FASER$\nu$2
      now as a function of $x$ after having integrated the event yields in the range $Q^2 \in [10,100]$ GeV$^2$
      for both the neutrino and antineutrinos.
      In addition to the event yield values, we also show the error bars corresponding to
      statistical errors only (left), systematic errors only (middle), and the
      sum in quadrature of the two (right panel).
      The bars in the horizontal direction indicate the width of the adopted $x$-bins.
      }
    \label{fig:error_plot_FASERv2_14}
\end{figure*}

The projected statistical uncertainties for muon-neutrino scattering
in the case of FASER$\nu$2 are displayed in the left panel
of Fig.~\ref{fig:error_plot_FASERv2_14}, which corresponds
to the same event yields as in
Fig.~\ref{fig:fasernu2_muon} for 
now as a function of $x$ after having integrated the event
yields in the range $Q^2 \in [10,100]$ GeV$^2$.
The error bar in the vertical direction indicates the statistical uncertainties, while
that in the horizontal direction corresponds to the width of the $x$-bins.
Except for the bins with the smallest values of $x$, statistical uncertainties indeed
are at the percent level or smaller for this experiment.

In addition to the statistical uncertainties evaluated from Eq.~(\ref{eq:statistical_uncertainties}),
one needs to also estimate the systematic uncertainties associated to the
finite precision in the reconstruction
of the final state leptonic and hadronic variables listed in Table~\ref{tab:FPF_experiments}. 
For instance, an event which would be classified into a given bin in $(x,Q^2,E_\nu)$ in the case
of a perfect detector may end up being
mis-classified into a different bin in the presence of systematic
shifts associated to the lepton energy $E_\ell$, lepton scattering angle $\theta_\ell$, and
hadronic energy $E_h$, as indicated by  Eq.~(\ref{eq:dis_kinematic_mapping}).

For each independent source of systematic uncertainty, which in this analysis
consists of  $\delta E_\ell$, $\delta E_h $,
and $\delta\theta_\ell$, we 
quantify its impact at the event yield level
\be
\label{eq:event_yields_systematic_error}
\delta_{\rm sys}^{(E_\ell)} N_{\rm ev}^{(i)} \, ,\quad
\delta_{\rm sys}^{(E_h)} N_{\rm ev}^{(i)}
\, ,\quad
\delta_{\rm sys}^{(\theta_\ell)} N_{\rm ev}^{(i)} \, ,\qquad i=1,\ldots,N_{\rm bin} \, ,
\ee
by extending the calculation delineated in Sect.~(\ref{sec:pseudo-data_generation}).
First, we generate a Monte Carlo set of events, denoted by $\mathcal{D}_0$,
composed by $N_{\rm mc} \approx 10^7$ samples and determine the assignment of each event
to a point in the $\lp x,Q^2,E_{\nu}\rp$ space.
We then take each event in $\mathcal{D}_0$ and smear it with Gaussian distributions whose variances are given by Table~\ref{tab:FPF_experiments} to produce a set of new samples $\{\mathcal{D}_k\}$.
The smeared events are subjected to the same DIS cuts from Eq.~(\ref{eq:DISconditions}) and acceptances from Table~\ref{tab:FPF_experiments}.
The bin assignment of the events in the smeared samples $\mathcal{D}_k$ will in general be different from those of the baseline sample $\mathcal{D}_0$.

We define the fractional uncertainty associated
to a given systematic source, say $\delta E_\ell$, for bin $i$
to be the mean of the absolute difference between the number of events in this bin for the smeared samples $\{\mathcal{D}_k\}$  and the number of events in this bin for ${\mathcal{D}_0}$:
\be
\delta_{\rm sys}^{(E_\ell)} =\la \left|  \frac{{N}^{(i)}_{E_\ell-{\rm smeared},k} -N_0^{(i)}}{N_0^{(i)}}\right|\ra \, .
\ee
The absolute systematic uncertainty in event yield caused by $\delta E_\ell$ is then $\delta_{\rm sys}^{(E_\ell)} N_{\rm ev}^{(i)}$.
Individual sources of systematic errors are treated as uncorrelated among them, and hence
by producing samples where only one source of error is varied at a time
we can determine the systematic errors, Eq.~(\ref{eq:event_yields_systematic_error}), in each bin
for each of the considered experiments.
This approach has the benefit that rescaling individual sources of systematic
uncertainties, say to assess the impact of improved detector performance,
becomes straightforward. 

\begin{figure*}[t]
  \centering
  \includegraphics[width=\textwidth]{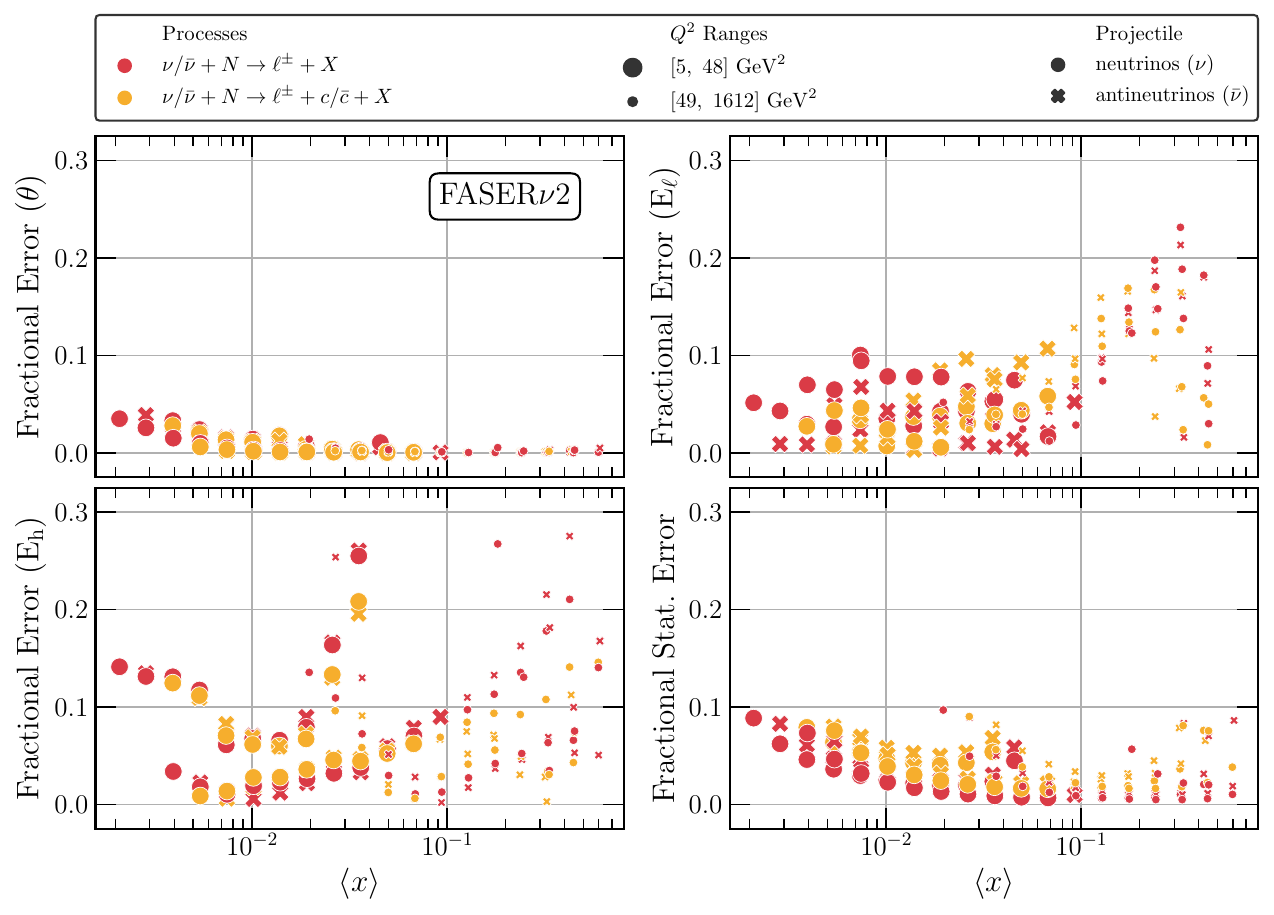}
  \caption{\small Estimated systematic uncertainties for the  measurements
    of the double-differential
    muon-neutrino scattering cross-section at FASER$\nu$2.
    We consider the systematic errors
    associated to the charged-lepton energy $E_\ell$ and scattering angle $\theta_\ell$
    and to the hadronic energy $E_h$.
    The size of each source of systematic error is plotted as a function
    of the average momentum fraction per bin $\la x\ra$
    in two different bins of $Q^2$.
    We indicate separately the results for neutrino and antineutrino projectiles as well as
    those associated to inclusive and to charm production measurements.
    For completeness, we display in the bottom-right panel the corresponding
    statistical uncertainties in the same bins.
  }
  \label{fig:percentage_uncertainties_overview}
\end{figure*}

Fig.~\ref{fig:percentage_uncertainties_overview}
displays the projected systematic uncertainties associated
to $E_\ell$, $\theta_\ell$, and $E_h$ 
for the  measurements of the double-differential
muon-neutrino scattering  at FASER$\nu$2. We note that there is a further systematic uncertainty associated to the overall neutrino flux, but we do not include this uncertainty in our estimation.
The magnitude of each systematic error is plotted as a function
of the average momentum fraction per bin $\la x\ra$
in two different bins of $Q^2$.
We indicate separately the results for neutrino and antineutrino projectiles as well as
those associated to inclusive and to charm production measurements.
For completeness, we also display in the bottom-right panel the corresponding
statistical uncertainties in the same bins.

In the case of FASER$\nu$2, the systematic uncertainties associated to the final state energies $E_\ell$ and $E_h$ are comparable and are in general at the $10\%$ level, ranging
from $1\%$ to  $30\%$ across bins.
The uncertainties associated with $\theta_\ell$ are typically at or well below the $1\%$ level
for these experiments, which can be attributed to their good spatial
resolution. 
On the other hand, for the FLArE experiment, uncertainties associated with the scattering angle $\theta_\ell$ dominate, being at the $ 40\%$ level and ranging from a few percent to ${\cal O}(1)$,
whereas the uncertainties associated with final-state energy reconstruction
are smaller, around the $10\%$ level, and range from $\lesssim 1\%$ to up to around $80\%$
depending on the bin.
For all experiments, statistical uncertainties are subdominant in most of the bins and are below the 10\% level,
especially at large-$x$ which is the kinematic region benefitting from the largest event rates.

Fig.~\ref{fig:error_plot_FASERv2_14} also displays
the integrated event yields for FASER$\nu$2 as a function of $x$ for only
systematic errors and for the sum in quadrature of statistical and systematic errors
(central and right panels respectively).
For most of the bins, for the baseline performance assumptions the total systematic
uncertainty dominates over the statistical uncertainties, with the possible exception
of the small-$x$ region where statistical and systematic errors are of comparable size.

The end result of the procedure is an estimate of the statistical and systematic uncertainties
for each bin of the measurement, from which an experimental covariance matrix can be constructed as
\begin{align}
	\label{eq:covmat_definition}
	{\rm cov}_{ij} =\delta_{ij} & \lp \delta^{\rm (stat)}  N_{\rm ev}^{(i)}\rp^2
	+ \sum_{k=1}^{n_{\rm sys}}\lp \delta_{\rm sys}^{(k)} N_{\rm ev}^{(i)} \rp \lp \delta_{\rm sys}^{(k)} N_{\rm ev}^{(j)} \rp,
	\nonumber\\ 
	& \text{ for } i,j=1,\ldots,N_{\rm bin} \, ,
\end{align}
and the same for the associated correlation
matrix of the measurement
\be
\label{eq:corrmat_definition}
\rho_{ij} =  \frac{{\rm cov}_{ij}}{\sqrt{ {\rm cov}_{ii} }\sqrt{ {\rm cov}_{jj} } } \, . 
\ee
The relative covariance matrix, $ {\rm cov}_{ij}/( N_{\rm ev}^{(i)}N_{\rm ev}^{(j)})$, is
independent of the considered observable and would also apply
for the double-differential cross-sections Eqns.~(\ref{eq:neutrino_DIS_xsec_FL}) and~(\ref{eq:antineutrino_DIS_xsec_FL}) which are related to the event yields by a constant prefactor.

The experimental covariance matrix constructed as per
Eq.~(\ref{eq:covmat_definition}) assumes that each source of systematic
uncertainty is 100\% correlated across all the bins of the measurement.
In the real experiment, the actual covariance matrix will be
composed by a large number of uncertainty sources, with typical
HERA and LHC precision measurements characterised by up to hundreds
of different sources of systematic error.
In particular, the assumption that a single source of systematic error, say $\delta E_\ell$,
is fully correlated among all the bins in $(x,Q^2)$ is unlikely to be accurate.

For this reason, following the HL-LHC projection strategy of~\cite{AbdulKhalek:2018rok},
here we neglect bin-by-bin correlations
and add in quadrature statistical and systematic errors,
\begin{align}
	\label{eq:covmat_definition_v2}
	{\rm cov}_{ij} &= \delta_{ij} \lp \delta^{\rm (stat)}  N_{\rm ev}^{(i)}\rp^2
	+ \delta_{ij}\lp f_{\rm corr}\rp^2\sum_{k=1}^{n_{\rm sys}} \lp f_{\rm red}^{(k)}\rp^2 \times
	\nonumber \\
	& \lp \delta_{\rm sys}^{(k)} N_{\rm ev}^{(i)} \rp^2 \text{ for } i,j=1,\ldots,N_{\rm bin} \, .
\end{align}
In Eq.~(\ref{eq:covmat_definition_v2}) we have introduced a parameter $f_{\rm red}^{(k)}\le 1$
that gauges the impact of a possible reduction of the $k$-th systematic error
as compared to the default experiment performance (reproduced with $f_{\rm red}=1$).
Furthermore, $f_{\rm corr}$ represents an effective correction factor that accounts for the fact that data with correlated
systematics may be more constraining than the same data where each source of error is simply
added in quadrature as we do here.
A value of $f_{\rm corr} \sim 0.5$, obtained from the inspection of available measurements
 for which the full information
 on correlated systematics is available, was estimated in ~\cite{AbdulKhalek:2018rok}
 and we adopt the same choice here.

 In  Sect.~\ref{sec:protonPDFs} we present results both for $f_{\rm red}=1$ (conservative)
 and $f_{\rm red}=0.5$ (optimistic scenario), though we note that it would be straightforward to revisit
 the projections for other assumptions of the detector performance.
 Specifically, $f_{\rm red}=0.5$ is assumed for the Hessian profiling
 of PDF4LHC21 and EPPS21 while $f_{\rm red}=1$ for the NNPDF4.0-based fits.

 \subsection{Pseudo-data generation}

 In order to generate pseudo-data for double-differential
 LHC neutrino scattering cross-sections, we follow the procedure
 used for the HL-LHC projections of~\cite{AbdulKhalek:2018rok} which was
 also adopted in~\cite{Ethier:2021ydt} and~\cite{Greljo:2021kvv} for SMEFT impact projections
 of vector-boson scattering and high-mass Drell-Yan data at the HL-LHC, respectively.
 The starting point are predictions for inclusive and charm-tagged
 differential neutrino scattering cross-section, denoted generically by
 \be
 \label{eq:theory_dis_projections}
 \mathcal{O}_i^{{\rm (th)}} \equiv \frac{d^2\sigma^{\nu N}(x_i,Q^2_i,y_i)}{dxdy} \, ,\quad
 i=1,\ldots,N_{\rm bin} \, ,
 \ee
 with $(x_i,Q^2_i,y_i)$ labeling the corresponding bin centres.
 The observables $\mathcal{O}_i $ in Eq.~(\ref{eq:theory_dis_projections})
are evaluated with 
{\sc\small YADISM}~\cite{yadism,Candido:2023utz}
interfaced to {\sc\small PineAPPL}~\cite{Carrazza:2020gss, christopher_schwan_2023_7995675}
to return a fast interpolation grid admitting a generic PDF input,
and with DGLAP evolution effects provided by {\sc\small EKO}~\cite{Candido:2022tld}.
DIS structure functions are evaluated at NNLO in the QCD expansion
and account both for heavy quark and target mass effects.
No higher-twists corrections are included.
In particular,
charm structure functions are evaluated in the FONLL general-mass variable-flavour-number
scheme~\cite{Forte:2010ta,Ball:2011mu,Faura:2020oom} at $\mathcal{O}\lp \alpha_s\rp$
accuracy.
For the proton PDF fits, we assume a free isoscalar target $N$, while
for the nuclear PDF one we allow for deviations from isoscalarity relevant
for a tungsten nucleus.

To ensure consistency, the PDF set and other theory settings, such as the perturbative
order and heavy quark scheme, adopted for the evaluation of
Eq.~(\ref{eq:theory_dis_projections}) should be the same as those
used in the fitting framework assessing their impact.
For instance, when using the {\sc\small xFitter} profiling of PDF4LHC21, one needs
to generate  neutrino structure functions also using PDF4LHC21 as input.
This ensures that the generated pseudo-data is coherent with the prior PDF
set used as baseline and avoids introducing artificial inconsistencies 
compromising the validity of the projection studies.

The central values for the pseudo-data, denoted by $\mathcal{O}_i^{{\rm (exp)}} $, are obtained
by fluctuating the reference theory prediction Eq.~(\ref{eq:theory_dis_projections})
by the corresponding fractional statistical and systematic
uncertainties,
\begin{equation}
	\label{eq:pseudo_data_v2}
	\mathcal{O}_i^{{\rm (exp)}}
	= \mathcal{O}_i^{{\rm (th)}}
	\left( 1+ r_i \delta_i^{\rm tot}
	\right) \,
	, \qquad i=1,\ldots,N_{\rm bin} \, ,
\end{equation}
where the total experimental uncertainty is the sum in quadrature of
statistical and systematic errors, accounting for a possible reduction
factor in the latter,
\begin{align}
	\delta_{i}^{\rm tot}
	= & \left( \left( \delta_i^{\rm stat}\right)^2 + \sum_{k=1}^{n_{\rm sys}}
	\left( f_{\rm corr} \times f_{\rm red}^{(k)} \times \delta_{i,k}^{\rm sys} \right)^2\right)^{1/2} \, , \nonumber\\
	& \text{ for } i=1,\ldots,N_{\rm bin} \, ,
\end{align}
and with $r_{i}$ being univariate Gaussian random numbers. 
As mentioned above, $f_{\rm red}^{(k)}$ is a reduction factor modelling
improvements in the experimental performance as compared to the baseline
settings summarised in Sect.~\ref{tab:FPF_experiments}.
The pseudo-data generated by means of Eq.~(\ref{eq:pseudo_data_v2}),
together with the corresponding covariance matrix computed according to Eq.~\eqref{eq:covmat_definition_v2},
define then the inputs of the subsequent proton and nuclear PDF determinations.

 \subsection{PDF impact assessment}
 \label{subsec:pdf_impact_assessment}

We consider two complementary approaches to assess the
impact of the projected LHC neutrino data on the proton and nuclear PDFs.
First, the Hessian profiling\cite{Paukkunen:2014zia, Schmidt:2018hvu, AbdulKhalek:2018rok, HERAFitterdevelopersTeam:2015cre} of prior proton and
nuclear PDF sets, taken to be PDF4LHC21~\cite{PDF4LHCWorkingGroup:2022cjn} and
EPPS21~\cite{Eskola:2021nhw} respectively.
Second, the direct inclusion 
in the NNPDF global analysis framework~\cite{NNPDF:2021uiq,NNPDF:2021njg}.

The profiling method applied to Hessian PDF fits is based
on minimising a goodness-of-fit error function defined as
\begin{align}
\chi^2 = 
\sum_{i=1}^{N_{\textrm{bin}}} &
\frac{\left(  \mathcal{O}_i^{\rm (exp)}
            + \Gamma_i^{\alpha,\textrm{exp}}
              b_\alpha^{\textrm{(exp)}}
            - \mathcal{O}_i^{\rm (th)}
            - \Gamma_i^{\beta,\textrm{th}}
              b_\beta^{(\textrm{th})}
     \right)^2
     }{ \left(\delta^{{\rm (stat)}}\mathcal{O}_i^{\rm (th)}\rp^2 } \nonumber\\
+ &\sum_\alpha \lp b_\alpha^{(\textrm{exp})}\rp^2
+ T^2 \sum_\beta  \lp b_\beta^{\textrm{(th)}}\rp ^2 \, ,
\label{eq:profilingchi2}
\end{align}
with the pseudodata 
$\mathcal{O}_i^{\rm (exp)}$ defined in  Eq.~(\ref{eq:pseudo_data_v2}).
The correlated uncertainties for the pseudodata and for the theoretical prediction 
are contained in the nuisance parameter vectors $b^{(\textrm{exp})}$ and $b^{(\textrm{th})}$, respectively, with $T$ the tolerance factor, and the total uncorrelated uncertainty is $\delta^{{\rm (stat)}}\mathcal{O}_i^{\rm (th)}$.

The effect of the nuisance parameters
on the observables $\mathcal{O}_i^{\rm (exp)}$ and $\mathcal{O}_i^{\rm (th)}$
is described by the matrices $\Gamma_i^{\textrm{exp}}$ and $\Gamma_i^{\textrm{th}}$.
The indices $\alpha$ and $\beta$ then run over the uncertainty nuisance parameters for the pseudodata and the theoretical prediction, respectively.
The nuisance parameter values $b_\beta^{\textrm{(th,min)}}$ that minimize Eq.~\eqref{eq:profilingchi2} give the central PDFs $f'_0$ optimized to the profiled dataset in the form
\begin{equation}
f_0' = f_0
      + \sum_\beta b_\beta^{\textrm{(th,min)}} 
        \left(  \frac{f_\beta^+   -  f_\beta^- }{2}
              -    b_\beta^{\textrm{(th,min)}}
                \frac{f_\beta^+ + f_\beta^- - 2f_0}{2}
        \right),
\end{equation}
where $f_0$ is the original central PDF and the up and down variation eigenvectors are given by $f^+, f^-$.
The reduction in the uncertainties of the profiled PDFs indicate the impact
of the projected data with respect to the assumed prior PDF set.

The profiling studies carried out in this work are performed using version 2.2.1
of the 
{\sc\small xFitter} open-source QCD analysis framework~\cite{Alekhin:2014irh, Bertone:2017tig, xFitter:2022zjb, xFitter:web}.
To this end, a new interface between  {\sc\small PineAPPL} and {\sc\small xFitter} has been developed and is made available in {\sc\small xFitter}.
All the experimental and theoretical data files used in the analysis, including
the  {\sc\small PineAPPL}  grids, are available
from the public {\sc\small xFitter} repository.
For the proton PDF profiling, a tolerance of $T^2 = 10$ is adopted,
which  corresponds approximately to the average tolerance
used in the CT18~\cite{Hou:2019efy} and MSHT20~\cite{Bailey:2020ooq} determinations,
the two Hessian sets entering the PDF4LHC21 combination~\cite{PDF4LHCWorkingGroup:2022cjn}, for
one-sigma PDF uncertainties.
For the Hessian profiling of EPPS21, a value of $T^2 = 33$ is used consistently with~\cite{Eskola:2021nhw}, and the resulting uncertainties are scaled down by a factor of 1.645 to obtain 68\% confidence level intervals.

Concerning the inclusion of the LHC neutrino pseudo-data
in the NNPDF proton analysis framework, we follow the procedure
outlined in~\cite{NNPDF:2021uiq}.
Fast interpolation tables (FK-tables)~\cite{Ball:2010de} combining {\sc\small EKO}
DGLAP evolution with {\sc\small YADISM} DIS coefficient functions
are computed using {\sc\small PineAPPL}, see also~\cite{Barontini:2023vmr}.
Predictions for the fitted observables Eq.~(\ref{eq:theory_dis_projections}) are evaluated
with NNPDF4.0 NNLO and included in the corresponding PDF determination
alongside all other datasets already present in the global fit.
We verify that in all cases $\chi^2/n_{\rm dat}\sim 1$ after the fit,
as expected given the built-in consistency between the prior PDF fit
and the generated pseudo-data.

\section{Constraints on proton and nuclear structure}
\label{sec:protonPDFs}

By following the strategy outlined in Sect.~\ref{sec:dis_pseudodata}, we 
quantify the impact on the proton and nuclear PDFs of differential  DIS
cross-section measurements obtained with the  LHC neutrino beam. 
Here we present results first for the Hessian profiling of the PDF4LHC21,
then for the Monte Carlo fit NNPDF4.0, and finally for the nuclear PDFs of EPPS21, also
by means of profiling.

First of all, we study the constraints on the PDFs 
provided by FASER$\nu$ during the LHC Run III data taking period,
showing that they are not able to improve the determination of PDFs.
We then move to study the impact of the FPF experiments, focusing
on FASER$\nu$2.
We study the stability of the results with respect to the inclusion of systematic uncertainties,
charm-tagged data, and lepton-charge separation.
We also compare the impact of the different FPF experiments separately and provide
results for their combination.

\begin{figure*}[ht]
	\centering
	\includegraphics[width=0.32\textwidth]{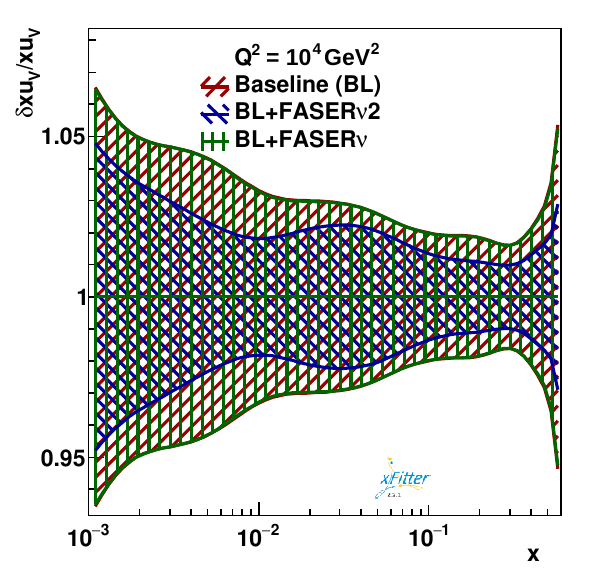}
	\includegraphics[width=0.32\textwidth]{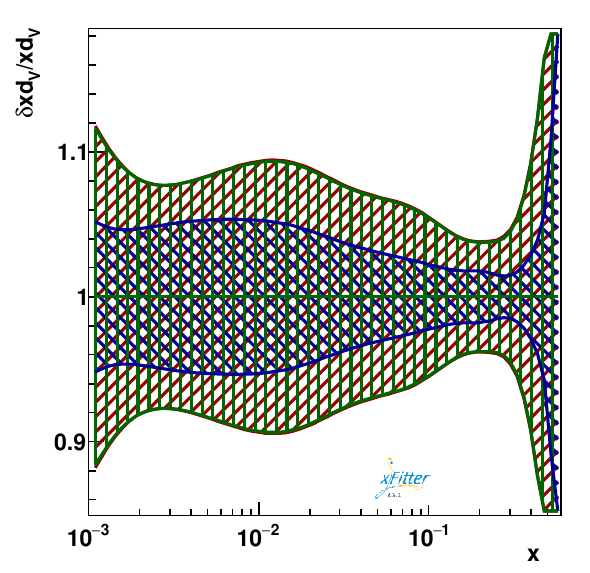}
	\includegraphics[width=0.32\textwidth]{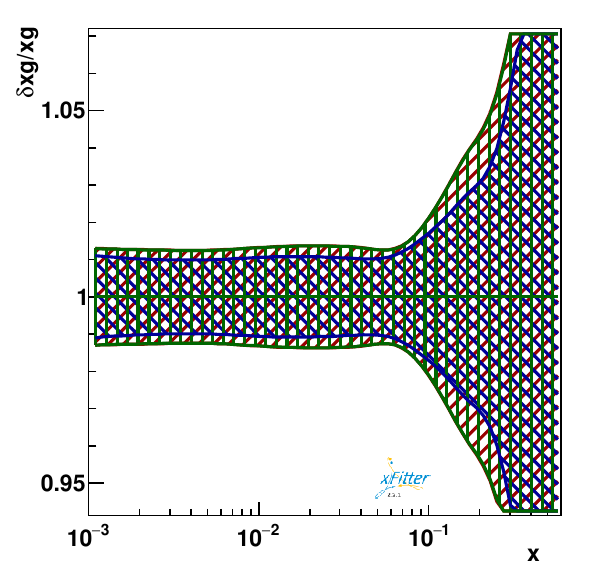}\\
	\includegraphics[width=0.32\textwidth]{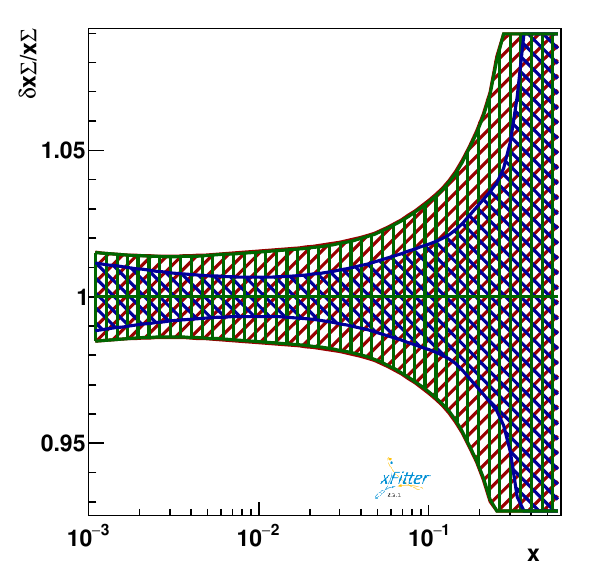}
	\includegraphics[width=0.32\textwidth]{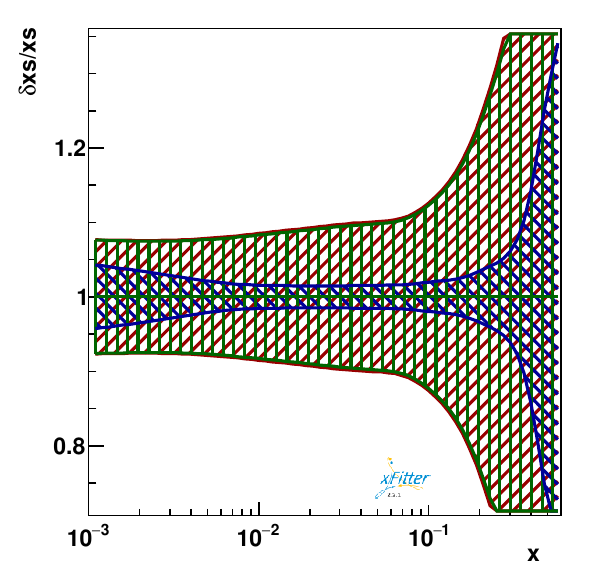}
	\caption{
The fractional PDF uncertainties (at the 68\% CL) at $Q^2 = 10^4 \, \textrm{GeV}^2$ 
for the up and down valence quarks, gluon, total quark singlet, and total strangeness PDFs
in the PDF4LHC21 baseline, compared to the results obtained once the FASER$\nu$
and FASER$\nu$2 structure functions are included in the fit.
In both cases we include charm-tagged structure functions
and assume final-state lepton-charge separation.
The FASER$\nu$ projections are based on a Run III integrated
luminosity of $\mathcal{L}=150$ fb$^{-1}$.
	}
	\label{fig:FASERnu2_vs_FASERnu}
\end{figure*}

\subsection{Proton PDFs: impact on PDF4LHC21}
\label{sec:pdf4lhc21}

We begin with the Hessian profiling of
the PDF4LHC21 set.
This proton PDF set is a Monte Carlo combination~\cite{Watt:2012tq,Carrazza:2015hva} of three global PDF sets, CT18~\cite{Hou:2019efy},
MSHT20~\cite{Bailey:2020ooq}, and NNPDF3.1~\cite{NNPDF:2017mvq}.
Its Hessian representations are obtained by means of the reduction methodologies developed in~\cite{Gao:2013bia,Carrazza:2015aoa,Carrazza:2016htc}.
Being based on the combination of three modern global PDF fits, PDF4LHC21 provides a conservative estimate
of  current uncertainties associated to our understanding of proton PDFs.
We profile PDF4LHC21 with pseudodata from various LHC neutrino experiments,
and study the stability of the results with respect to variations in the profiling inputs.

\paragraph{Impact of the FASER$\nu$ Run III measurements.}
Fig.~\ref{fig:FASERnu2_vs_FASERnu} shows the
fractional uncertainties (at the 68\% confidence level) at $Q^2 = 10^4 \, \textrm{GeV}^2$ 
for the up and down valence quarks, gluon, total quark singlet, and total strangeness PDFs
in the PDF4LHC21 baseline, compared to the results obtained once the  FASER$\nu$
and FASER$\nu$2 structure functions are
added by means of Hessian profiling.
In both cases we include charm-tagged structure functions
and assume final-state lepton-charge separation.
The FASER$\nu$ projections are based on a Run III integrated
luminosity of $\mathcal{L}=150$ fb$^{-1}$.
As indicated by Table~\ref{tab:integrated_rates}, by the end of Run III, one expects
that FASER$\nu$ will have recorded around 600 and 1800 electron- and muon-neutrinos
respectively corresponding to deep-inelastic scattering events, of which around 100
and 300 respectively are associated to charm-tagged events.
We display results for the profiling in which the experimental covariance matrix
considers only statistical uncertainties.
We restrict the comparisons to the region $10^{-3}\lsim x \lsim 0.7$ covered
by the LHC neutrino experiments (see also Fig.~\ref{fig:Kin_nNNPDF30_EIC_FPF}).

From Fig.~\ref{fig:FASERnu2_vs_FASERnu}
we find that neutrino DIS measurements at FASER$\nu$ are unable to improve
PDF uncertainties as compared to the baseline scenario encapsulated by PDF4LHC21.
The reason is two-fold: the smaller event rates as compared to FASER$\nu$2,
and the reduced coverage of the $(x,Q^2)$ phase space shown in Fig.~\ref{fig:fasernu2_muon}.
The differences in PDF sensitivity between FASER$\nu$ and FASER$\nu$2 in
Fig.~\ref{fig:FASERnu2_vs_FASERnu} illustrate
the  importance of realising the FPF in order to exploit the full physics potential enabled 
by LHC neutrinos for QCD and hadron structure studies.

It should be emphasized that the lack of PDF sensitivity displayed in Fig.~\ref{fig:FASERnu2_vs_FASERnu}
does not imply that measuring DIS structure functions at FASER$\nu$ will not provide useful information.
First of all, our procedure assumes perfect compatibility between current data and the projected DIS
neutrino measurements at the LHC, something which remains to be demonstrated experimentally.
Second, FASER$\nu$ still covers a region of neutrino energies unexplored by previous experiments, hence providing
a new validation of our QCD calculations and of neutrino interactions at the TeV scale.
Third, it would represent a non-trivial proof-of-concept that LHC neutrino differential measurements can be
unfolded to the cross-section level to be used in  theoretical
interpretations, paving the way and demonstrating the feasibility of subsequent neutrino DIS
measurements at the FPF experiments.

\begin{figure*}[t]
\centering
\includegraphics[width=0.32\textwidth]{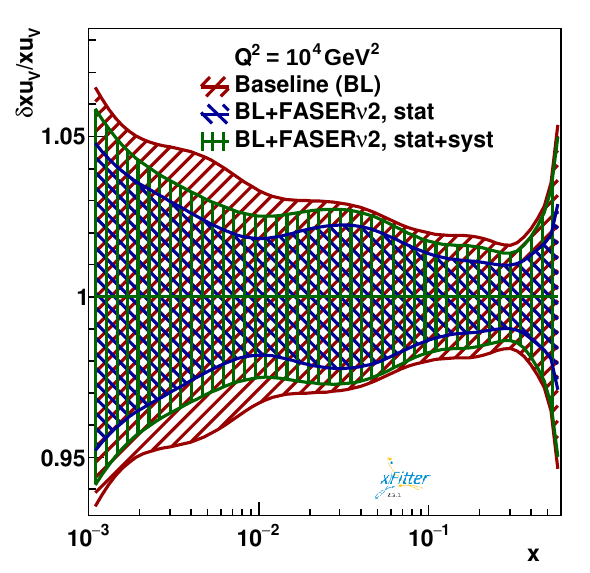}
\includegraphics[width=0.32\textwidth]{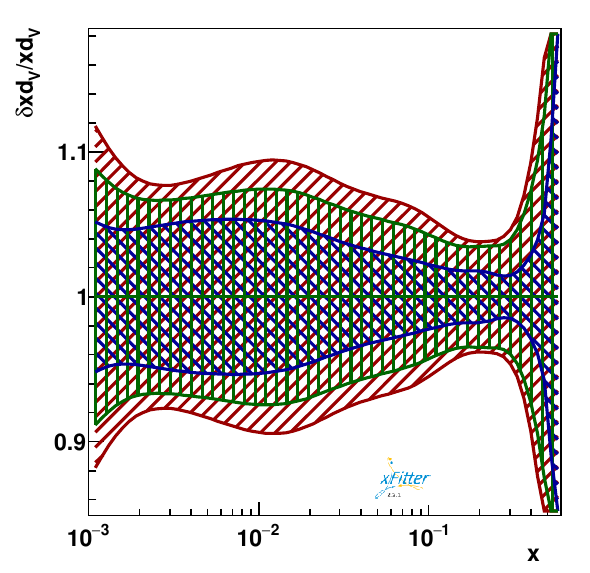}
\includegraphics[width=0.32\textwidth]{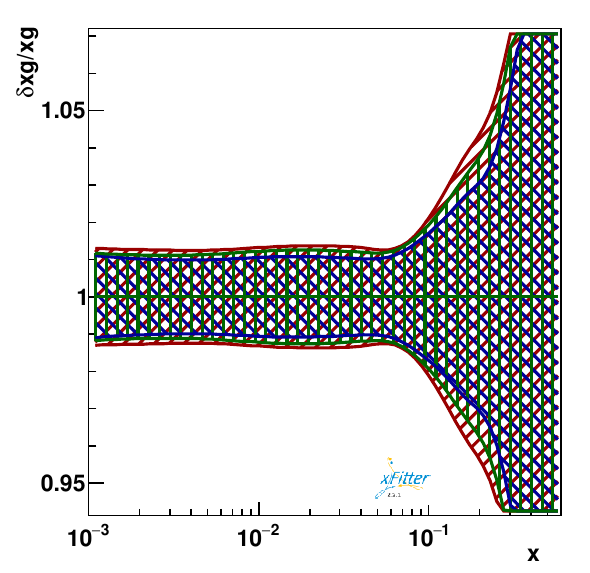}\\
\includegraphics[width=0.32\textwidth]{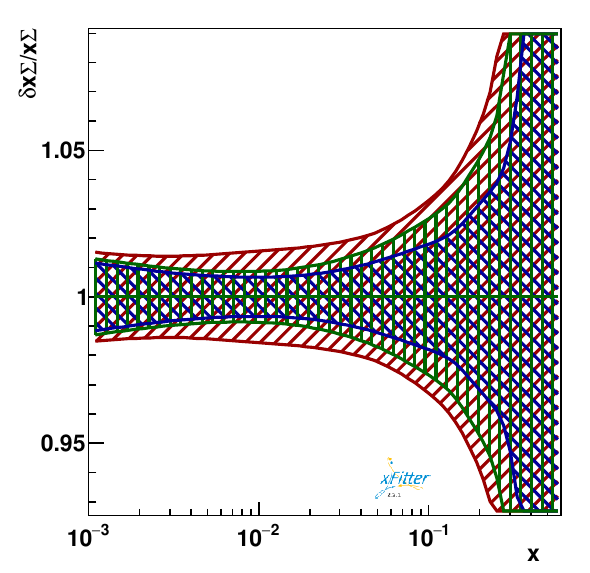}
\includegraphics[width=0.32\textwidth]{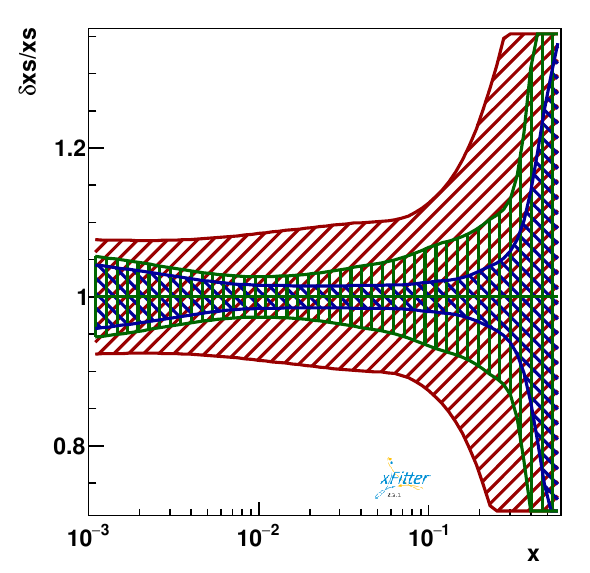}
\caption{Same as in Fig.~\ref{fig:FASERnu2_vs_FASERnu} comparing
  the FASER$\nu$2 impact projections without and with systematic
uncertainties accounted for in the experimental covariance matrix.
}
\label{fig:FASERnu2_baseline}
\end{figure*}

\paragraph{Constraints from FASER$\nu$2.}
Fig.~\ref{fig:FASERnu2_baseline} shows the
same comparison as in as in Fig.~\ref{fig:FASERnu2_vs_FASERnu}, now for the
  the FASER$\nu$2 impact projections without and with systematic
  uncertainties accounted for in the experimental covariance matrix,
  following
the procedure spelled out in Sect.~\ref{subsec:uncertainties}.
The FASER$\nu$2 pseudo-data accounts for  both  inclusive and charm-tagged structure functions
and assumes outgoing lepton- charge identification.
We display results for the profiling in which the experimental covariance matrix
considers only statistical uncertainties, as well as for the scenario where
statistical and systematic errors are added in quadrature.
In addition of a reduction of PDF uncertainties, the Hessian profiling
also results in general in a shift in the PDF central values.
This shift is however arbitrary, since it 
depends on the fluctuations
entering the pseudo-data generation, and is hence ignored in the following.

Inspection of Fig.~\ref{fig:FASERnu2_baseline} reveals that, as opposed
to the FASER$\nu$ impact projections, measurements of DIS structure
functions at FASER$\nu$2 would reduce PDF uncertainties on the quark and antiquark
PDFs, while leaving
the gluon essentially unaffected.
As expected for a neutrino scattering experiment, its impact is most marked for
those PDF combinations sensitive to quark flavour separation such as
the up and down valence PDF as well as the total strangeness.
Indeed, the reduction of PDF uncertainties is particularly significant for the latter,
a consequence of the inclusion of charm-tagged structure functions in the fit.
Given that all PDF determinations entering PDF4LHC21 already include existing neutrino
DIS measurements, the fact that FASER$\nu$2 pseudo-data still manages to improve
uncertainties highlights the new information provided by the LHC neutrino experiments.

By comparing the impact of the FASER$\nu$2 structure functions
in the case where only statistical errors are considered with that
where also systematic uncertainties are accounted for,
one finds that the latter eventually become a limiting factor,
but also that they not modify the qualitative findings of the statistics-only scenario.
Indeed, while systematic uncertainties somewhat degrade the PDF sensitivity,
they do not wash it out for any of the quark PDFs.
Furthermore, in the projections presented in this work,
we assume the performance parameters of  Table~\ref{tab:FPF_experiments}, which
however could be improved in the actual realisation of the experiments,
using for instance detector improvements or different kinematic reconstruction techniques.
We also note that the availability of complementary experiments accessing the same neutrino
beam should allow their mutual cross-calibration, such that the combination of their
data brings in more information than just the naive statistics scaling.

\begin{figure*}[t]
\centering
\includegraphics[width=0.32\textwidth]{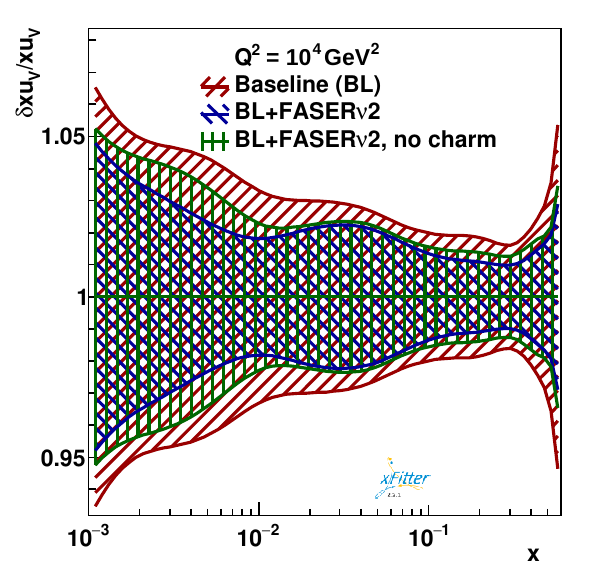}
\includegraphics[width=0.32\textwidth]{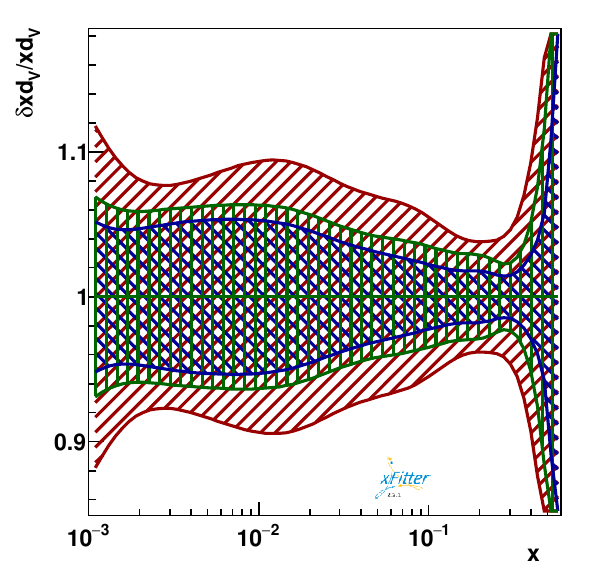}
\includegraphics[width=0.32\textwidth]{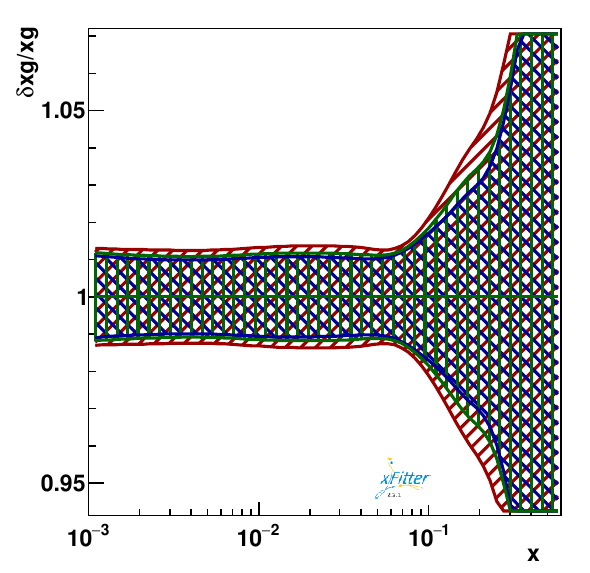}\\
\includegraphics[width=0.32\textwidth]{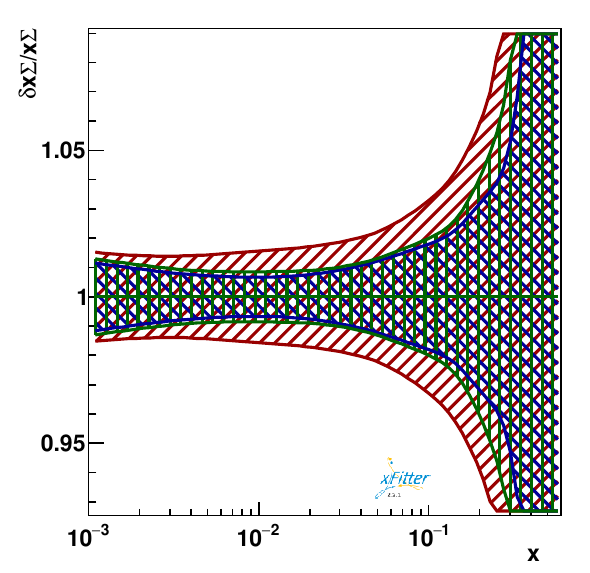}
\includegraphics[width=0.32\textwidth]{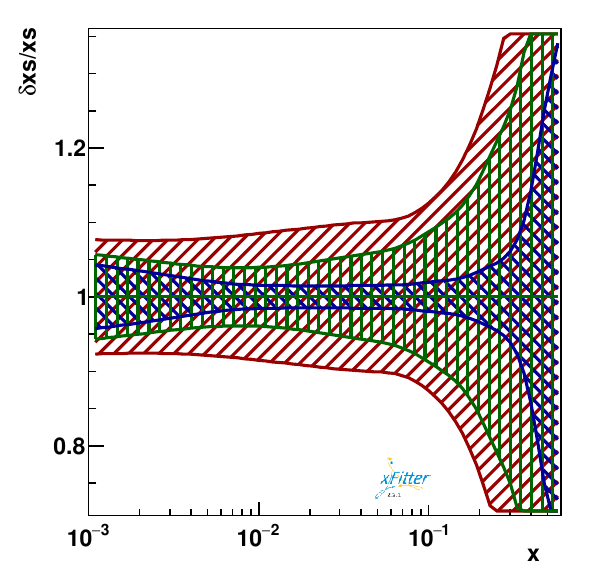}
\caption{Same as Fig.~\ref{fig:FASERnu2_baseline} (statistical uncertainties only),
  now showing results in the scenario where charm-tagged structure function measurements
  are excluded from the analysis.
}
\label{fig:FASERnu2_nocharm}
\end{figure*}

\paragraph{Relevance of charm-tagged measurements.}
The analysis of Fig.~\ref{fig:FASERnu2_baseline} highlights that LHC neutrino data is particularly
constraining for the poorly-known strange PDF, which is one of the quark flavour combinations
for which proton PDF fits differ the most~\cite{Faura:2020oom}.
To further investigate this point, Fig.~\ref{fig:FASERnu2_nocharm} compares the impact of the FASER$\nu$2 data shown in
Fig.~\ref{fig:FASERnu2_baseline}, for the case in which only statistical
uncertainties are considered, with the results of the same profiling once the charm-tagged
structure function data is excluded from the fit.
While differences are moderate for the up and down quark PDFs, the significant loss
of information resulting from this exclusion of charm-tagged data is clearly
visible for the strange PDF.
Specially in the region $x\gsim 0.01$, the constraints on strangeness shown in
Fig.~\ref{fig:FASERnu2_nocharm} are mostly washed out in the absence of charm-tagged data.
We thus establish that inclusive neutrino DIS measurements constrain predominantly
the up and down quark and antiquark PDFs (and thus also the quark singlet), while the charm-tagged
structure functions are responsible for most of the constraints provided on the total strangeness.
The PDF reach of the LHC neutrino experiments would thus be  markedly limited in experiments without
charm-identification capabilities.

\begin{figure*}[t]
\centering
\includegraphics[width=0.32\textwidth]{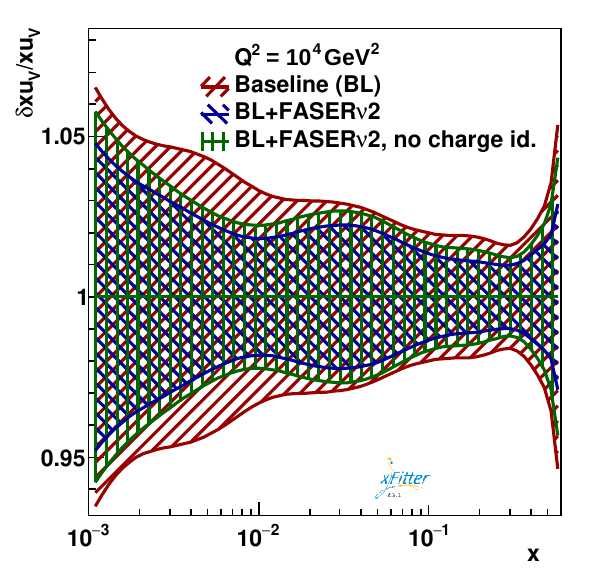}
\includegraphics[width=0.32\textwidth]{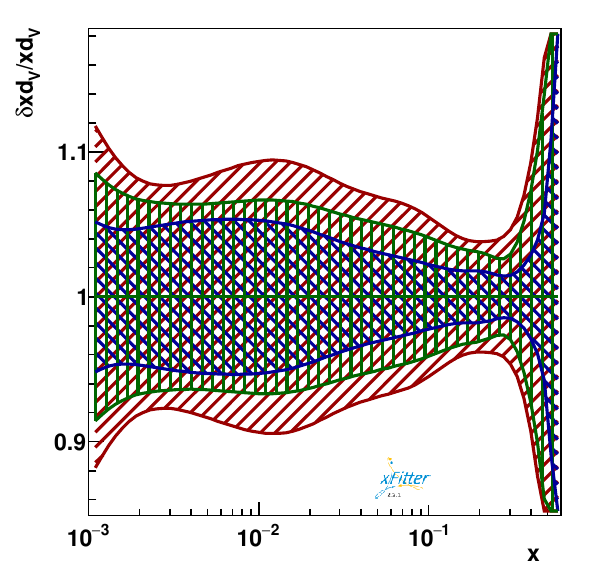}
\includegraphics[width=0.32\textwidth]{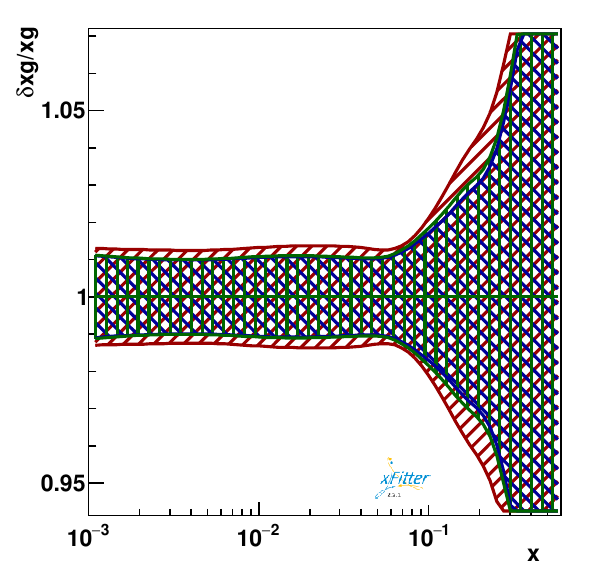}\\
\includegraphics[width=0.32\textwidth]{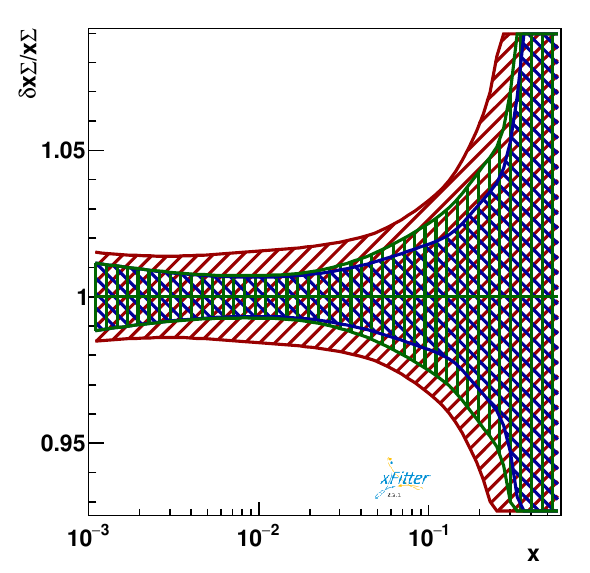}
\includegraphics[width=0.32\textwidth]{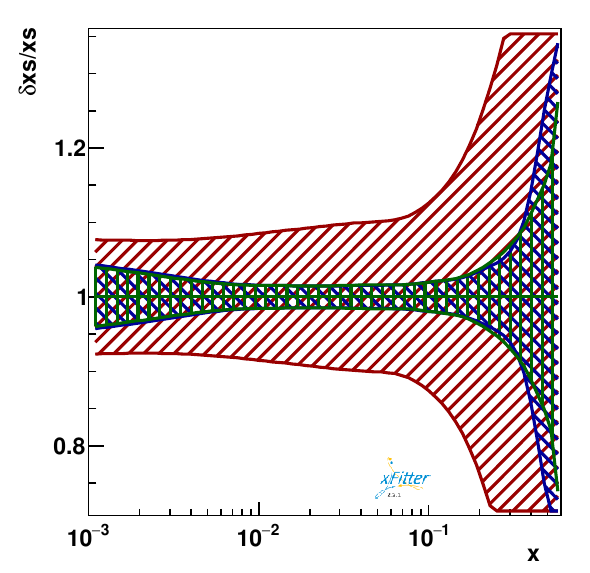}
\caption{Same as Fig.~\ref{fig:FASERnu2_baseline} (statistical uncertainties only),
  now showing results in the scenario where the charge of the final-state charged lepton
  cannot be identified.
 }
\label{fig:FASERnu2_nochargeID}
\end{figure*}

\paragraph{Lepton-charge identification.}
Being able to identify the charge of the produced final-state lepton in charged-current
neutrino scattering demands equipping an experiment with a powerful enough magnet suitable to
deflect this lepton within the detector fiducial volume.
Our baseline results for FASER$\nu$2 in Fig.~\ref{fig:FASERnu2_baseline} assume that this charge-identification
is possible, and therefore include separate structure function datasets for neutrino and anti-neutrino projectiles.
In order to ascertain to which extent the constraints provided by FASER$\nu$2 structure functions
depend on the availability of such a magnet,
Fig.~\ref{fig:FASERnu2_nochargeID} compares the reduction of the PDF uncertainties using
the FASER$\nu$2 data with and without assuming charged-lepton identification capabilities.

This analysis finds that the lack of charged-lepton identification actually does not degrade significantly the PDF
sensitivity of FASER$\nu$2.
Having access of the lepton charge information improves a bit the constraints for the down and (to a lesser
extent) the up valence quark PDFs,  while there are no differences for the total quark
singlet and for strangeness.
Such an improvement can be understood by inspecting the leading-order decomposition of neutrino DIS
structure functions in terms of the different PDF flavours for different targets, Eqns.~(\ref{eq:neutrinoSFs_proton})--(\ref{eq:neutrinoSFs_isoscalar}).
For an isoscalar target, as assumed here, structure functions are very similar for neutrino
and anti-neutrino projectiles, with differences restricted to the strangeness asymmetry.
Given that this strangeness asymmetry is quite small, this also explains why the impact
on strangeness, which is driven by the charm-tagged data, is the same irrespective of whether one identifies
the outgoing lepton charge.
We conclude that LHC neutrino DIS data exhibits competitive PDF sensitivity even for detectors which cannot
separate incoming neutrinos from antineutrinos.

\begin{figure*}[htbp]
\centering
\includegraphics[width=0.32\textwidth]{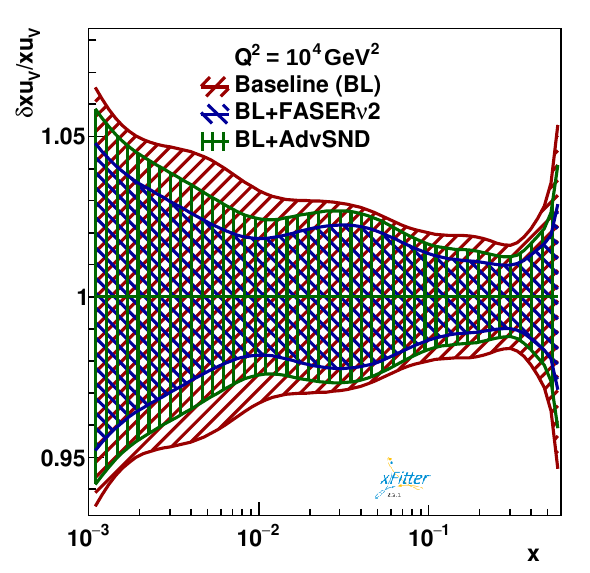}
\includegraphics[width=0.32\textwidth]{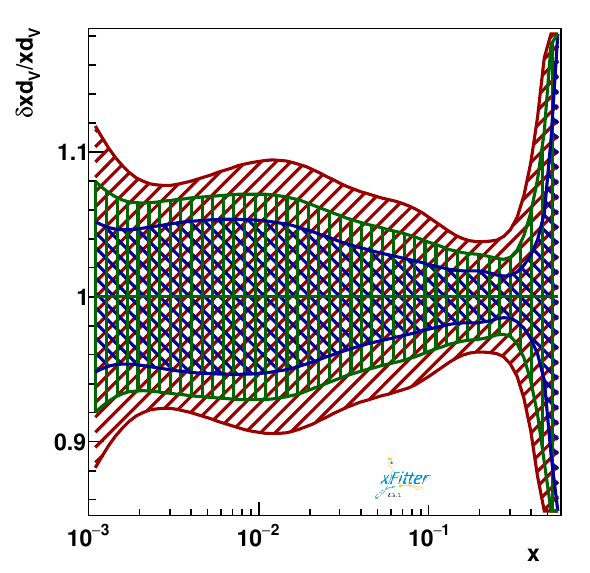}
\includegraphics[width=0.32\textwidth]{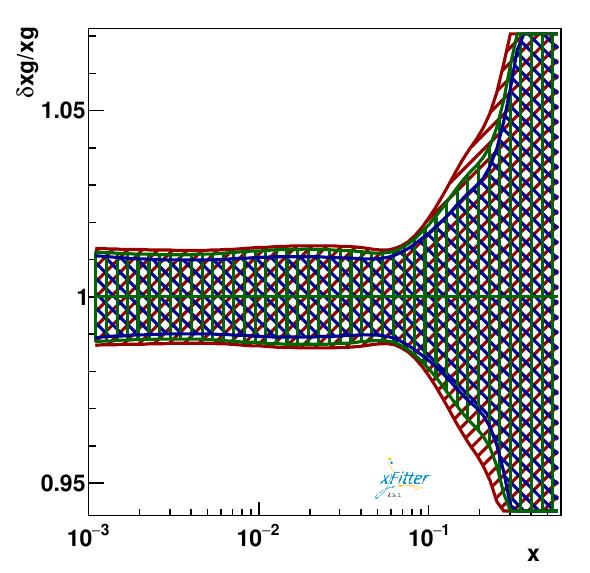}\\
\includegraphics[width=0.32\textwidth]{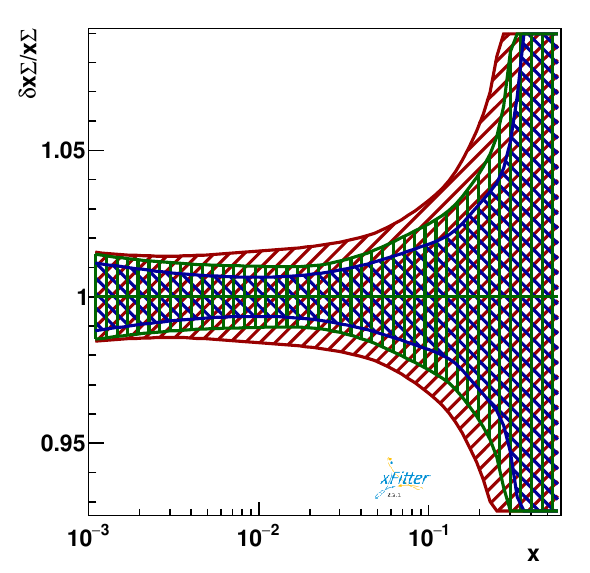}
\includegraphics[width=0.32\textwidth]{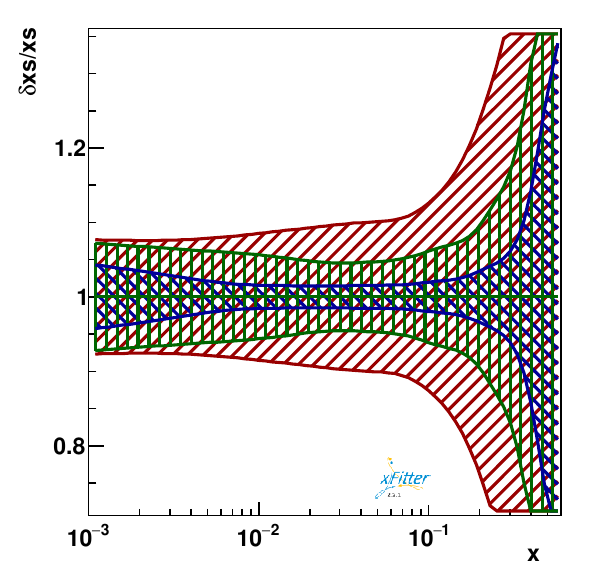}\\
\includegraphics[width=0.32\textwidth]{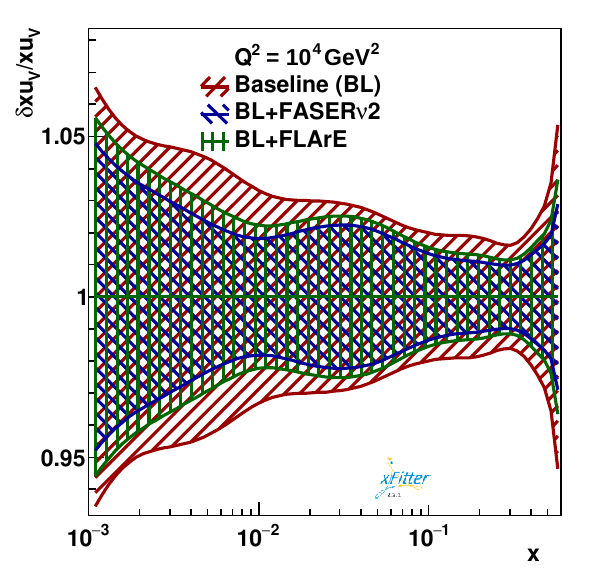}
\includegraphics[width=0.32\textwidth]{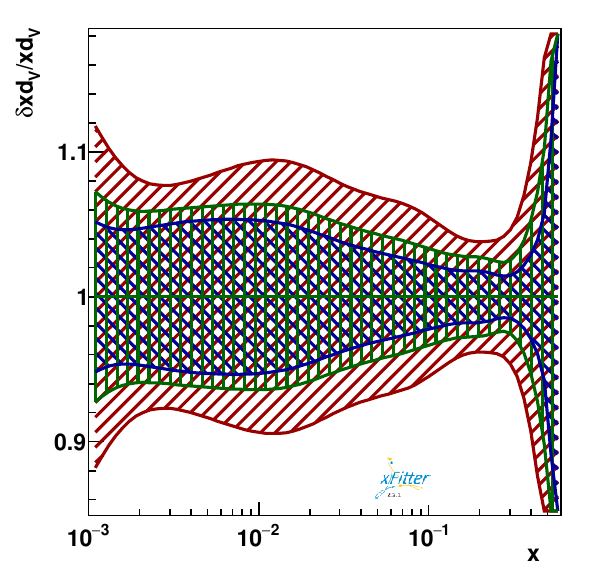}
\includegraphics[width=0.32\textwidth]{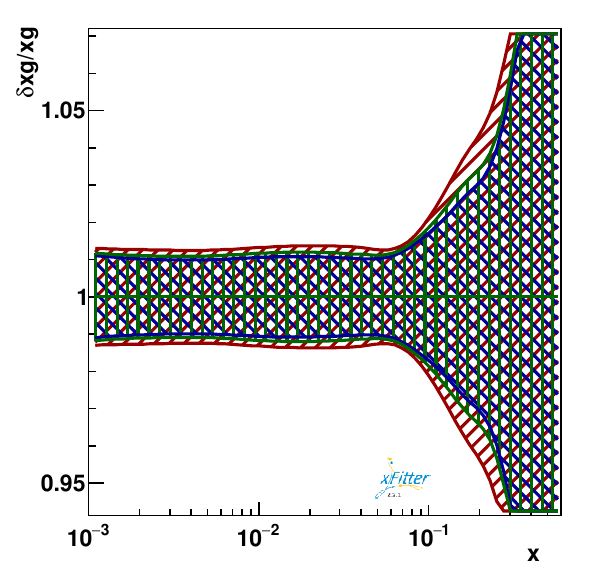}\\
\includegraphics[width=0.32\textwidth]{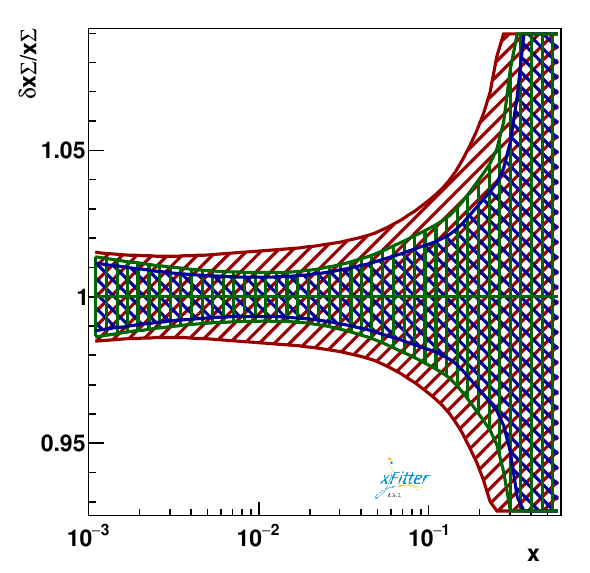}
\includegraphics[width=0.32\textwidth]{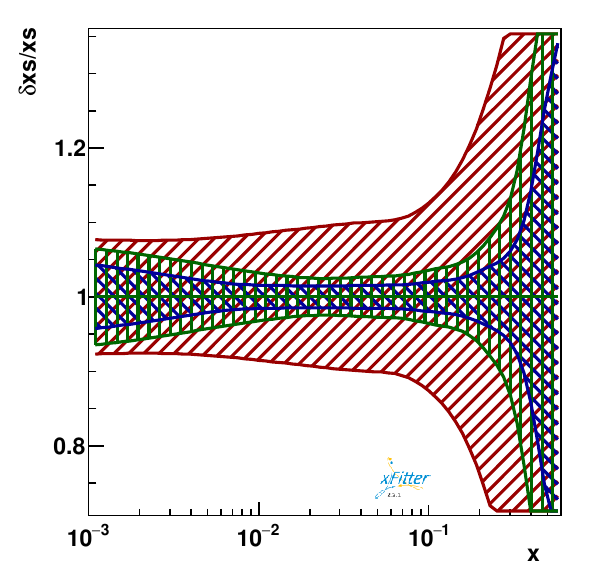}
\caption{
  Same as Fig.~\ref{fig:FASERnu2_baseline} (statistical uncertainties only),
  comparing
  the projected PDF impact of FASER$\nu$2 with that of  AdvSND (top panels) and FLArE10 (bottom panels). 
}
\label{fig:FASERnu2_FLAre10}
\end{figure*}

\paragraph{Experiment dependence.}
Together with the impact projections for FASER$\nu$2, we have also produced analogous results for
other proposed FPF experiments, specifically for AdvSND and FLArE.
In the latter case, we consider the 10 tonne variant, FLArE10.
Fig.~\ref{fig:FASERnu2_FLAre10} compares the PDF sensitivity
of FASER$\nu$2, in the scenario where systematic uncertainties are neglected, with the corresponding
results for AdvSND and FLArE10 respectively.
As summarised by Table~\ref{tab:integrated_rates}, each of these experiments
has associated different expected numbers of DIS events, namely 510k, 56k, and 110k (670k)
inclusive muon-neutrino events for FASER$\nu$2, AdvSND and FLArE10~(100) respectively,
with 76k, 7k, and 15k (90k) in the charm-tagged case.

In agreement with expectations, 
detectors with the largest event rates are those exhibiting the
best PDF sensitivity.
From Fig.~\ref{fig:FASERnu2_FLAre10} one can see that the three
FPF experiments studied lead to a reduction of the  uncertainties in the quark PDFs.
By comparing their relative impact, we find that the constraints
achieved by FASER$\nu$2 are somewhat better than for the other two experiments,
consistent with the larger event yields obtained in the former.
In the case of AdvSND, the  smaller sample of charm-tagged events is responsible for the reduction
in the constraints on strangeness.
Another consequence of the smaller event rates in AdvSND and FLArE10 as compared
to FASER$\nu$2 is the milder impact for the $x\lsim 10^{-2}$ region,
which can only be covered once the integrated statistics become large
enough,  as indicated by the differential bin-by-bin yields displayed in Fig.~\ref{fig:fasernu2_muon}.

While Fig.~\ref{fig:FASERnu2_FLAre10} compares fits based on FPF pseudo-data
with only statistical uncertainties, a more robust comparison between the PDF reach of
the various FPF experiments requires accounting also for the systematic uncertainties, which ultimately become
one of the limiting factors.
Also, as mentioned above, cross-calibration between experiments should provide
valuable input to enhance the PDF sensitivity.

\begin{figure*}[t]
\centering
\includegraphics[width=0.32\textwidth]{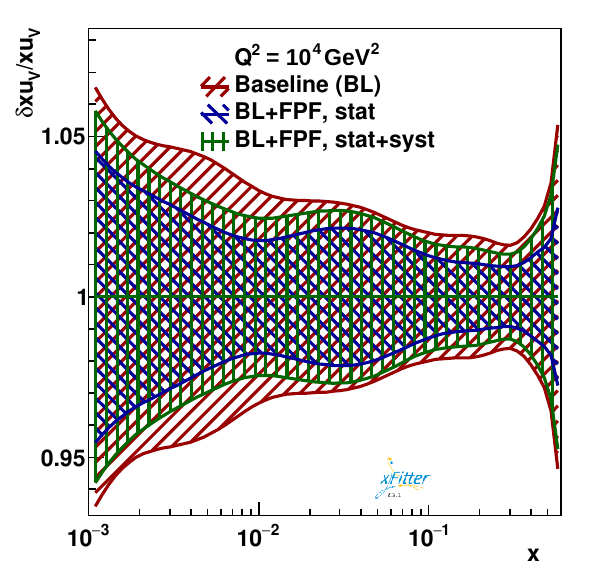}
\includegraphics[width=0.32\textwidth]{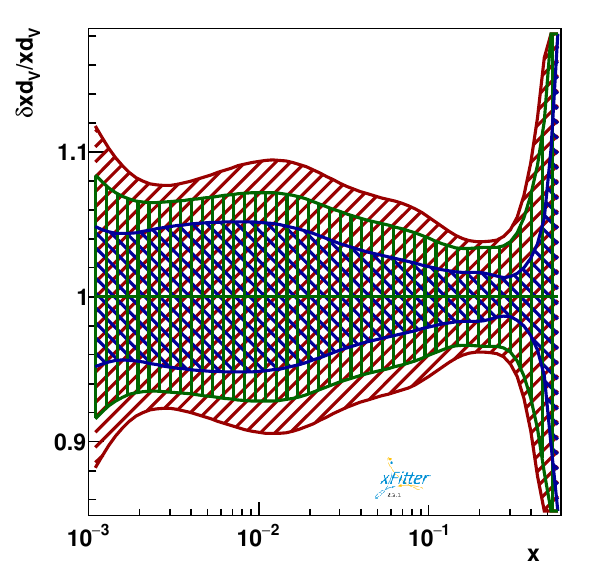}
\includegraphics[width=0.32\textwidth]{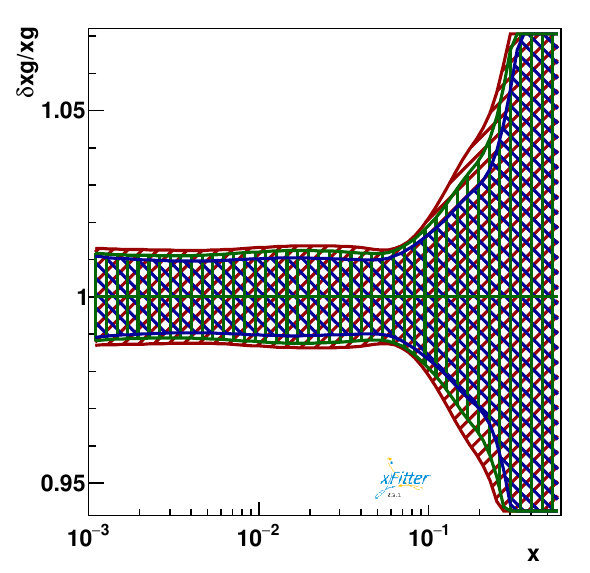}\\
\includegraphics[width=0.32\textwidth]{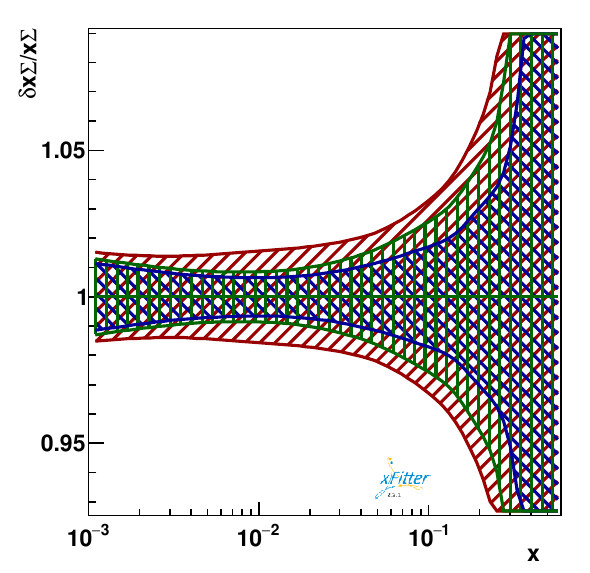}
\includegraphics[width=0.32\textwidth]{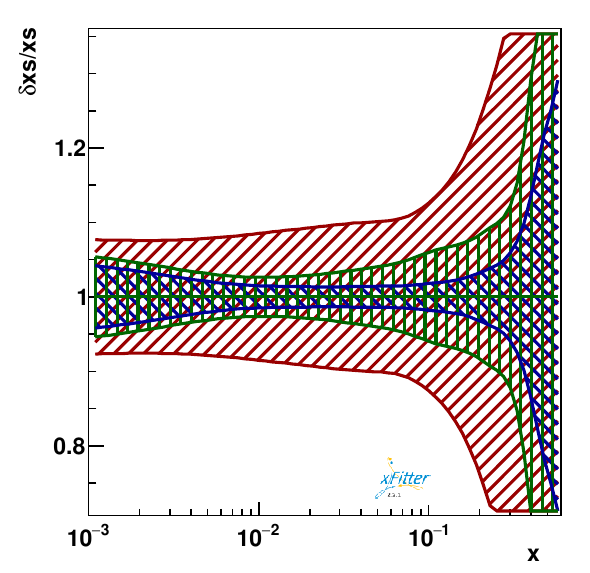}
\caption{
  Same as Fig.~\ref{fig:FASERnu2_baseline} now quantifying the
  projected PDF impact of the three FPF experiments added simultaneously to
  the analysis: FASER$\nu$2, AdvSND, and FLArE10. 
}
\label{fig:FPF_combined}
\end{figure*}

\paragraph{Combined impact of the FPF experiments.}
We have also carried out a further PDF analysis including simultaneously the three FPF experiments
considered here: FASER$\nu$2, AdvSND, and FLArE10.
Fig.~\ref{fig:FPF_combined} displays the combined impact of the FPF experiments when added
on top of the  PDF4LHC21 baseline by means of Hessian profiling, both in the statistics-only case and
when systematic and statistical uncertainties are added in quadrature.
Potential correlations between individual experiments are neglected in this exercise.
The combined impact of the three FPF experiments on PDF4LHC21 shown in Fig.~\ref{fig:FPF_combined}
is only moderately improved as compared to the results of the FASER$\nu$2-only profiling
in Fig.~\ref{fig:FASERnu2_baseline}.
This shows that in this combination of experiments the one with the largest individual
sensitivity dominates, in this case FASER$\nu$2.
Again, here we assume perfect consistency between theory and data, while
within the actual experiments the consistency (or lack thereof) between
individual measurements cannot be guaranteed.

The projections displayed in
Fig.~\ref{fig:FPF_combined}, and in particular the statistics-only case, represents
a best-case scenario for the reduction of PDF uncertainties which can be expected from
the analysis of neutrino DIS structure function data at the FPF,
under the assumption that PDF4LHC21 represents
our current knowledge about the quark and gluon structure of the proton.

\paragraph{The effect of neutrino flux uncertainties.}
The results presented so far assume the incoming far-forward neutrino flux as input, with vanishing uncertainties.
One can estimate the impact that uncertainties associated to this neutrino flux input have
on the PDF constraints by extending the procedure of~\cite{Kling:2023tgr}, 
based on the predictions provided in~\cite{Ahn:2009wx, Ahn:2011wt, Riehn:2015oba, Fedynitch:2018cbl, Pierog:2013ria, Buonocore:2023kna, Roesler:2000he, Fedynitch:2015kcn, Bai:2020ukz, Bai:2021ira, Bai:2022xad, Ostapchenko:2010vb, Bhattacharya:2023zei, Fieg:2023kld, Maciula:2022lzk},
by introducing an additional fit parameter to account for the overall normalisation of the neutrino flux (while assuming the shape is unchanged with respect to the baseline predictions).
The constraint for this flux normalisation parameter is obtained considering two scenarios: first, by using the expected Run 3 data from FASER$\nu$, and second, by using 50\% of the data expected at FASER$\nu$2 during the HL-LHC run.
In the latter case, only the remaining 50\% of FASER$\nu$2 data is then considered available for PDF determination, to avoid double counting.

\begin{figure*}[t]
\centering
\includegraphics[width=0.4\textwidth]{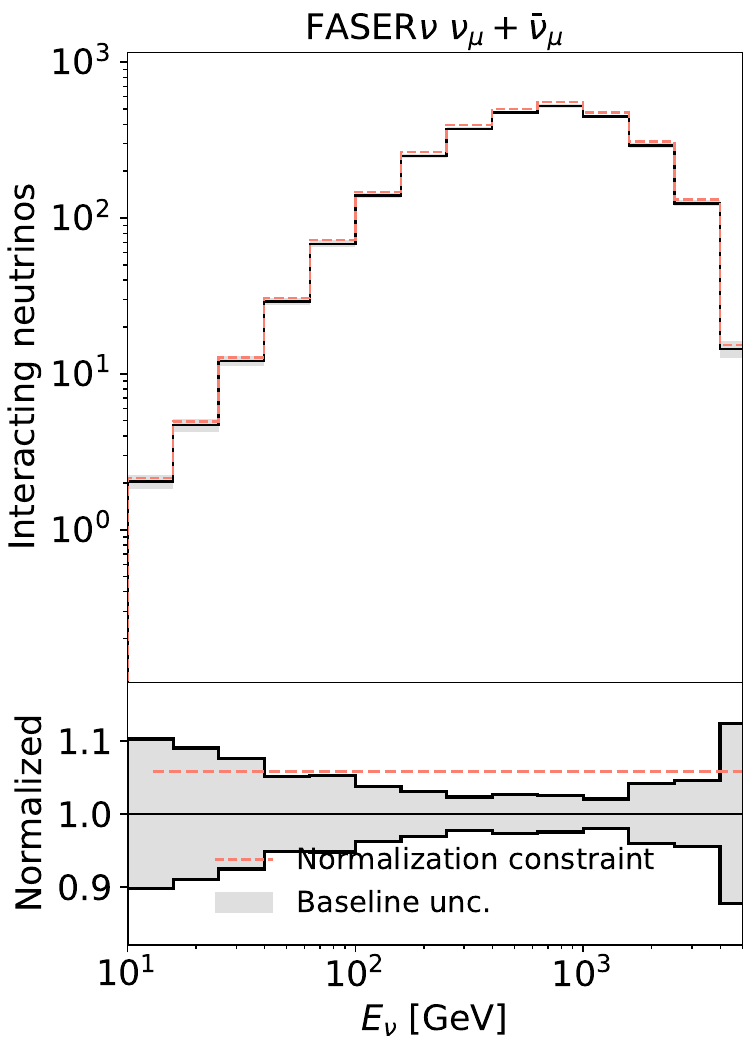}
\hspace{1cm}
\includegraphics[width=0.4\textwidth]{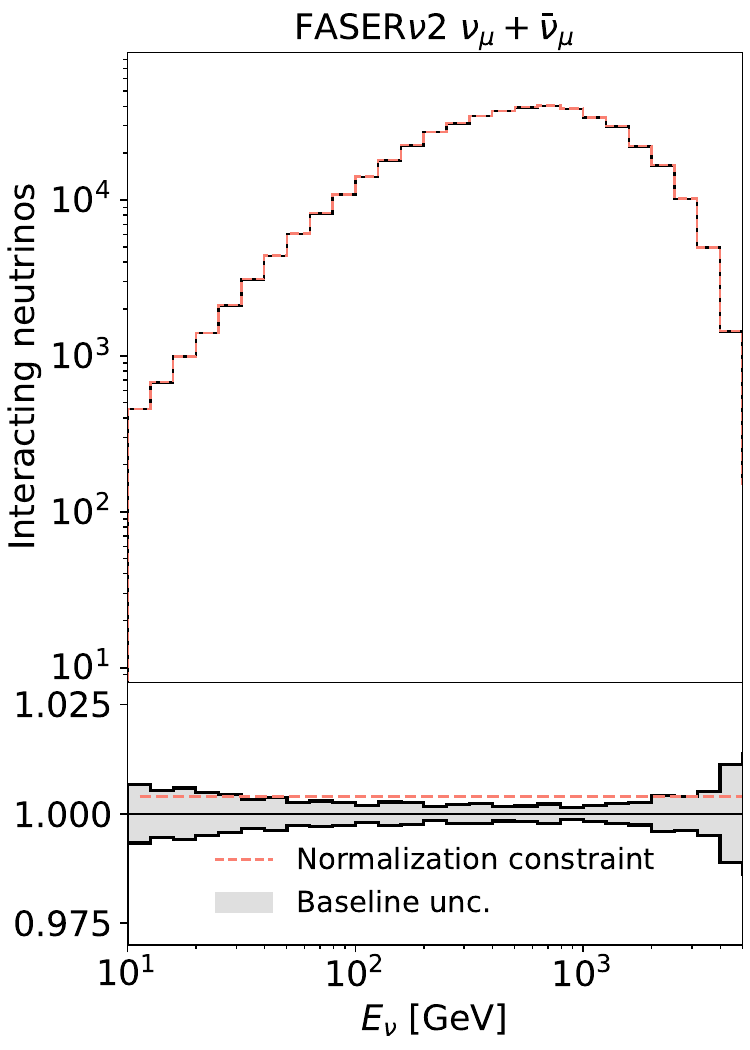}
\caption{The baseline uncertainty associated to the theoretical modelling of the far-forward muon neutrino fluxes~\cite{Kling:2023tgr} at FASER$\nu$ (left) and FASER$\nu$2 (right panel), compared to the results of constraining the overall normalisation of this flux from  FASER$\nu$ data and from 50\% of the expected FASER$\nu$2 data, respectively. 
}
\label{fig:flux_constraints}
\end{figure*}

\begin{figure*}[t]
\centering
\includegraphics[width=0.32\textwidth]{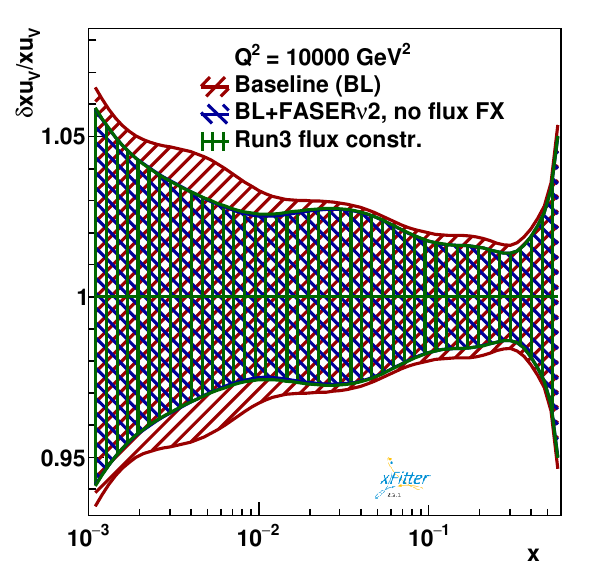}
\includegraphics[width=0.32\textwidth]{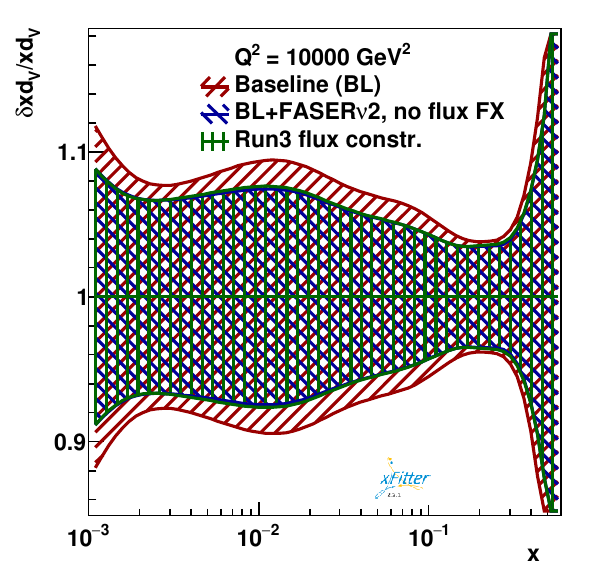}
\includegraphics[width=0.32\textwidth]{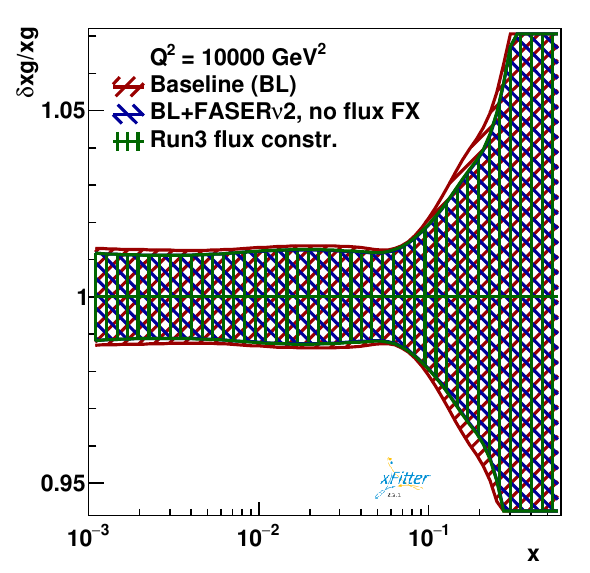}\\
\includegraphics[width=0.32\textwidth]{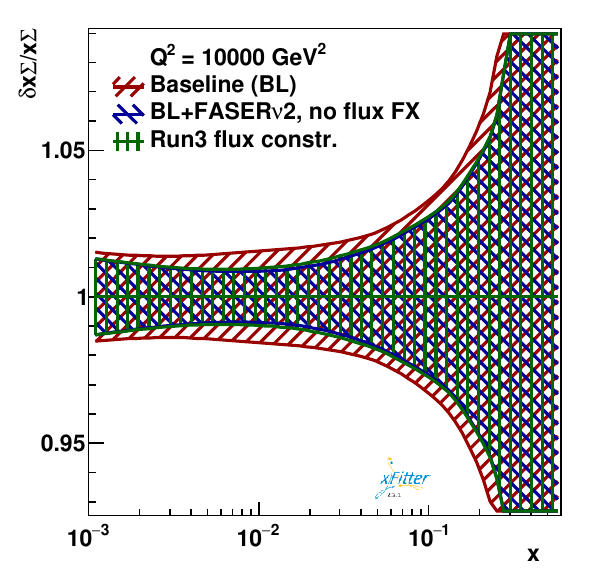}
\includegraphics[width=0.32\textwidth]{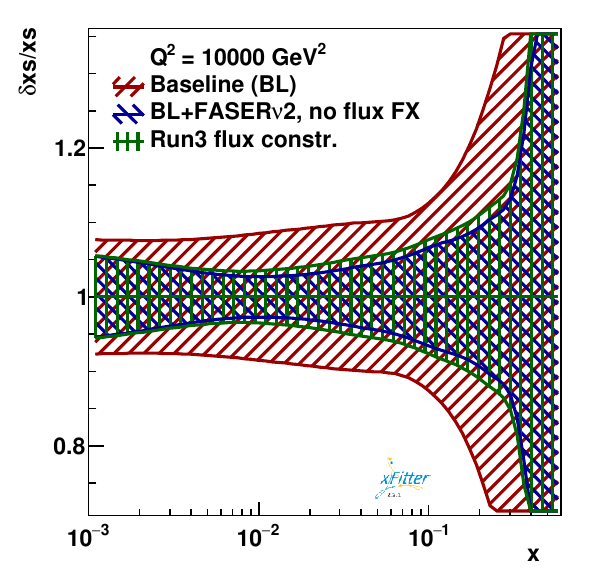}
\caption{
  Same as Fig.~\ref{fig:FASERnu2_baseline}, now quantifying the
  impact on the FASER$\nu$2 results of accounting for the muon neutrino flux uncertainties, assuming its
  overall normalisation has been constrained with FASER$\nu$ Run 3 data.
  See text for more details.
}
\label{fig:FASERv2_flux_unc_run3}
\end{figure*}

\begin{figure*}[t]
\centering
\includegraphics[width=0.32\textwidth]{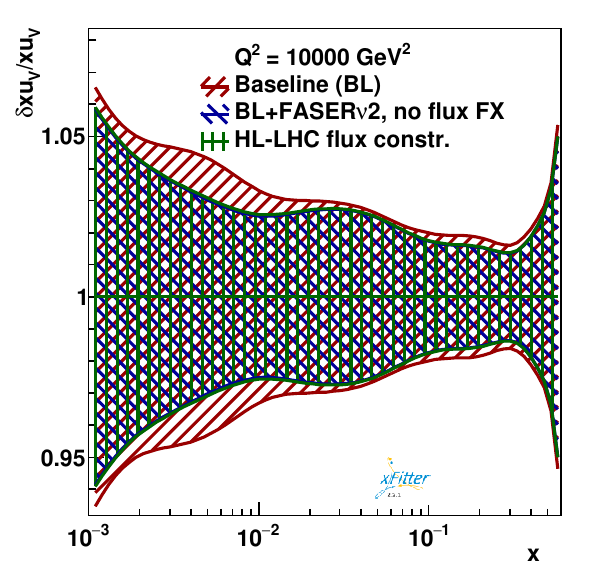}
\includegraphics[width=0.32\textwidth]{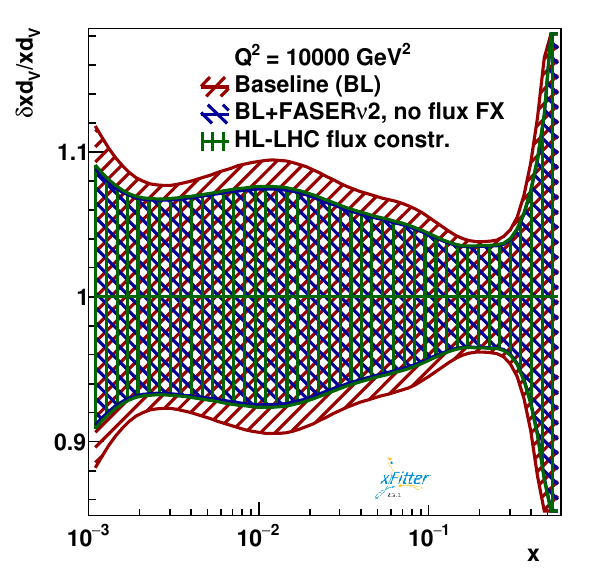}
\includegraphics[width=0.32\textwidth]{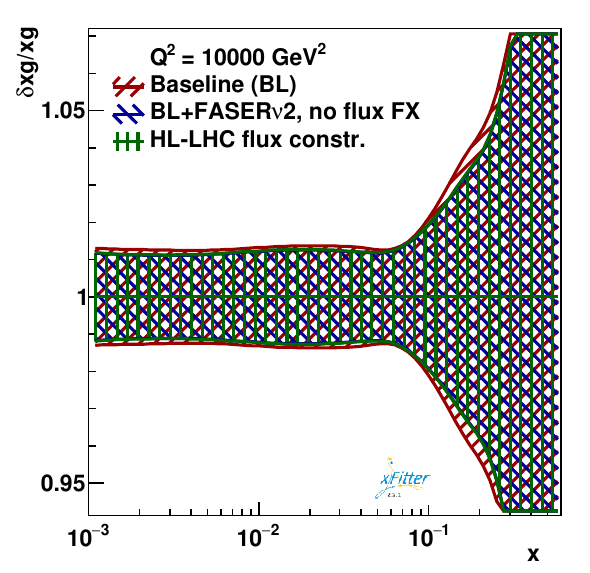}\\
\includegraphics[width=0.32\textwidth]{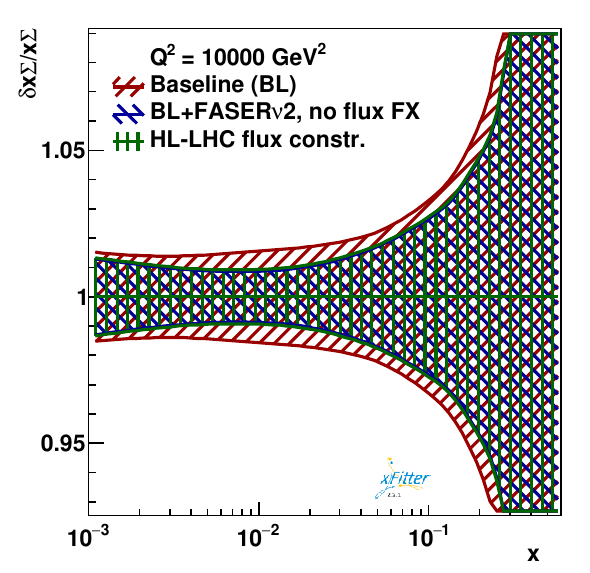}
\includegraphics[width=0.32\textwidth]{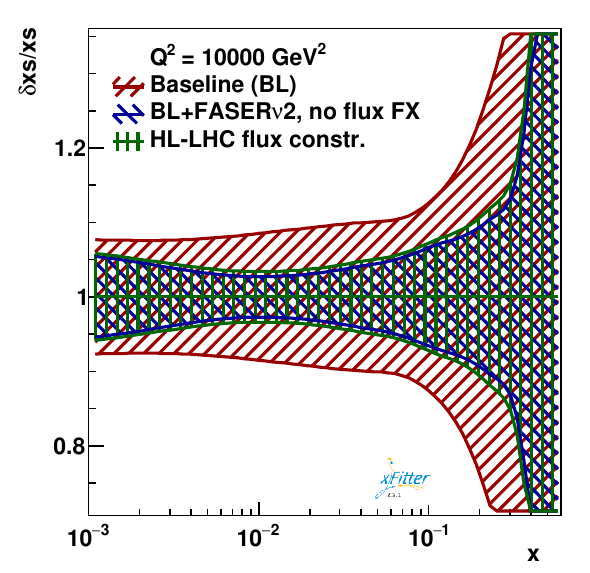}
\caption{
  Same as Fig.~\ref{fig:FASERnu2_baseline} now quantifying the
  effect of neutrino flux uncertainties in the analysis of FASER$\nu$2 pseudodata, assuming that the neutrino flux normalisation is constrained using 50\% of FASER$\nu$2 data to constrain the flux and the remainder for PDF determination. 
}
\label{fig:FASERv2_flux_unc_HL-LHC}
\end{figure*}

The resulting flux normalisation parameter can be constrained approximately to a precision of
6\% using FASER$\nu$
Run 3 data, and to
0.4\% using 50\% of FASER$\nu$2 data, as illustrated in Fig.~\ref{fig:flux_constraints}.
These constraints are then incorporated into
the profiling studies as additional fully correlated (across bins) uncertainties, also accounting for an increase in statistical uncertainty in the case where only 50\% of data is used.
The impact on the FASER$\nu$2 projections of the neutrino flux uncertainties, in the case of the 
PDFs resulting from using the Run 3 constraint are shown in Fig.~\ref{fig:FASERv2_flux_unc_run3}, and those obtained with the HL-LHC constraints in Fig.~\ref{fig:FASERv2_flux_unc_HL-LHC}.
In either case, the results indicate only a very small increase in the resulting PDF uncertainties.
One may conclude that neutrino flux uncertainties do not significantly degrade the PDF constraining potential of the FPF experiments presented in this work.


\subsection{Proton PDFs: impact on NNPDF4.0}
\label{sec:nnpdf40}

The PDF sensitivity studies based on the {\sc\small xFitter} profiling
of PDF4LHC21 are complemented by those based on their direct inclusion
in the NNPDF4.0 analysis.
Here we present results for the impact of the combined FPF neutrino
measurements on NNPDF4.0, namely the analogous study of that shown
in Fig.~\ref{fig:FPF_combined} in the case of PDF4LHC21.
We have also produced results for the different variations studied in
Sect.~\ref{sec:pdf4lhc21} with the NNPDF fitting methodology.
Given that the findings obtained with the NNPDF fits
are compatible with those obtained with {\sc\small xFitter}, it is not necessary to duplicate them here
and we only show results for the impact of the combined FPF pseudo-data.

\begin{figure*}[t]
\centering
\includegraphics[width=0.99\textwidth]{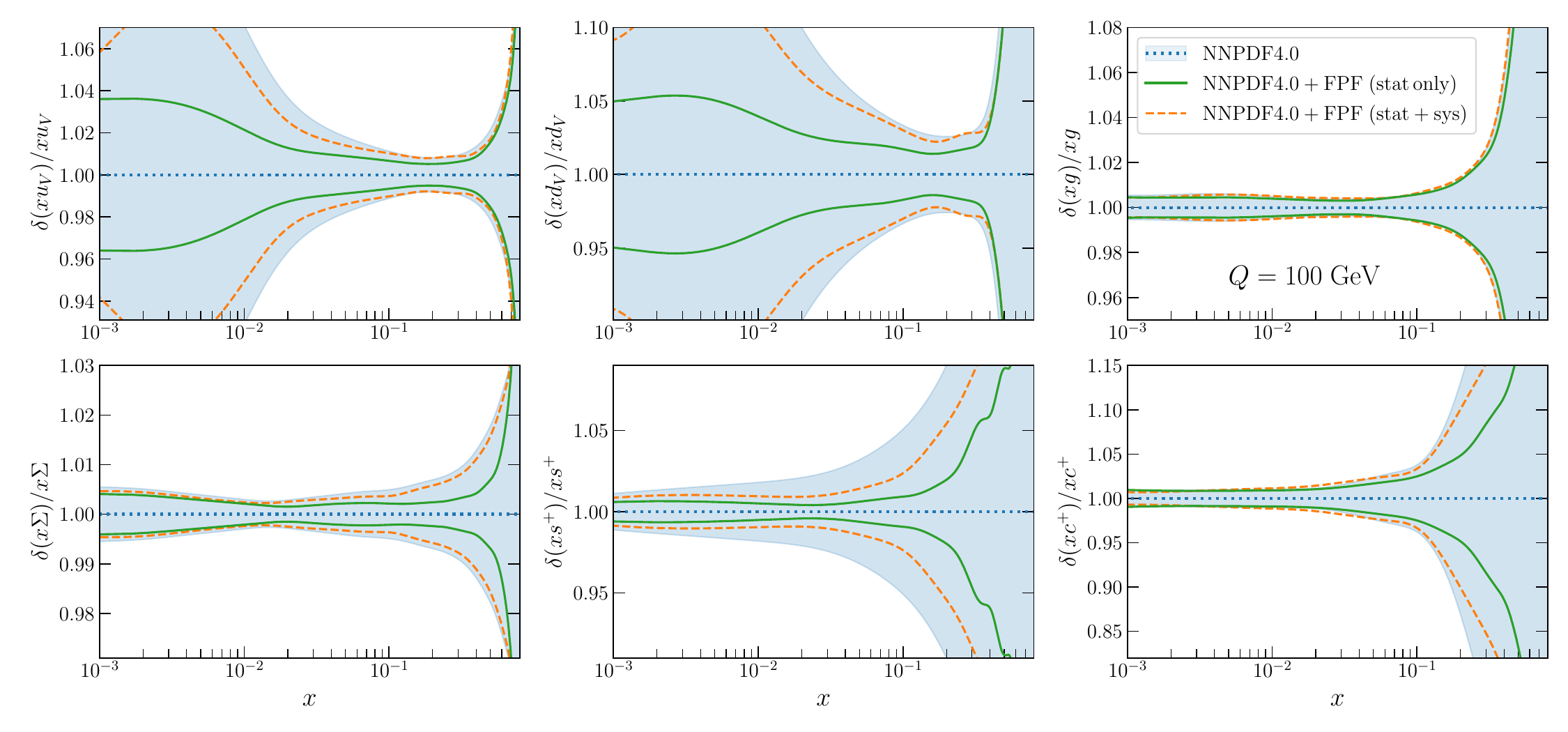}
\caption{
  Same as Fig.~\ref{fig:FPF_combined} for the results obtained
  using the NNPDF4.0 fitting methodology.
  The baseline NNPDF4.0 NNLO analysis is compared
  with the results of the fits which include the combined FPF dataset
  in both the statistics-only scenario and for the case
  where systematic uncertainties are also accounted for.
  As in the PDF4LHC21 case, the bands indicate the one-sigma PDF uncertainties.
  We now also show results for the charm PDF (bottom right panel), which
  in the NNPDF4.0 fit is determined from the data entering the fit.
}
\label{fig:NNPDF40_baseline}
\end{figure*}

Fig.~\ref{fig:NNPDF40_baseline} displays the
same results as Fig.~\ref{fig:FPF_combined} now in the case of the
fits obtained with the NNPDF4.0 fitting methodology.
The baseline NNPDF4.0 NNLO analysis is compared
with the results of the fits which include the combined FPF dataset,
in both the statistics-only scenario and for the case
where systematic uncertainties are also accounted for.
As in the PDF4LHC21 analysis, the bands indicate the one-sigma PDF uncertainties.
Note that we also show results for the charm PDF (bottom right panel), which
in the NNPDF4.0 fit is also determined directly from the data entering
the fit~\cite{Ball:2016neh}.
In this way, we can determine the possible constraints that neutrino
DIS measurements at the LHC may impose on the intrinsic charm
content of the proton~\cite{Ball:2022qks}.

From  Fig.~\ref{fig:NNPDF40_baseline} one finds that
the impact on the PDFs found by direct inclusion of the FPF structure
functions on the NNPDF4.0 fit is qualitatively consistent with
that found in the the Hessian profiling of PDF4LHC21.
In particular, the constraints are more significant for the up and down
quark valence PDFs as well as for the total strangeness.
As was the case with PDF4LHC21, the uncertainty reduction
for the strangeness is particularly striking.
Also consistently with the results of Sect.~\ref{sec:pdf4lhc21}, the gluon
PDF is unchanged, and the impact on the total quark PDF
is restricted to the large-$x$ region.
While one can also identify differences related to analysis choices
such as the PDF parametrisation and the method used to assess the PDF
error reduction, for example, concerning the small-$x$ behaviour of $xu_V$ and $xd_V$,
we conclude that by and large the projected impact of the FPF structure function
measurements is independent of the specific fitting methodology adopted.

\begin{figure*}[t]
\centering
\includegraphics[width=0.99\textwidth]{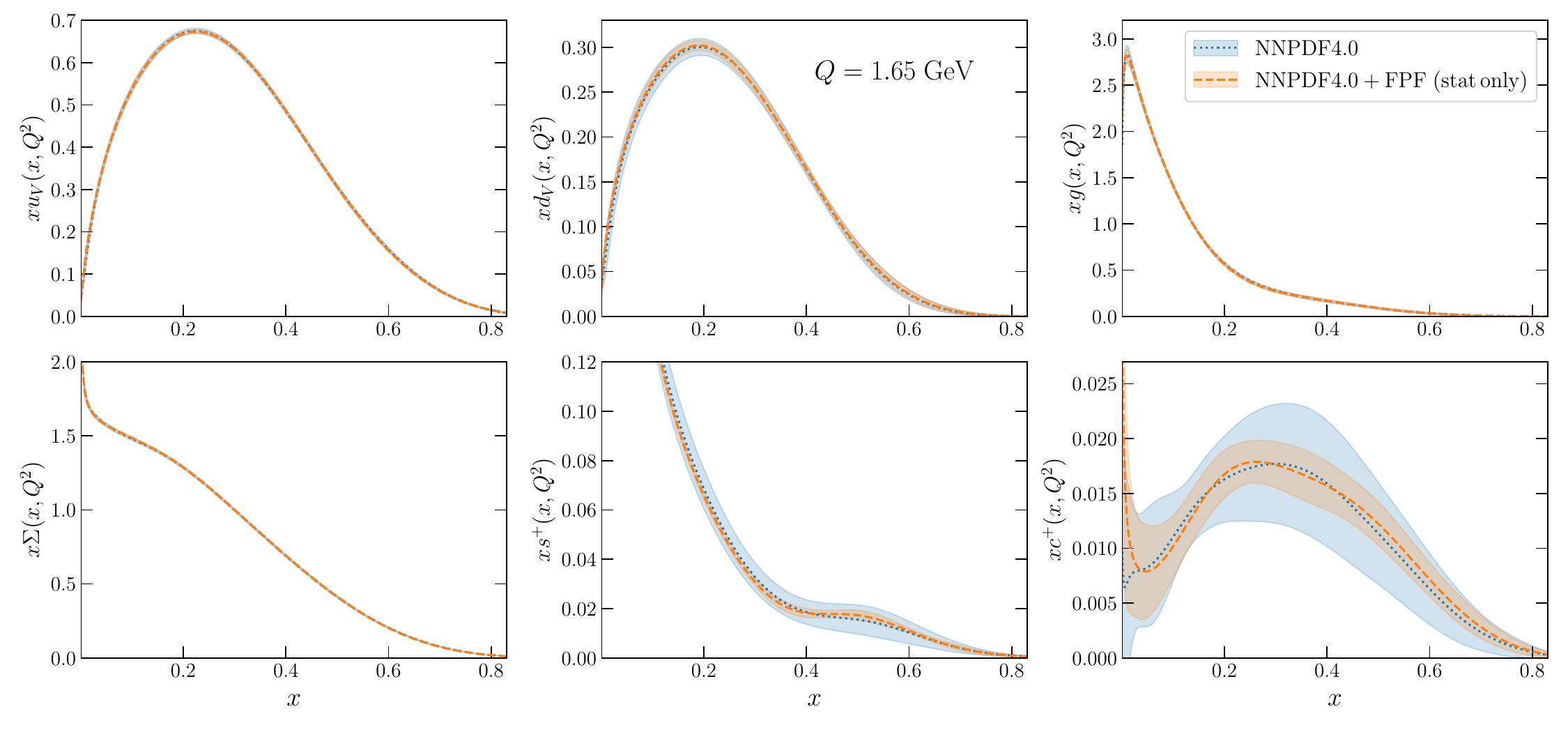}
\caption{
  Same as Fig.~\ref{fig:NNPDF40_baseline} now for the absolute PDFs
  at the initial parametrisation scale, $Q=1.65$ GeV, focusing
  on the large-$x$ region.
}
\label{fig:NNPDF40_lowQ_abs}
\end{figure*}

As indicated by Fig.~\ref{fig:NNPDF40_baseline}, the FPF data is also expected
to reduce the uncertainties associated to the fitted charm PDF in the large-$x$ region 
dominated by the non-perturbative component.
In order to highlight this impact on the large-$x$ region of the PDFs,
Fig.~\ref{fig:NNPDF40_lowQ_abs} presents the same comparison as
in  Fig.~\ref{fig:NNPDF40_baseline} now for the absolute PDFs
at the initial parametrisation scale, $Q=1.65$ GeV, using a linear scale, for the statistics-only
scenario.
One can observe the improved constraints on the charm PDF in the large-$x$ region,
further confirming previous studies~\cite{Anchordoqui:2021ghd,Feng:2022inv} that indicate the
sensitivity of the FPF to the intrinsic charm content of the proton.
This comparison also illustrates the excellent constraining power on the proton
strangeness enabled by the FPF charm-tagged structure function data.

\begin{table*}[!t]
  \centering
  \footnotesize
  \renewcommand{\arraystretch}{1.30}
  \begin{tabularx}{\textwidth}{X|l|c|c|c}
  \toprule
  & \multirow{3}{*}{Dataset}
  & \multicolumn{3}{c}{\bf NNPDF4.0 NNLO} \\
  &
  & \multirow{2}{*}{Baseline}
  & \multicolumn{2}{c}{with combined FPF data}  \\
  &
  &
  & Statistical-only  & Statistical + systematics  \\
  \midrule
  $\chi^2/n_{\rm dat}$ &  \multirow{4}{*}{Global}  & 1.17  &  1.15   & 1.14   \\
  $\la E_{\rm tr}\ra_{\rm rep}$
  &  &  2.26  &  2.24   & 2.21   \\
  $\la E_{\rm val}\ra_{\rm rep}$
  & & 2.36    &   2.33  & 2.30  \\
  $\la \chi^2/n_{\rm dat}\ra_{\rm rep}$
  & &   1.19   &    1.18    & 1.16  \\
  \midrule
  \multirow{8}{*}{$\chi^2/n_{\rm dat}$}
  & DIS neutral-current                     &  1.22    &  1.23    &   1.22   \\
  & DIS charged-current                     &   0.90  &   0.91    &  0.90    \\
  & Drell--Yan (inclusive and one-jet)      &  1.76   &   1.84    &   1.83   \\
  & Top-quark pair production               &  1.23   &   1.19    &  1.25     \\
  & Single-top production                   &  0.36    &   0.37   &   0.37    \\
  & Inclusive jet production                &  0.96    &  0.96      &   0.94    \\
  & Dijet production                        &  2.03     &    2.03    &   2.00    \\
  & Direct photon production                 &  0.74   &   0.74     &    0.73   \\
  & FPF (total)                 &  {\it 1.29  / 0.92}  &   1.10     &   0.89    \\
  \bottomrule
\end{tabularx}
\vspace{0.2cm}
\caption{\small Statistical estimators for the NNPDF4.0 NNLO
  baseline fit, compared to the variants including
  the combined FPF dataset shown in Figs.~\ref{fig:NNPDF40_baseline}
  and~\ref{fig:NNPDF40_lowQ_abs}.
  From top to bottom: total $\chi^2/n_{\rm dat}$, average
  over replicas of the training and validation figures of merit
  $\la E_{\rm tr}\ra_{\rm rep}$ and $\la E_{\rm val}\ra_{\rm rep}$,
  average $\chi^2/n_{\rm dat}$ over replicas $\la \chi^2/n_{\rm dat}\ra_{\rm rep}$,
  $\chi^2/n_{\rm dat}$ for datasets grouped by process.
  The last row displays the $\chi^2/n_{\rm dat}$ obtained for the combined FPF dataset.
  For the baseline, we indicate the pre-fit values of $\chi^2/n_{\rm dat}$  for the  FPF dataset (in italics)
  computed without
  and with systematic errors in the covariance matrix.
  \label{tab:chi2_nnpdf40_baseline}
}
\end{table*}

The statistical estimators for the NNPDF4.0 NNLO
baseline fit, compared to the variants including
the combined FPF dataset shown in Figs.~\ref{fig:NNPDF40_baseline}
and~\ref{fig:NNPDF40_lowQ_abs}, are reported in Table~\ref{tab:chi2_nnpdf40_baseline}.
The last row displays the $\chi^2/n_{\rm dat}$ obtained for the combined FPF dataset,
for which one finds $\chi^2/n_{\rm dat} \sim 1$ up to fluctuations, consistent
with the fact that the FPF pseudo-data
is produced by construction to be compatible with the NNPDF4.0 baseline.
For the same reason, the description of the other experiments entering
the NNPDF4.0 global fit is not distorted by the inclusion of the FPF data.
In particular, the $\chi^2/n_{\rm dat}$ of the available (non-FPF) DIS neutral-current and charged-current
datasets is unchanged (1.22 and 0.90, respectively) from the baseline to the fit with FPF pseudo-data.
A moderate increase is found to the $\chi^2/n_{\rm dat}$ of Drell-Yan processes, from 1.76 to 1.83, possibly
because of the enhanced weight that DIS data carries in this fit variant.

\begin{figure*}[htbp]
\centering
\includegraphics[width=0.99\textwidth]{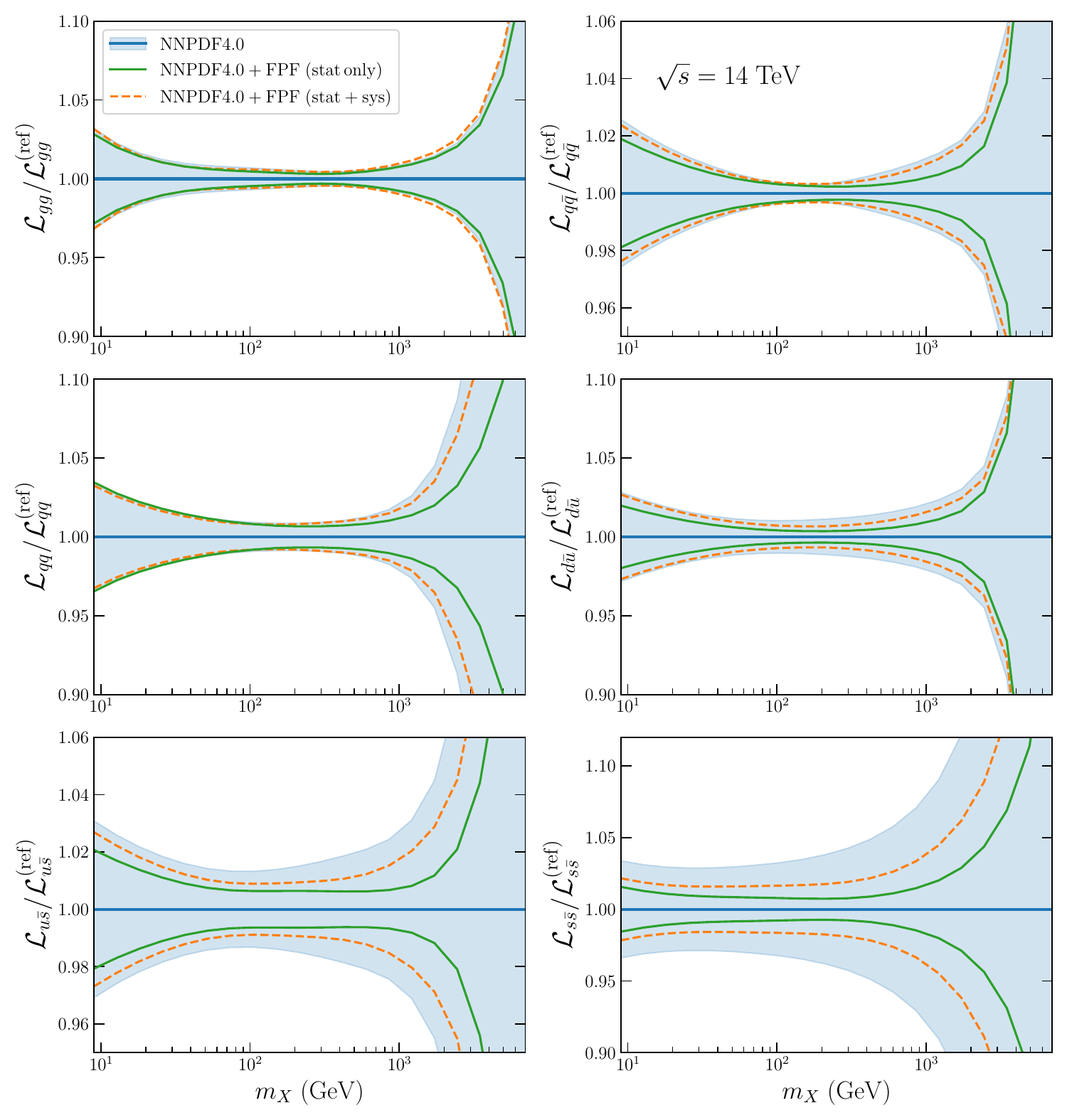}
\caption{Comparison of the PDF luminosities evaluated at $\sqrt{s}=14$ TeV
  as a function of the final-state invariant mass $m_X$ between
  the baseline NNPDF4.0 NNLO determination and its variants which include
  the full set of FPF pseudo-data, see Figs.~\ref{fig:NNPDF40_baseline}
  for the corresponding PDF comparison.
  Results are shown normalised to the central value of the NNPDF4.0 baseline.
  From left to right and top to bottom we show the gluon-gluon, quark-antiquark,
  quark-quark, down-antiup, up-antistrange, and strange-antistrange luminosities.
  The bands in the NNPDF4.0 baseline indicate the one-sigma PDF uncertainties.
}
\label{fig:nnpdf40_fpf_lumis}
\end{figure*}

Fig.~\ref{fig:nnpdf40_fpf_lumis} displays a comparison of the PDF luminosities at $\sqrt{s}=14$ TeV
as a function of the final-state invariant mass $m_X$ between
the baseline NNPDF4.0 fit and its variants including the combined FPF dataset,
see Fig.~\ref{fig:NNPDF40_baseline} for the corresponding PDF comparison.
Results are shown normalised to the central value of the NNPDF4.0 baseline.
Specifically, we show the gluon-gluon, quark-antiquark, 
quark-quark, down-antiup, up-antistrange, and strange-antistrange  luminosities.
The reduction of PDF uncertainties found in  Figs.~\ref{fig:NNPDF40_baseline}
and~\ref{fig:NNPDF40_lowQ_abs} is also visible
for the partonic luminosities, specially in the best-case scenario where only statistical
uncertainties in the FPF pseudo-data are considered.
The improved quark-antiquark luminosities at $Q\sim 100$ GeV
benefit theoretical predictions for gauge-boson production at the LHC.
It is also worth noting the PDF error reduction for the quark-quark luminosity at
high invariant masses,
which will feed into searches for new heavy particles at the TeV scale
driven by quark-initiated scattering.
Unsurprisingly, luminosities related to the strange PDF are those
which improve the most, such as the $u\bar{s}$ channel which provides a significant
contribution to inclusive $W^+$ production.

\subsection{Nuclear PDFs: impact on EPPS21}
\label{sec:nuclearPDFs}

The studies presented in Sects.~\ref{sec:pdf4lhc21} (for PDF4LHC21)
and~\ref{sec:nnpdf40} (for NNPDF4.0) treat, from the point of view
of PDF constraints, the neutrino scattering target
as composed by isoscalar free-nucleons.
They hence neglect both nuclear modifications and non-isoscalar effects.
We now revisit these analyses but accounting for the fact that the target material
in the FASER$\nu$2 and AdvSND experiments is tungsten, with $A=184$.
In this exercise we do not include the FLArE structure function data,
corresponding to a different target material (liquid argon, with $A=40$), although
in an actual nuclear PDF fit all FPF datasets would be included simultaneously.
Nuclear modifications associated to a tungsten target, when compared
to a free isoscalar nucleon, are not necessarily small, and  will affect the event rate predictions for
the FPF experiments.
This fact also implies that  measurements of differential neutrino cross-sections
on heavy nuclear targets provide direct constraints on these nuclear modifications
without relying on assumptions on their $A$ dependence.

We quantify the impact of the FPF structure function measurements
on the nuclear PDFs of tungsten by applying the same Hessian profiling
of Sects.~\ref{sec:pdf4lhc21} to EPPS1, a global determination of nuclear PDFs
that accounts for the constraints of existing datasets involving nuclei as target or projectiles,
and based on the CT18 set as proton boundary condition.
We note that EPPS21 already includes information from neutrino DIS on nuclear targets
from the CHORUS and NuTeV experiments.
EPPS21 is generally in good agreement with other recent nPDF fits such as nNNPDF3.0~\cite{AbdulKhalek:2022fyi}
and nCTEQ15HQ~\cite{Duwentaster:2022kpv}.

The application of profiling to EPPS21 follows the same strategy as that
for PDF4LHC21, with the caveat that its Hessian error sets also include the contribution
from the uncertainties  associated to their reference proton PDF set, in this case CT18.
Given that the measured event rates depend on both the proton PDFs and the nuclear modifications,
when profiling EPPS21 we also account for the Hessian sets associated to the proton
PDF dependence.
Also relevant for the subsequent discussion, EPPS21 adopts a tolerance
factor of $T^2 = 20$ when defining their  68\%  confidence level uncertainties,
to be contrasted with  $T^2 = 10$ used in PDF4LHC21.
 
\begin{figure*}[t]
\centering
\includegraphics[width=0.32\textwidth]{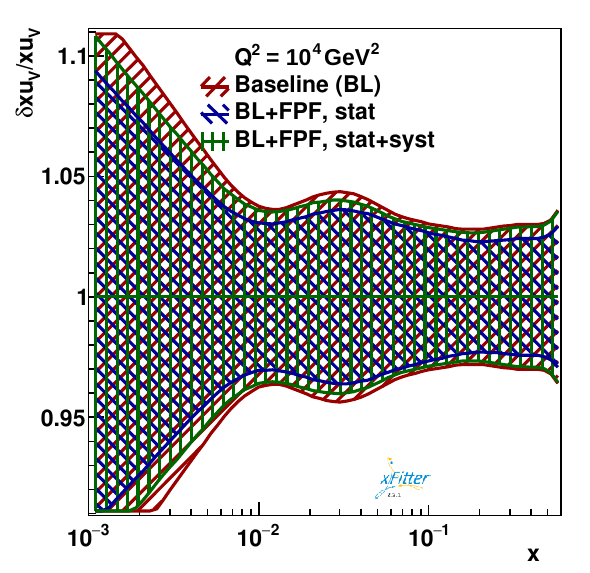}
\includegraphics[width=0.32\textwidth]{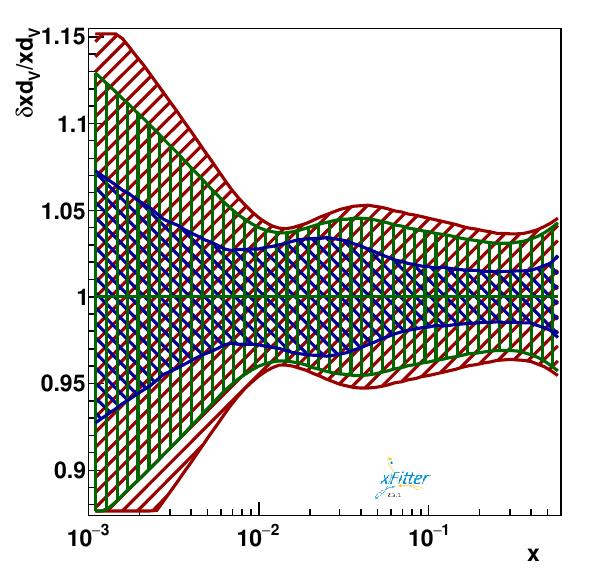}
\includegraphics[width=0.32\textwidth]{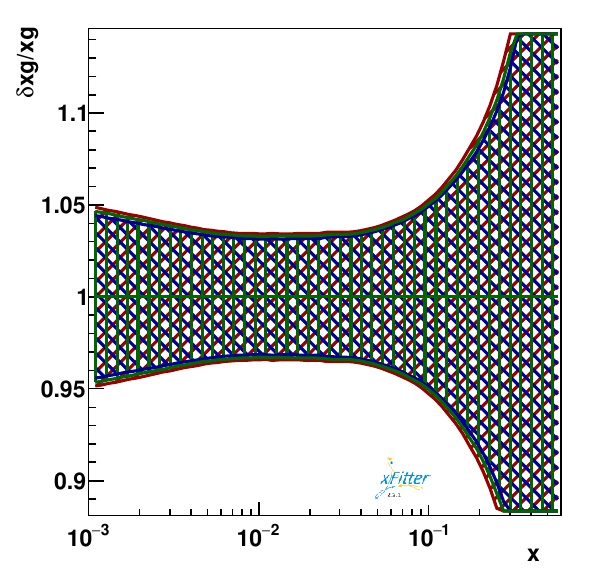}\\
\includegraphics[width=0.32\textwidth]{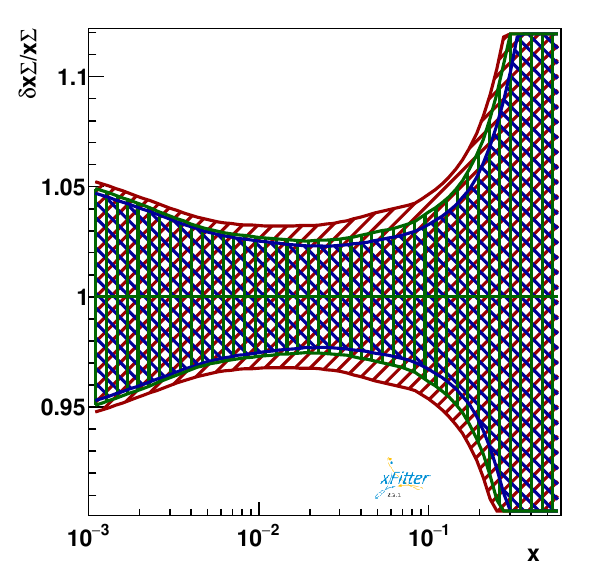}
\includegraphics[width=0.32\textwidth]{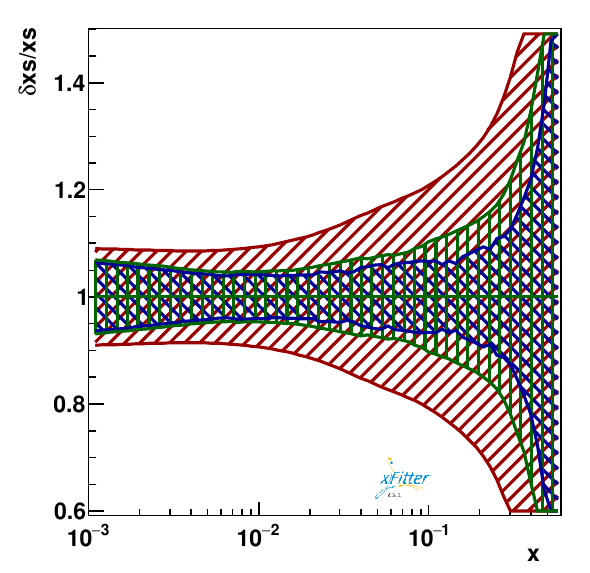}
\caption{Same as Fig.~\ref{fig:FPF_combined} now corresponding to the Hessian
 profiling of EPPS21 for a tungsten nucleus, and excluding FLArE contributions.
}
\label{fig:profiling_FPF_nuclear}
\end{figure*}

Fig.~\ref{fig:profiling_FPF_nuclear} displays a similar comparison as that of
Fig.~\ref{fig:FPF_combined}, now corresponding to the Hessian
profiling of EPPS21 for a tungsten nucleus.
The FPF dataset is in this case composed by FASER$\nu$2 and AdvSND.
One finds that the impact of FPF structure function measurements
is most important for the strange PDF, and to a lesser extent for the up and down
valence quark PDFs.
For the latter, the projected impact of FPF measurements appears to be somewhat milder
as in the case of PDF4LHC21, especially once systematic uncertainties are taken into account.

While the main qualitative features of the nPDF profiling are consistent with the free-proton
case, one reason explaining the observed differences may be the usage of a larger tolerance factor $T$ in EPPS21
as compared to PDF4HC21, which effectively reduces the impact of new data in the fit.
Another potential reason is the more restrictive functional form adopted in EPPS21 as compared
to the PDF sets entering PDF4LHC21, a consequence of the smaller dataset available
for hard scattering on nuclear targets.
Given that these  functional forms are fixed when applying Hessian profiling,
assessing the impact on a nuclear PDFs by a direct refit, as was done for NNPDF4.0,
may be required to establish the impact of FPF structure function data
on nPDFs.
Additional results for the profiling of EPPS21 are presented in App.~\ref{app:nPDF_impact_appendix}.

\section{Implications for Higgs and weak boson production}
\label{sec:pheno}

The reduction of PDF uncertainties made possible by neutrino DIS measurements
at the FPF, discussed in Sect.~\ref{sec:protonPDFs},
enables more precise theoretical predictions for core processes at the
HL-LHC.
Here we present an initial study of the phenomenological implications
of the PDFs enhanced with LHC neutrino data
for hard-scattering processes at proton colliders.
Specifically, we present results for 
single and double gauge boson production,  Higgs production in
vector-boson fusion, and Higgs production and in association with a vector boson.
We focus on processes sensitive to the quark-quark and quark-antiquark initial
states, given that LHC neutrinos do not provide information on the gluon
PDF and hence they cannot inform  gluon-initiated
processes, such as top quark pair production or Higgs production in gluon fusion.

We adopt the same calculational settings as in the LHC phenomenology analysis considered
in the PDF4LHC21 combination study~\cite{PDF4LHCWorkingGroup:2022cjn} and provide predictions
both for inclusive cross-sections, integrated in the fiducial
region, and for differential distributions.
We evaluate these cross-sections using matrix elements
which include NLO corrections both in the
QCD and electroweak coupling using
{\sc\small mg5\_amc@nlo}~\cite{Frederix:2018nkq}
interfaced to {\sc\small PineAPPL}~\cite{Carrazza:2020gss}.
For all processes, realistic selection and acceptance cuts on the final state particles
have been applied.
No further theory uncertainties are considered in this
analysis, given that we don't aim to compare with experimental data.

\begin{figure*}[htbp]
\centering
\includegraphics[width=0.32\textwidth]{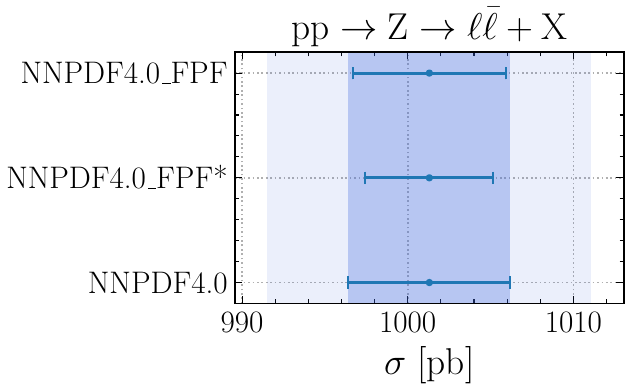}
\includegraphics[width=0.32\textwidth]{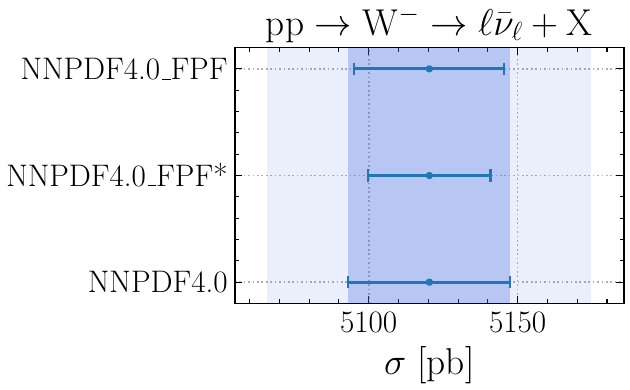}
\includegraphics[width=0.32\textwidth]{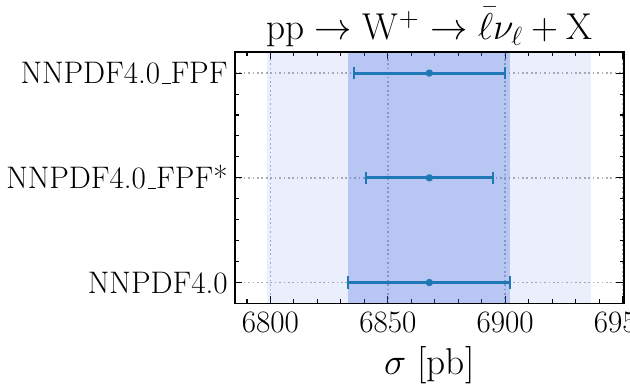}
\includegraphics[width=0.32\textwidth]{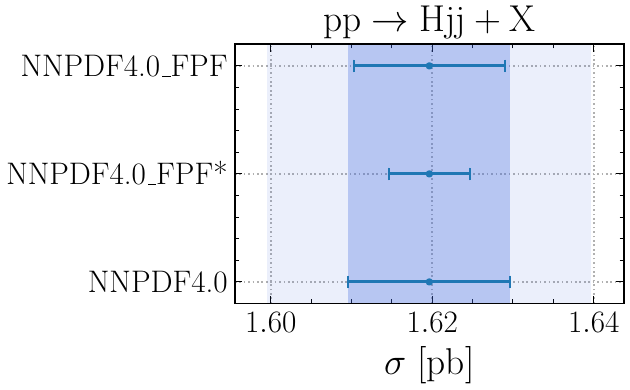}
\includegraphics[width=0.32\textwidth]{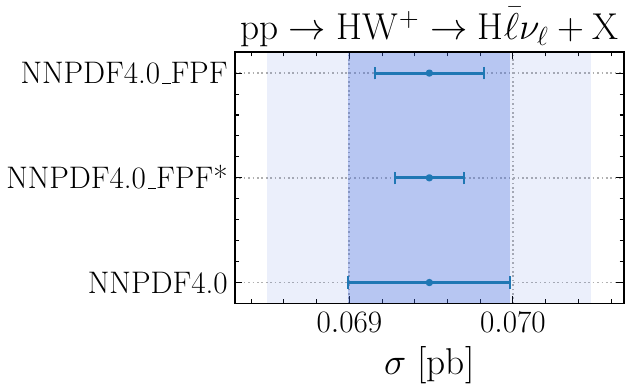}
\includegraphics[width=0.32\textwidth]{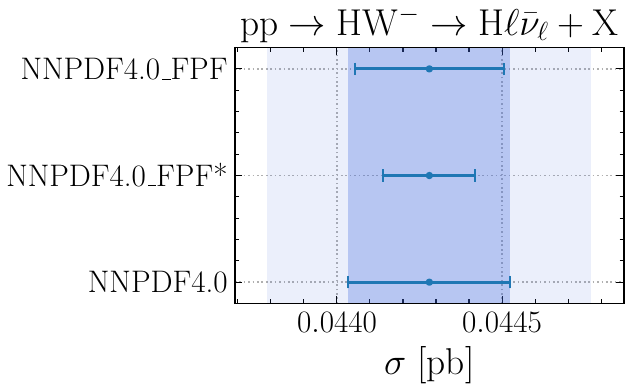}
\includegraphics[width=0.32\textwidth]{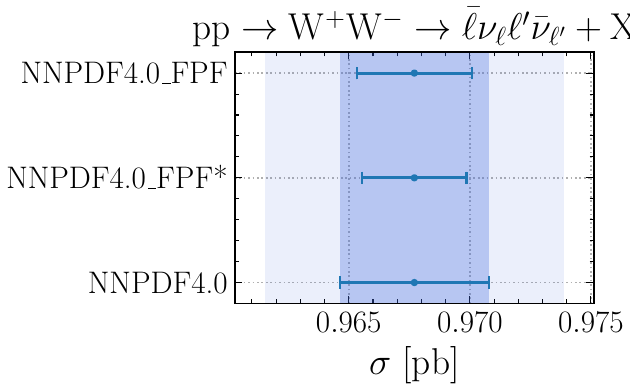}
\includegraphics[width=0.32\textwidth]{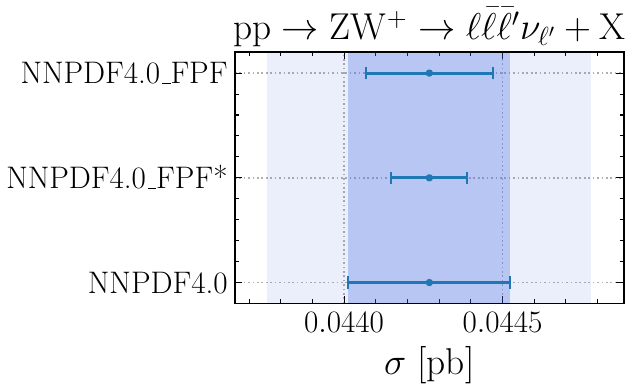}
\includegraphics[width=0.32\textwidth]{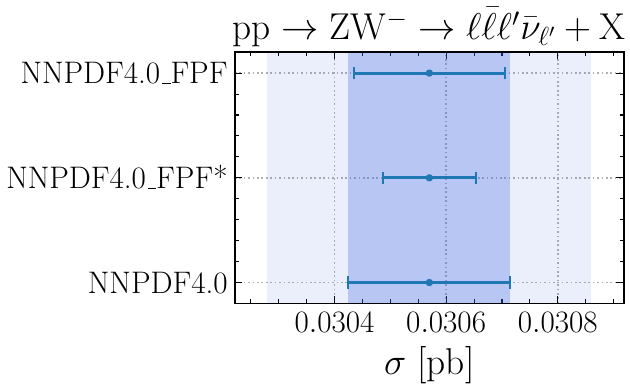}
\includegraphics[width=0.32\textwidth]{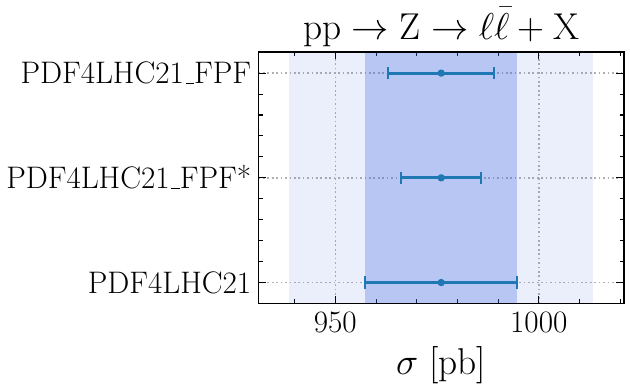}
\includegraphics[width=0.32\textwidth]{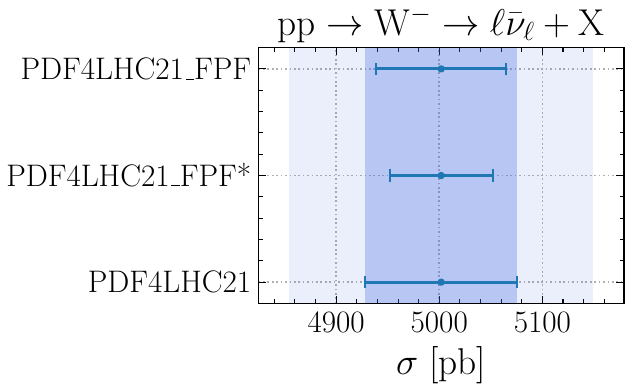}
\includegraphics[width=0.32\textwidth]{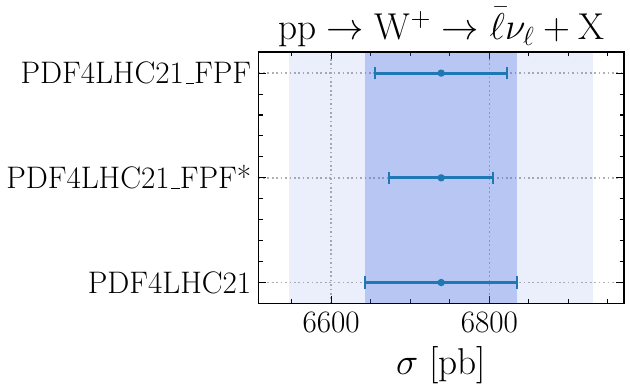}
\includegraphics[width=0.32\textwidth]{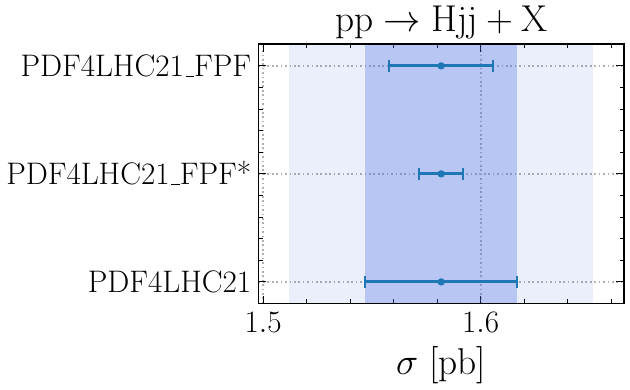}
\includegraphics[width=0.32\textwidth]{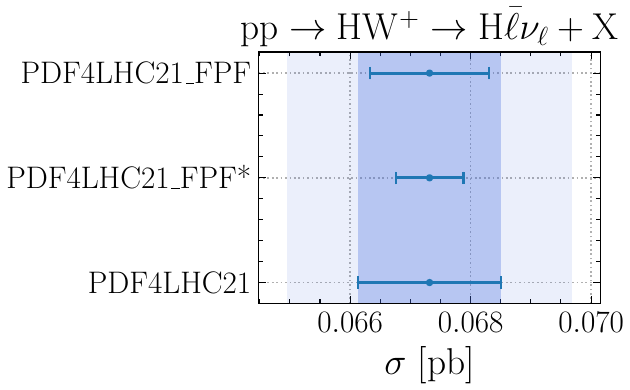}
\includegraphics[width=0.32\textwidth]{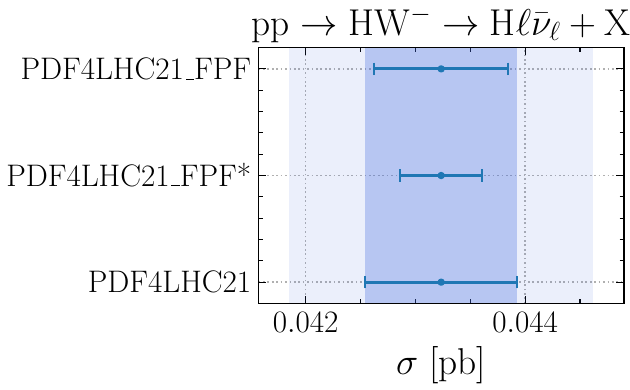}
\includegraphics[width=0.32\textwidth]{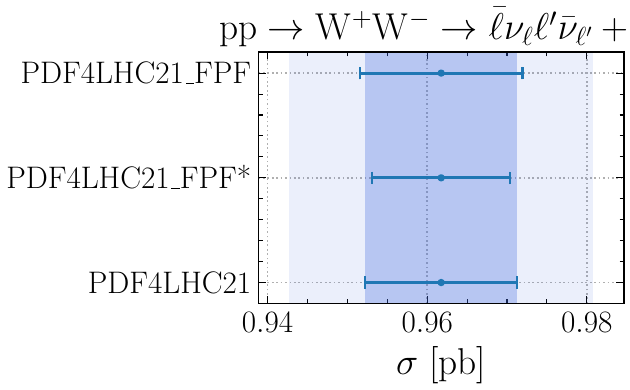}
\includegraphics[width=0.32\textwidth]{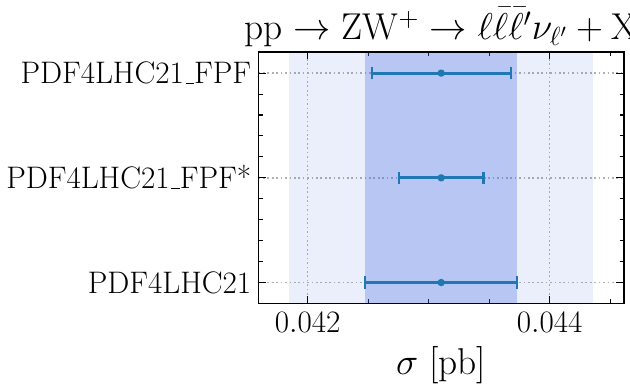}
\includegraphics[width=0.32\textwidth]{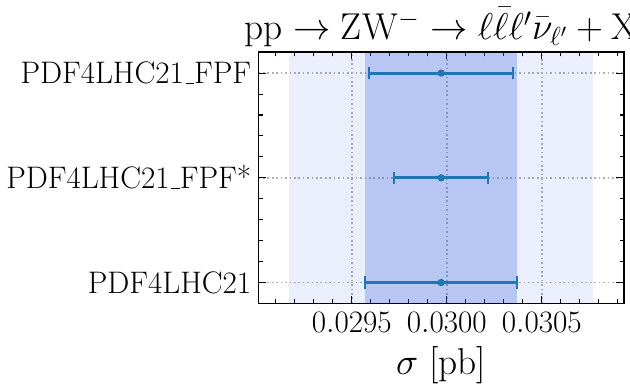}
\caption{Fiducial cross-sections for representative LHC processes at $\sqrt{s}=14$ TeV
  evaluated with NNPDF4.0 (upper panels) and PDF4LHC21 (bottom panels),
  compared with the fits including the FPF structure function projections.
  See~\cite{NNPDF:2021njg,PDF4LHCWorkingGroup:2022cjn} for the calculational settings
  used to evaluate these cross-sections.
For the baseline
predictions, the dark (light) bands indicate the 68\% (95\%) CL uncertainties.
The fits labelled as ``\_FPF$^*$'' are the ones considering statistical uncertainties,
while those labelled as ``\_FPF'' also include systematic errors.
In the fits with FPF data, the central values are set to be the same as
in the corresponding baseline.
We provide predictions for inclusive Drell-Yan production ($Z, W^+, W^-$), Higgs production
in vector-boson fusion, Higgs associated
production, and diboson production ($W^+W^-$, $W^+Z$, $W^-Z$).
}
\label{fig:NNPDF40_pheno_integrated}
\end{figure*}

\begin{figure*}[htbp]
\centering
\includegraphics[width=0.49\textwidth]{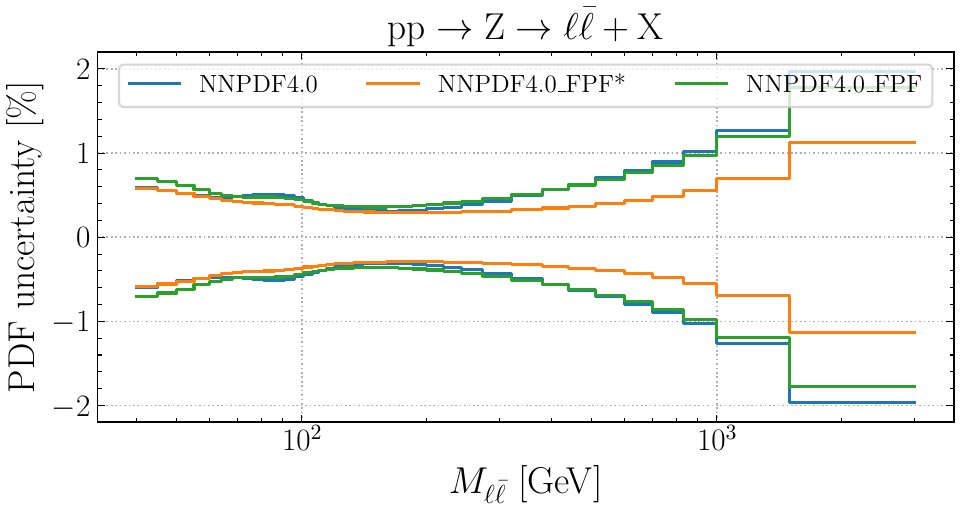}
\includegraphics[width=0.49\textwidth]{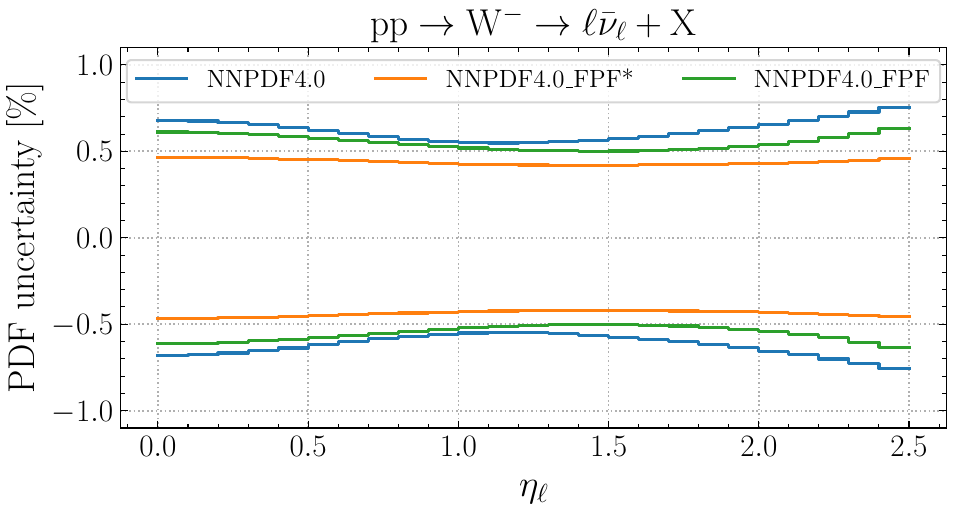}
\includegraphics[width=0.49\textwidth]{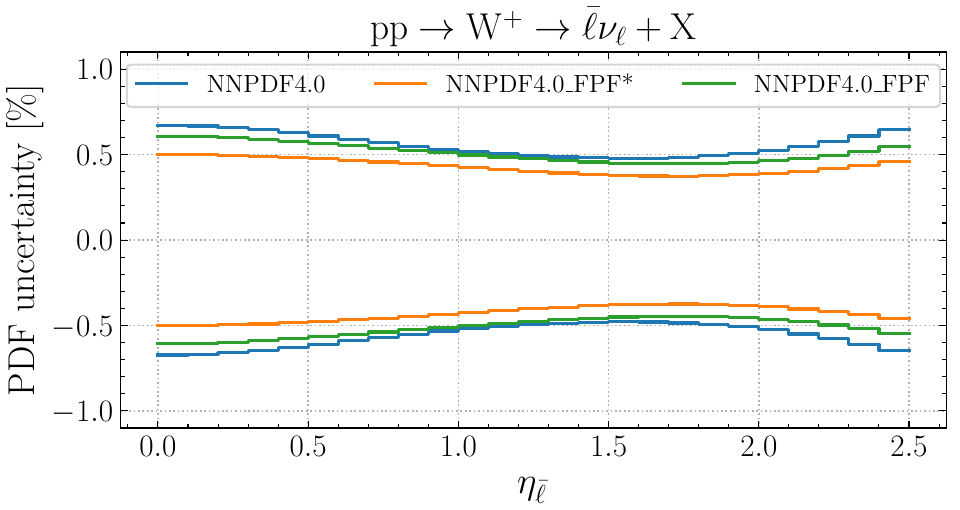}
\includegraphics[width=0.49\textwidth]{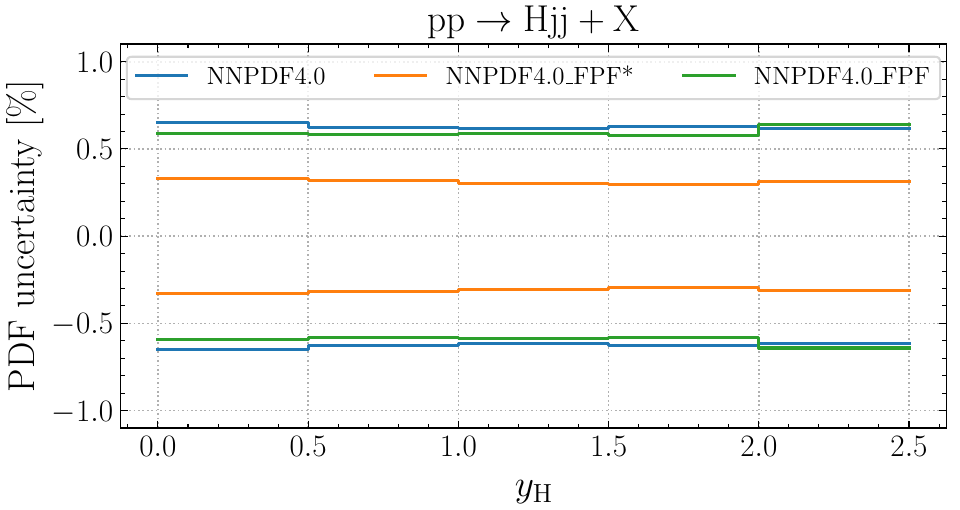}
\includegraphics[width=0.49\textwidth]{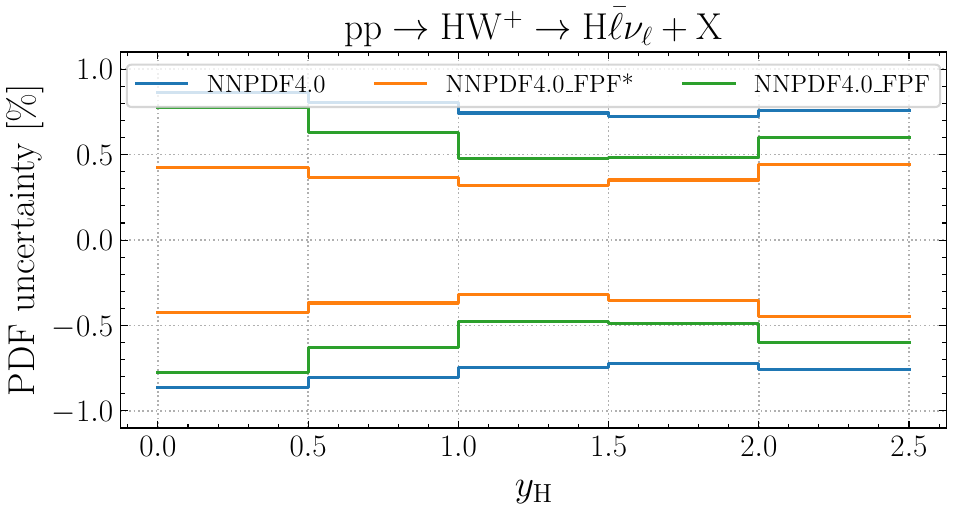}
\includegraphics[width=0.49\textwidth]{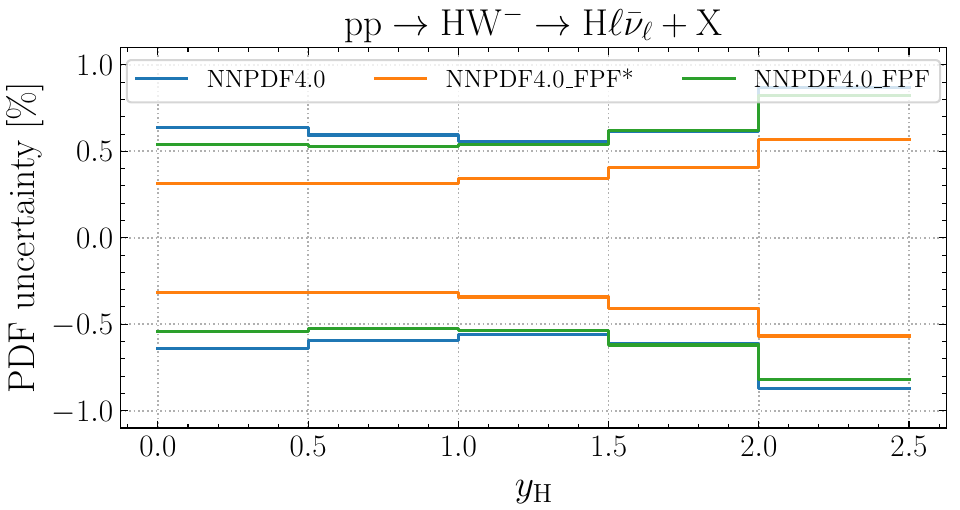}
\includegraphics[width=0.49\textwidth]{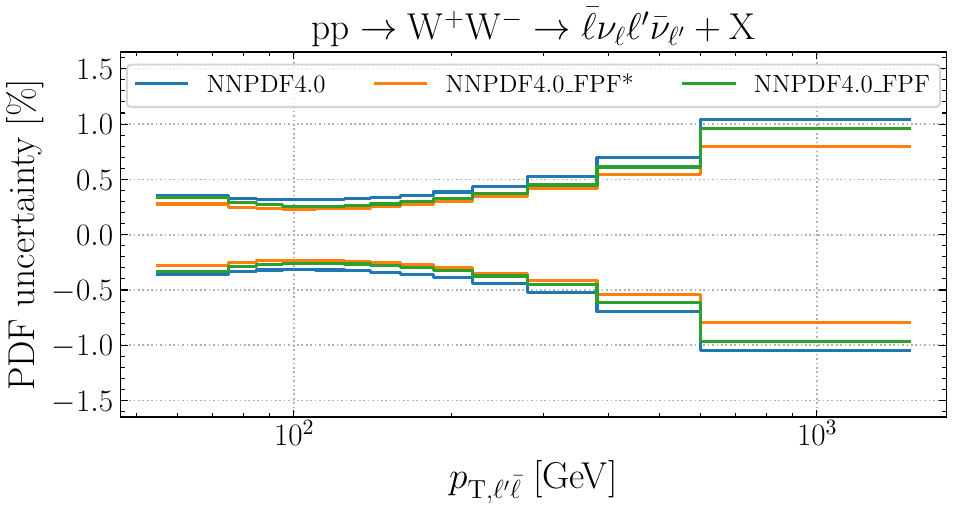}
\includegraphics[width=0.49\textwidth]{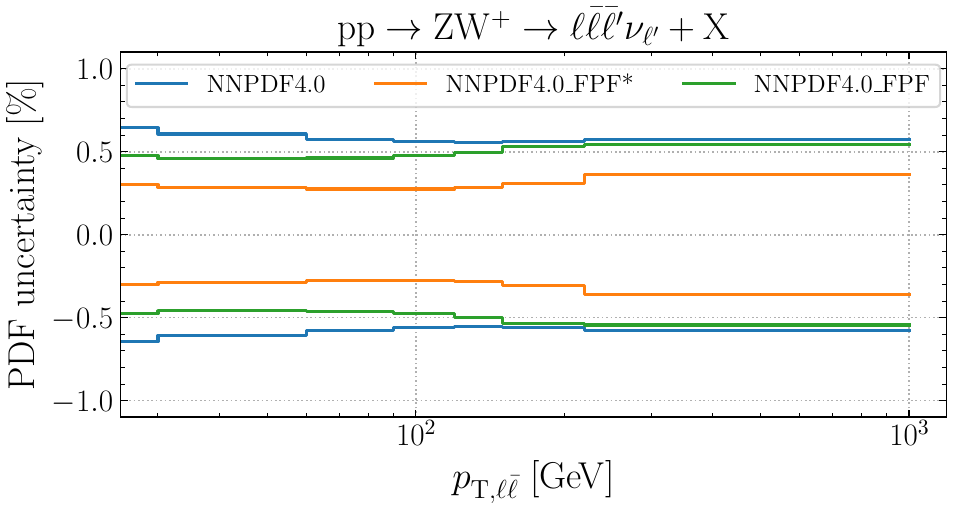}
\includegraphics[width=0.49\textwidth]{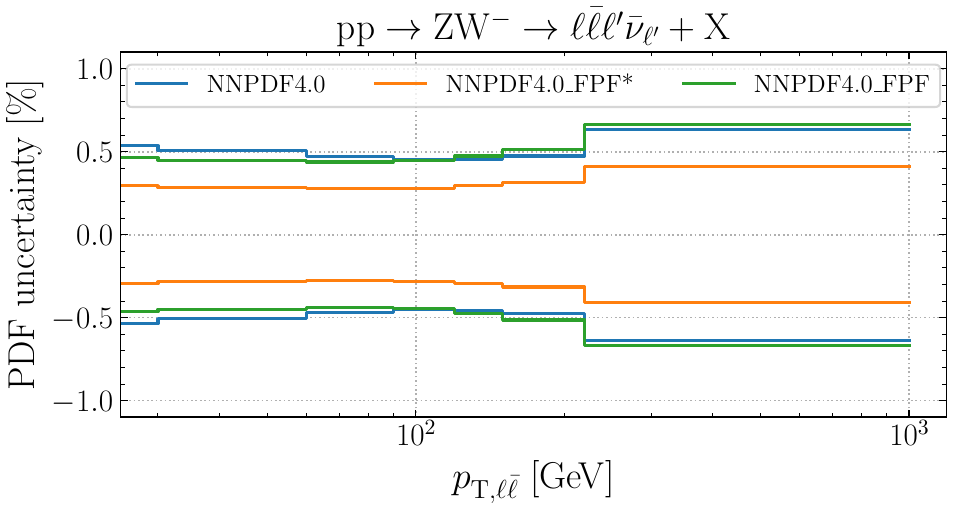}
\caption{Same as Fig.~\ref{fig:NNPDF40_pheno_integrated}
for the corresponding differential distributions in the case of NNPDF4.0
}
\label{fig:NNPDF40_pheno_differential}
\end{figure*}

\begin{figure*}[htbp]
	\centering
	\includegraphics[width=0.49\textwidth]{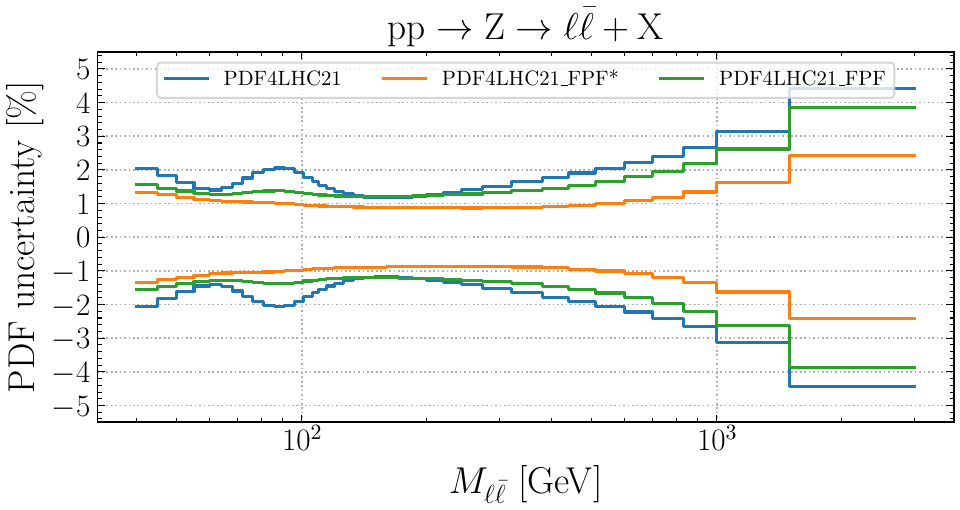}
	\includegraphics[width=0.49\textwidth]{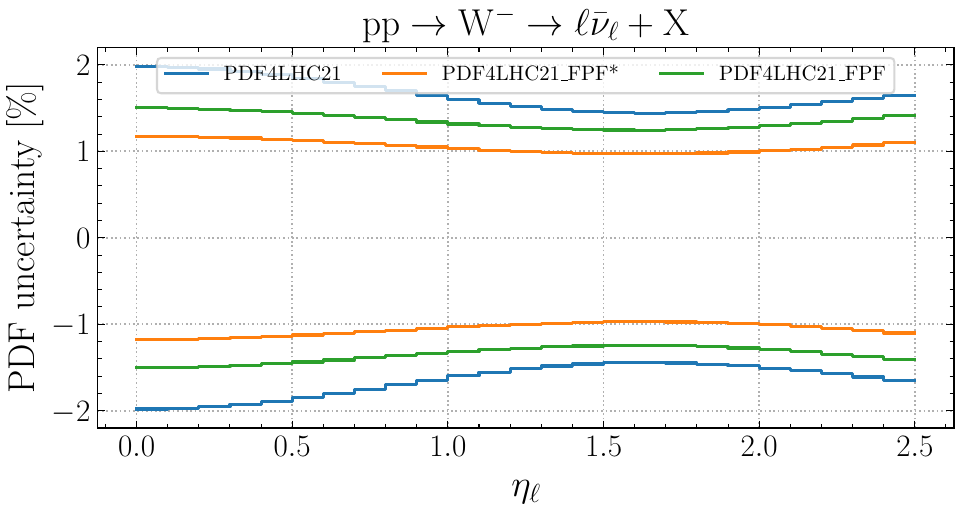}
	\includegraphics[width=0.49\textwidth]{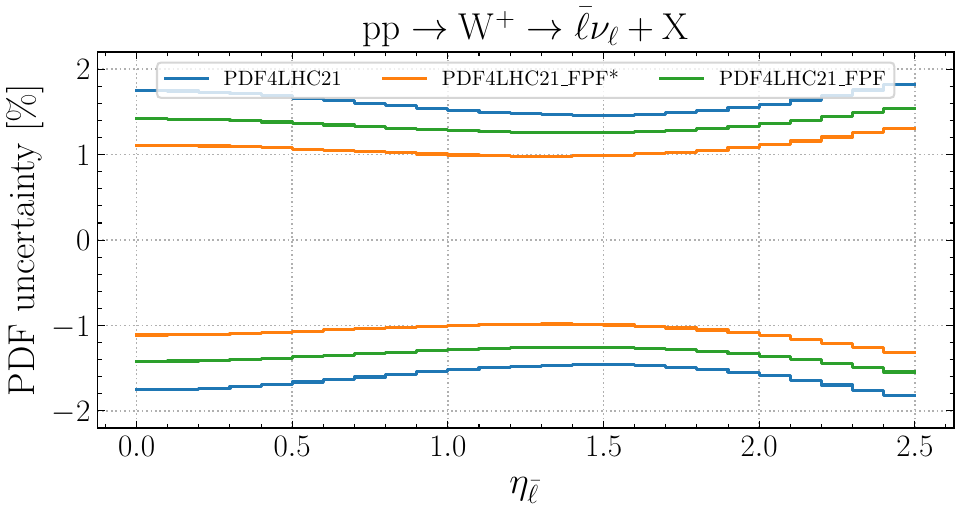}
	\includegraphics[width=0.49\textwidth]{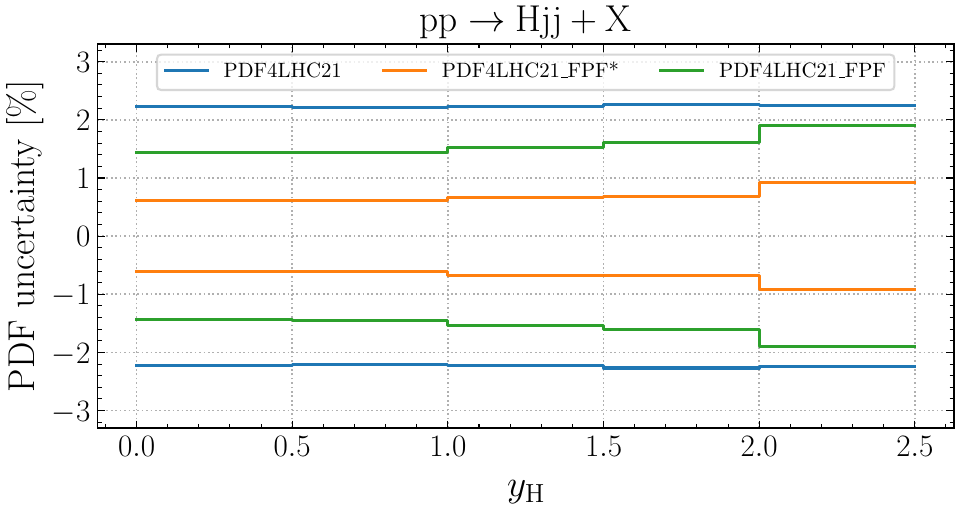}
	\includegraphics[width=0.49\textwidth]{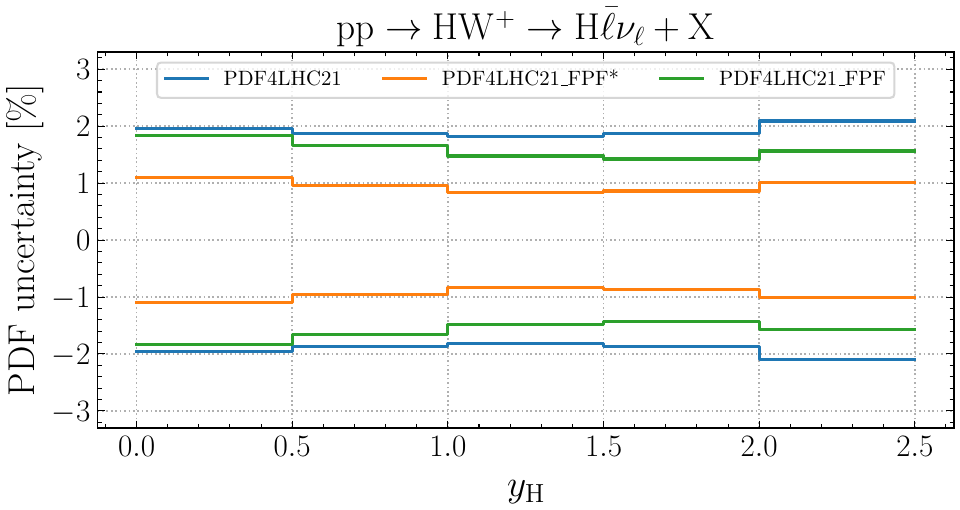}
	\includegraphics[width=0.49\textwidth]{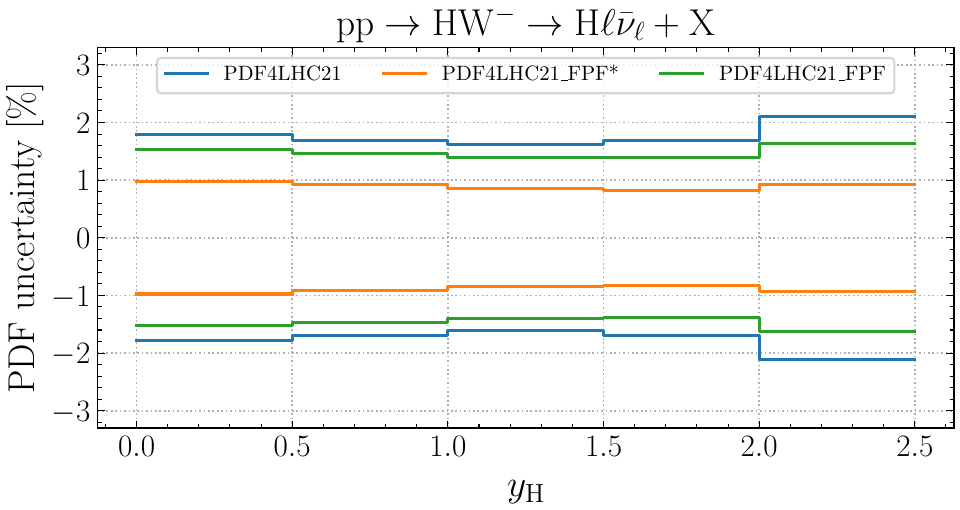}
	\includegraphics[width=0.49\textwidth]{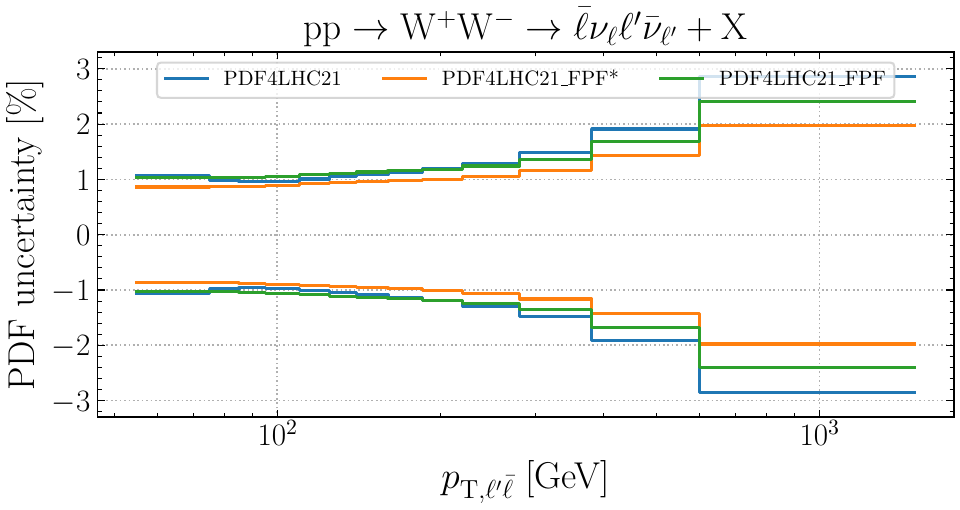}
	\includegraphics[width=0.49\textwidth]{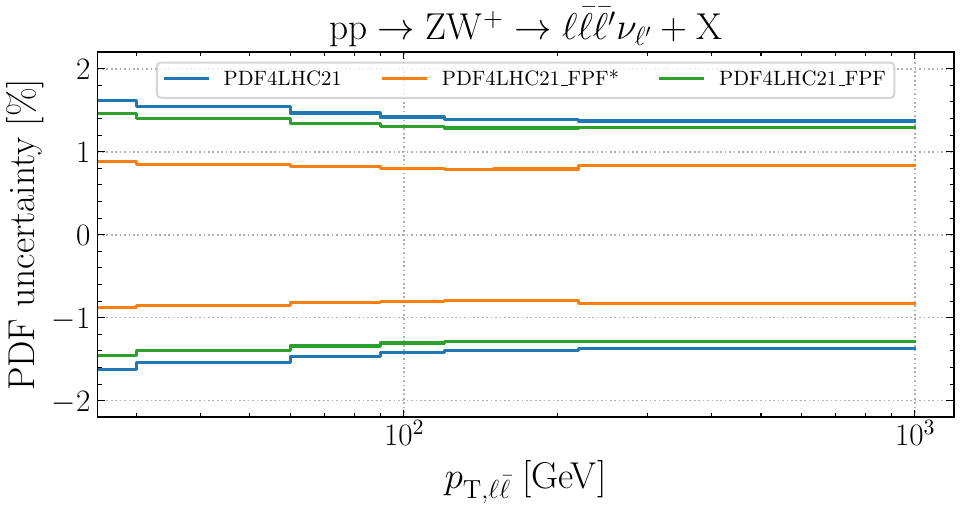}
	\includegraphics[width=0.49\textwidth]{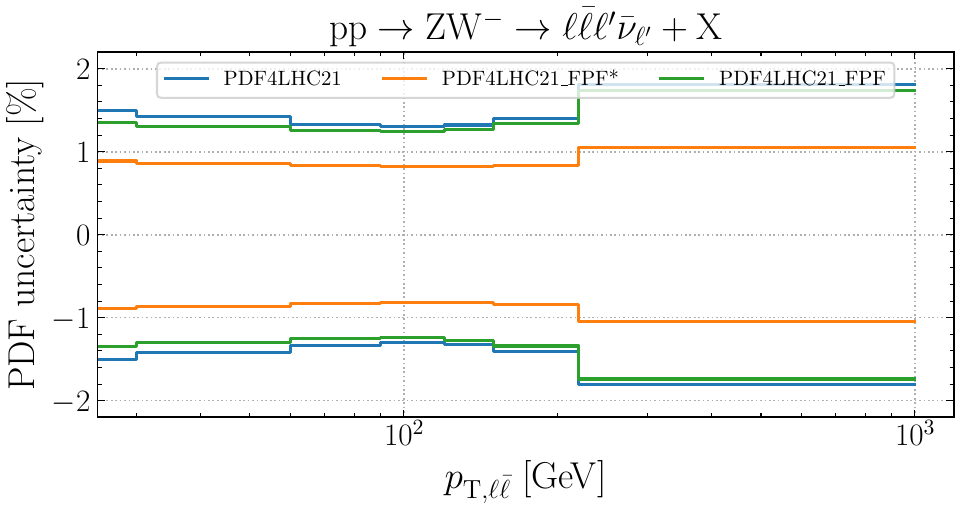}
	\caption{
		Same as Fig.~\ref{fig:NNPDF40_pheno_differential} but with PDF4LHC21.
	}
	\label{fig:PDF4LHC21_pheno_differential}
\end{figure*}

Fig.~\ref{fig:NNPDF40_pheno_integrated} displays
fiducial cross-section for representative LHC processes at $\sqrt{s}=14$ TeV
evaluated with NNPDF4.0 (top panels)
and with PDF4LHC21 (bottom panels),
compared with the results obtained from
the corresponding fits including the FPF structure function projections.
For the fits with FPF data we display the variants where the covariance matrix
consists only of statistical errors (``\_FPF$^*$'') and where also
systematic errors are accounted for (``\_FPF'').
For the baseline
predictions, the dark (light) bands indicate the 68\% (95\%) CL uncertainties.
The central values are set to be the same as in the baseline calculation in all cases.
From top to bottom, we show inclusive Drell-Yan production in the $Z$, $W^+$, and $W^-$
channels, Higgs production
in vector-boson fusion, Higgs associated
production, and diboson production in the $W^+W^-$, $W^+Z$, and $W^-Z$ final-states.
The corresponding comparisons at the level of differential distributions
are shown in Figs.~\ref{fig:NNPDF40_pheno_differential} and~\ref{fig:PDF4LHC21_pheno_differential}
for NNPDF4.0 and PDF4LHC21 respectively.
As done for the fiducial cross-sections of Fig.~\ref{fig:NNPDF40_pheno_integrated},
we only indicate the relative PDF uncertainty in each fit.

Inspection of Figs.~\ref{fig:NNPDF40_pheno_integrated}--\ref{fig:PDF4LHC21_pheno_differential}
confirms the potential of LHC neutrino structure function measurements
to improve theoretical predictions for Higgs, electroweak and high-mass
processes at the HL-LHC.
In general, qualitative consistency between the results based on NNPDF4.0
and on PDF4LHC21 is found, as was already the case at the PDF level.
As  compared to the best-case scenario where only statistical uncertainties
are taken into account, the reduction of PDF errors in the LHC cross-sections
obtained thanks for the FPF structure functions becomes less marked,
but still visible for most of the processes under consideration,
upon the inclusion of systematic uncertainties in the fit covariance matrix.

Concerning the fiducial integrated cross-sections, a decrease in PDF
uncertainties is observed for all processes,
including for single gauge boson production relevant for core
HL-LHC analyses such as the $m_W$ and $\sin^2\theta_W$ measurements.
The specific improvement in precision depends
on the underlying scattering reaction as well as on the range
of $x$ and $Q^2$ covered by each process.
In the case of  Higgs associated production with vector bosons and in vector-boson
scattering,
driven by the quark-antiquark and quark-quark initial states respectively,
one observes that PDF uncertainties can  be reduced
by up to a factor two thanks to the FPF measurements, for instance
in the case of the $hW^+$ and $hW^-$ cross-sections,
in the most optimistic scenario in which systematic uncertainties
are neglected.
Likewise for the diboson cross-sections, with in this case the largest
improvement observed for the $ZW^+$  and  $ZW^-$ final states, with a reduction
of up to $\sim 40\%$ in the PDF uncertainties.

In the case of the differential cross-sections displayed in Figs.~\ref{fig:NNPDF40_pheno_differential}
and~\ref{fig:PDF4LHC21_pheno_differential},
one observes
how the impact of the FPF structure functions on LHC observables depends
on the hard-scattering scale.
For instance, searches for heavy resonances in the high-mass tail of the Drell-Yan
distributions could be improved by the addition of the FPF data.
Similar considerations apply for diboson prediction, and in the case of the $ZW^{\pm}$ channel we observe
an improvement specially in the low $p_{T,\ell\bar{\ell}}$ region which
represents the bulk of the fiducial cross-sections.
For the Higgs production processes, the PDF uncertainty in the theory predictions is relatively
stable as a function of the rapidity.
The effects of accounting for systematic uncertainties in the fit covariance
matrix are somewhat more visible here as compared to the inclusive cross-sections,
indicating that they affect mostly
the tails, rather than the bulk, of the distributions, and in particular the large-$x$
behaviour of the PDFs.
This observation emphasizes the importance of reducing systematic errors
in the FPF measurements in order to enhance the cross-talk with HL-LHC analyses.

For completeness, App.~\ref{app:comparisons_with_HLLHC} revisits
the phenomenological studies  presented in this section for the fiducial cross-sections
in terms of the  HL-LHC PDF impact projections of~\cite{AbdulKhalek:2018rok}.
This comparison highlights the complementarity between the two experiments
in terms of PDF constraints, being fully orthogonal and arising from different
scattering process.
Furthermore, the FPF constraints benefit from the valuable property
that an eventual contamination from new physics on the PDF fits 
can be neglected, since this process is driven by $Q^2$ values outside 
the possible presence of BSM physics, at least concerning new heavy states.
Therefore, the ultimate sensitivity on high-precision observables being
obtained by integrating the constraints both from the HL-LHC and from the FPF
in the same global determination of PDFs.

\section{Summary and outlook}
\label{sec:summary}

In this work we have, for the first time,
quantified the impact that measurements of high-energy
neutrino DIS structure functions at the LHC would have on the quark
and gluon structure of the nucleon.
Our analysis requires the generation of
DIS pseudo-data fully differential in $x$, $Q^2$, and $E_\nu$
for the various ongoing and proposed far-forward
neutrino LHC experiments, including the estimate
of their associated systematic uncertainties.
Consistent results are obtained from both the Hessian profiling
of PDF4LHC21 and from the direct inclusion of LHC neutrino
structure functions in the NNPDF4.0 global
fit, revealing a reduction of PDF uncertainties in the
light quark sector, in particular concerning strangeness, as well
as for the large-$x$ charm PDF in the case of NNPDF4.0.
We have assessed the  robustness of these results upon
removing charm-tagged data and final-state lepton-charge
identification, as well as upon the combination
of all FPF experiments within a single analysis.

We have also demonstrated the rich interplay between
far-forward and central measurements
at the HL-LHC by providing predictions
for a range of Higgs and gauge boson production processes,
both for integrated cross-sections in the fiducial region and for
single-differential distributions.
This phenomenological analysis suggests that a reduction
of the  PDF uncertainty by up to a factor two
may be within reach for some of these cross-sections
in the most optimistic scenario.
As was the case at the PDF level, also for 
the HL-LHC projections results based on  PDF4LHC21 and NNPDF4.0
are in qualitative agreement.

Several avenues extending the results of this work may be foreseen.
First of all, as the design of the proposed FPF experiments
 advances, it will be possible to derive
 more accurate estimates of the systematic
uncertainties (and of their correlations) 
which eventually become the limiting factor.
This will allow studying whether  improved detection methods,
novel reconstruction techniques i.e. based
on deep learning, or combining information from different
experiments can  push down the systematic uncertainties
affecting the measurements.
Second, the modelling of neutrino scattering at the LHC would benefit from
the use of Monte Carlo event generators accounting
for higher-order QCD corrections.
As compared to the currently used LO generators,
these would improve the description of the final-state
kinematics, which in turn determine the acceptance rate
of the reconstructed events.
Furthermore, the predictions from such precise Monte Carlo generators
should be folded with a full-fledged detector simulation
in order to robustly determine selection efficiencies, e.g.
such as those related to charm and $D$-meson tagging.

Third, while here we focus on the DIS
region, ongoing and future LHC neutrino experiments also provide
important information on shallow-inelastic scattering (SIS)
 at lower values of $Q$~\cite{Jeong:2023hwe,Candido:2023utz},
 which in turn are relevant for inclusive cross-sections
entering atmospheric and oscillation neutrino experiments.
By following the approach presented in this work, it should be possible to quantify
the improvements that LHC data provides on models
of neutrino scattering in this poorly-understood SIS region.
Finally, here we have taken the incoming neutrino fluxes
as an external input and neglected any associated uncertainties.
However, measurements of these fluxes provide
unique information on light and heavy forward hadron
production in QCD, and in particular open
a new window to the small-$x$ gluon PDF.
For this reason, ultimately one needs to simultaneously constrain 
the incoming fluxes and the neutrino scattering cross-sections
from the measured event rates.
Such joint interpretation would require extending the present
analysis with a data-driven parametrisation of the neutrino fluxes,
which could subsequently be compared with different theoretical predictions.

Our findings highlight how exploiting the LHC neutrino beam for hadron structure
studies effectively provides CERN with a ``Neutrino-Ion Collider'', a charged
current-counterpart of the EIC, without  changes in
its accelerator infrastructure or additional energy costs.
In addition for their intrinsic interest for hadronic science,
measurements of neutrino structure function at the LHC
provide a novel, and until now ignored, handle to inform theoretical
predictions of hard-scattering cross-sections  at the HL-LHC.

\begin{center}
\rule{5cm}{.1pt}
\end{center}
\bigskip

Together with this paper, we 
make public in Zenodo~\cite{cruz_martinez_juan_2023_8355209}
the corresponding {\sc\small LHAPDF} grids~\cite{Buckley:2014ana}
for the PDF4LHC21 and NNPDF4.0 fits including FPF data.
We also release
the generated neutrino structure function pseudo-data
and the corresponding theory
calculations, for the different scenarios considered in this work.
These projections for neutrino structure functions
should be of relevance for a broad range of
applications related to forward
neutrino scattering at the LHC, from tests
of lepton flavour university at the TeV scale
in the neutrino sector to probes of anomalous neutrino
interactions and searches for sterile neutrinos
distorting  oscillation patterns.

\subsection*{Acknowledgments}
We are grateful to many colleagues involved in the Forward
Physics Facility initiative for illuminating
discussions and encouragement along the course of this project,
in particular Jamie Boyd, Jonathan Feng, and Albert de Roeck.
We thank specially Felix Kling for many useful discussions
and for providing updated predictions for the neutrino fluxes.
We thank Akitaka Ariga and Tomoko Ariga for discussion
concerning FASER$\nu$2, Milind Diwan,
Wenjie Wu, and Steven Linden concerning FLArE,
and Antonia di Crescenzo for information concerning AdvSND.
We thank Christopher Schwan, Simone Amoroso and Xiaomin Shen for useful 
discussions regarding the implementation and testing of the 
{\sc\small PineAPPL} interface to {\sc\small xFitter}.
We also thank Emanuele R. Nocera for providing us with the EIC pseudodata.

The work of M.~F. was supported by NSF Grant PHY-2210283 and was also supported by NSF Graduate Research Fellowship Award No. DGE-1839285.
The work of T.~G., G.~M., and J.~R. is partially supported by NWO, the Dutch Research Council.
The work of T.~R. and J.~R. is partially supported by an ASDI2020
Fellowship from the Netherlands eScience Center.
The work of T.~M is supported by the National Science Centre, Poland, research grant No. 2021/42/E/ST2/00031.

\appendix
\section{Comparison with HL-LHC PDF projections}
\label{app:comparisons_with_HLLHC}

\begin{figure*}[!h]
	\centering
	\includegraphics[width=0.32\textwidth]{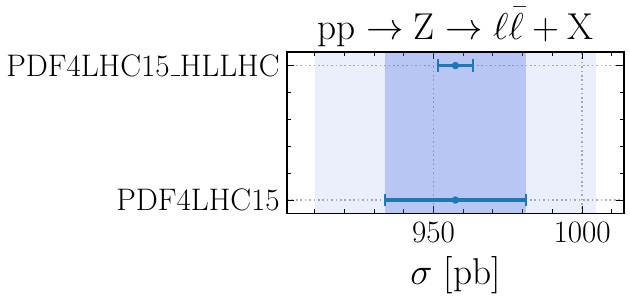}
	\includegraphics[width=0.32\textwidth]{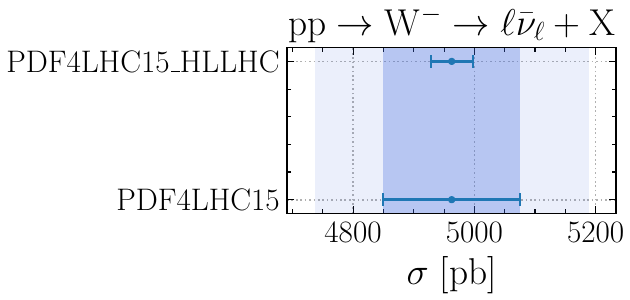}
	\includegraphics[width=0.32\textwidth]{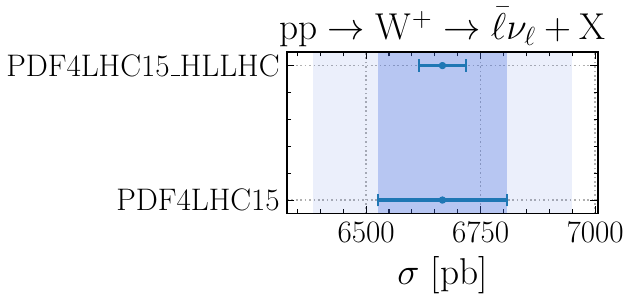}
	\includegraphics[width=0.32\textwidth]{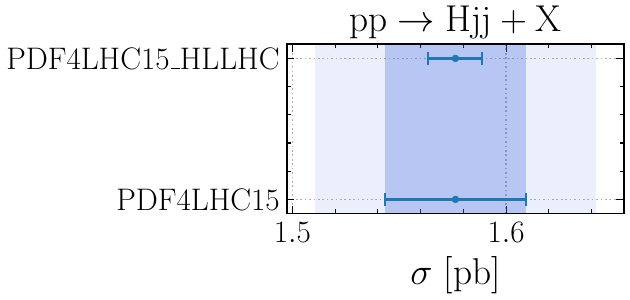}
	\includegraphics[width=0.32\textwidth]{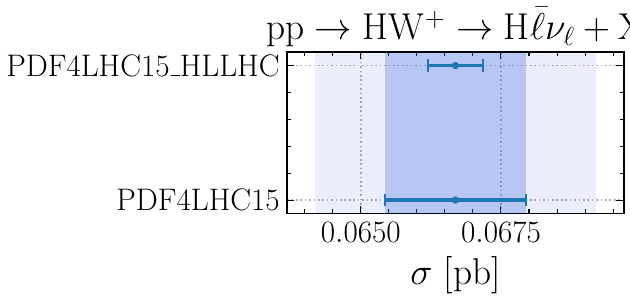}
	\includegraphics[width=0.32\textwidth]{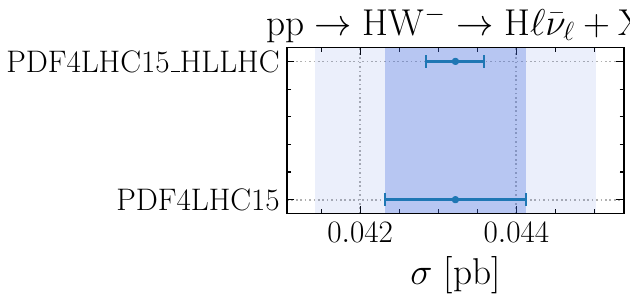}
	\includegraphics[width=0.32\textwidth]{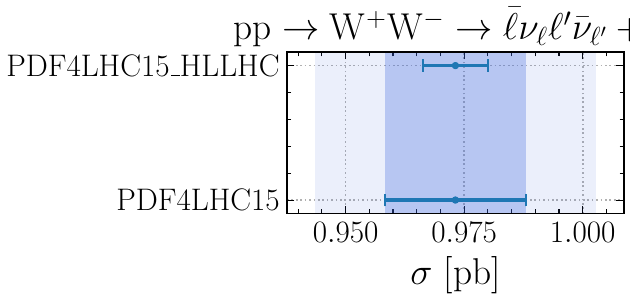}
	\includegraphics[width=0.32\textwidth]{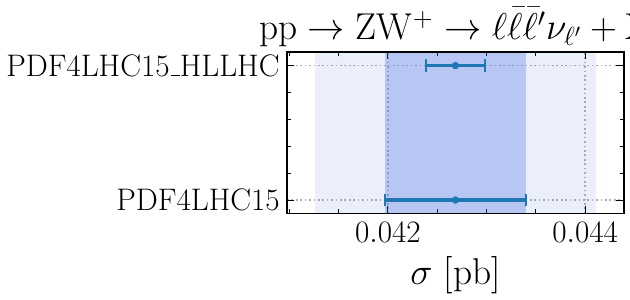}
	\includegraphics[width=0.32\textwidth]{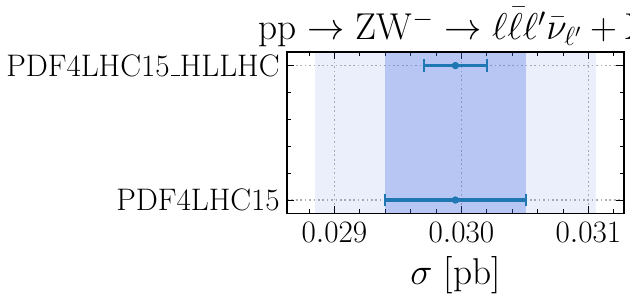}
	\caption{The same LHC fiducial cross-sections as in Fig.~\ref{fig:NNPDF40_pheno_integrated},
          now comparing the PDF4LHC15 baseline predictions with those based
          on the PDFs including HL-LHC pseudo-data from~\cite{AbdulKhalek:2018rok}.
          Specifically, we consider the HL-LHC PDF projections
          from ``Scenario B'', corresponding to an intermediate scenario for the expected reduction
          of systematic uncertainties. }
	\label{fig:HLLHC_pheno_integrated}
\end{figure*}

Here we revisit the phenomenology studies of
Sect.~\ref{sec:pheno} using the HL-LHC PDF impact projections
presented in~\cite{AbdulKhalek:2018rok}.
These projections were obtained with the same Hessian profiling strategy
as discussed in Sect.~\ref{sec:dis_pseudodata}, with the important
difference that the prior PDF set was PDf4LHC15 rather than
PDF4LHC21.
Therefore, while a direct comparison with the results presented in
Sect.~\ref{sec:pheno} is not possible due to the use of a different prior,
we can assess the relative reduction of PDF uncertainties in both cases,
and hence compare the reach on the PDFs of the FPF neutrino structure functions
and that of the HL-LHC measurements considered in the analysis of~\cite{AbdulKhalek:2018rok}.

This comparison of the relative PDF sensitivity of the FPF and
the HL-LHC data is interesting given that the  constraints from the 
two experiments are fully orthogonal and arise from completely different
scattering process.
Furthermore, the FPF constraints benefit from the valuable property
that an eventual contamination from new physics on the PDF fits 
can be neglected, since this process is driven by $Q^2$ values outside 
the possible presence of BSM physics (at least concerning new heavy particles).
Specially, should anomalies be revealed at the HL-LHC, having the independent
validation of the large-$x$ PDFs provided by the FPF would be extremely valuable 
for its interpretation.

Fig.~\ref{fig:HLLHC_pheno_integrated} displays the
same LHC fiducial cross-sections as in Fig.~\ref{fig:NNPDF40_pheno_integrated},
now comparing the PDF4LHC15 baseline predictions with those based
on the PDFs including HL-LHC pseudo-data from~\cite{AbdulKhalek:2018rok}.
Specifically, we consider the HL-LHC PDF projections
from ``Scenario B'', corresponding to an intermediate scenario for the expected reduction
of systematic uncertainties.
In general, the relative reduction on PDF uncertainties provided by the HL-LHC experiments
is more marked as in the case of the FPF pseudo-data, with the important caveat that
PDF4LHC21 already includes much more LHC data in comparison with its predecessor,
PDF4LHC15.
Another possible limiting assumption used in~\cite{AbdulKhalek:2018rok} is that
correlated uncertainties can be reliable estimated at the few-permille level
relevant for the intepretation of the HL-LHC data, something that is proving quite challenging
even for Run I and II measurements.

Nevertheless, the message obtained by comparing Fig.~\ref{fig:HLLHC_pheno_integrated} 
 with Fig.~\ref{fig:NNPDF40_pheno_integrated} is that of complementarity,
with the ultimate sensitivity on high-precision observables being
obtained by integrating the constraints both from the HL-LHC and from the FPF
in the same global determination of PDFs.

\section{Additional nPDF impact studies}
\label{app:nPDF_impact_appendix}

\begin{figure*}[htbp]
	\centering
	\includegraphics[width=0.32\textwidth]{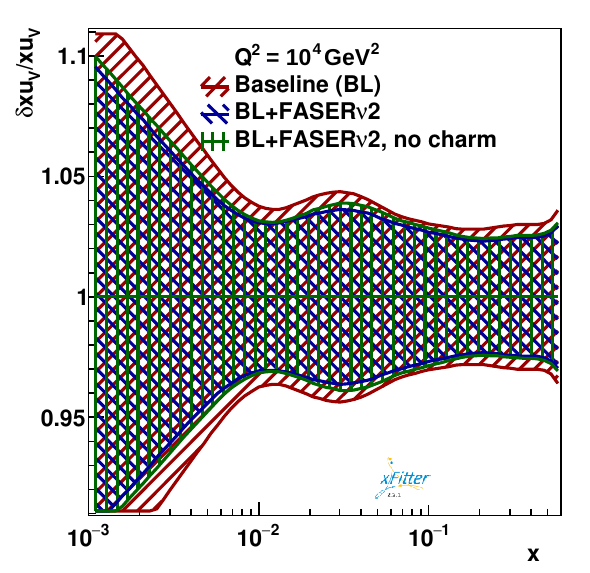}
	\includegraphics[width=0.32\textwidth]{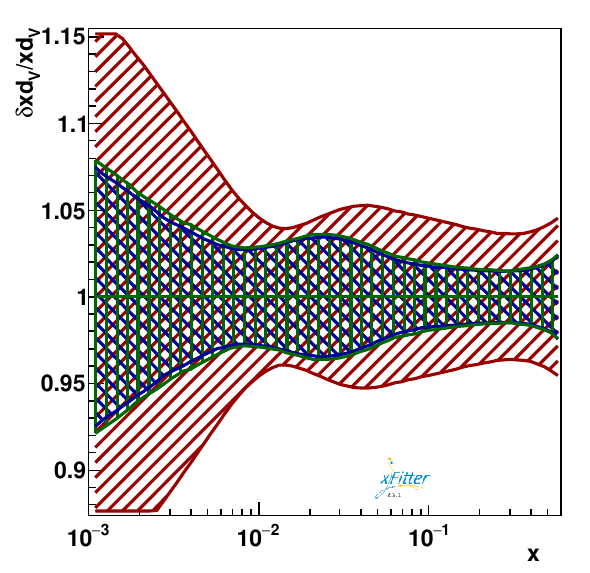}
	\includegraphics[width=0.32\textwidth]{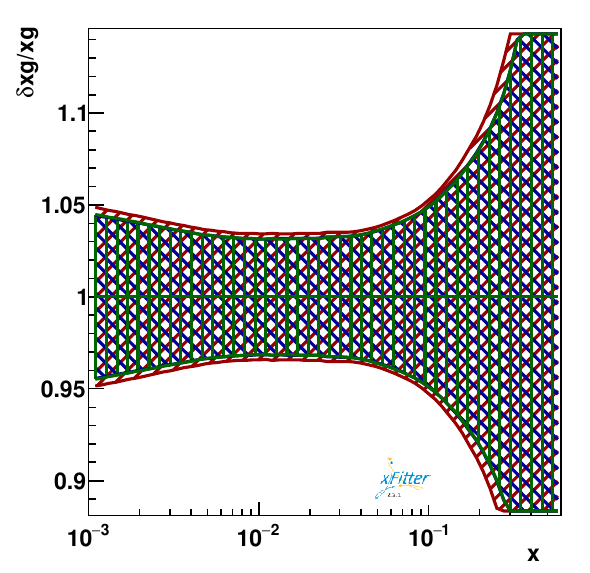}\\
	\includegraphics[width=0.32\textwidth]{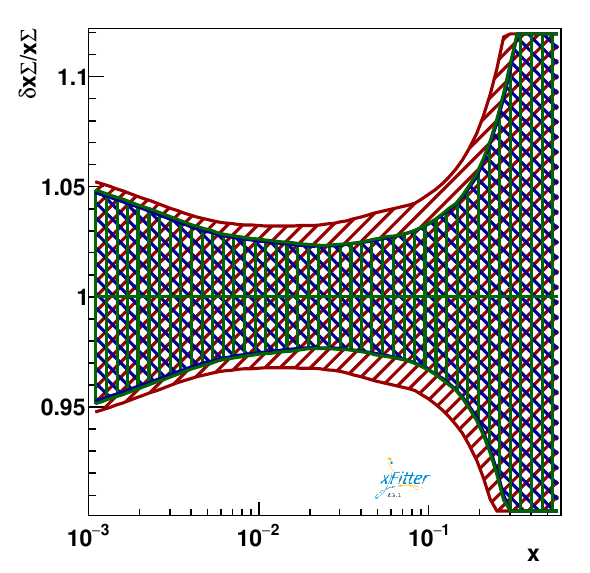}
	\includegraphics[width=0.32\textwidth]{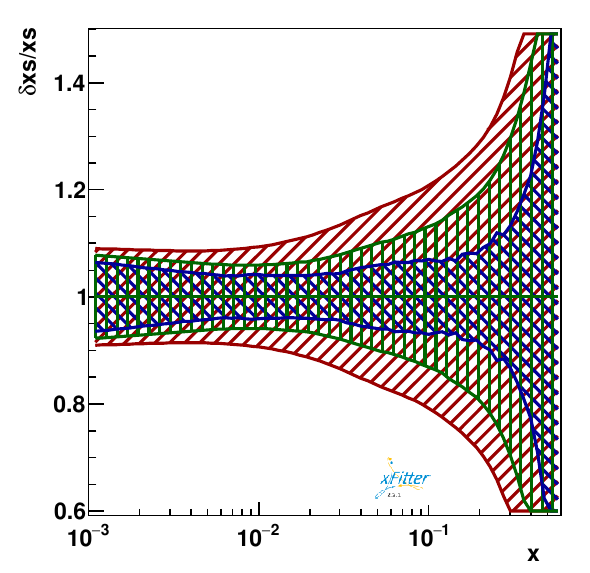}\\
	\includegraphics[width=0.32\textwidth]{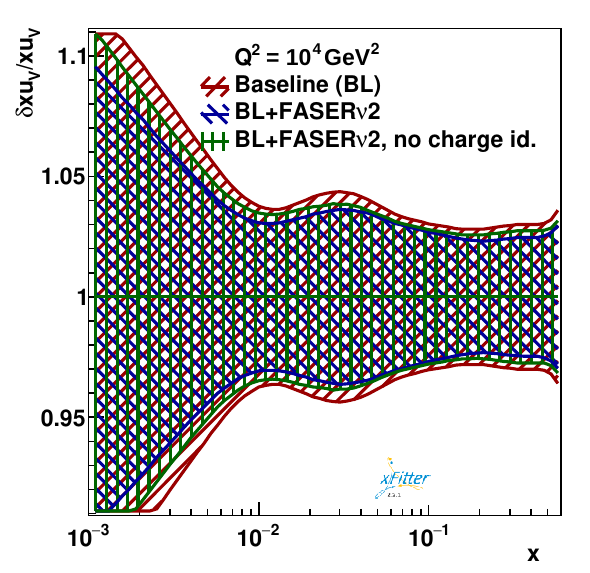}
	\includegraphics[width=0.32\textwidth]{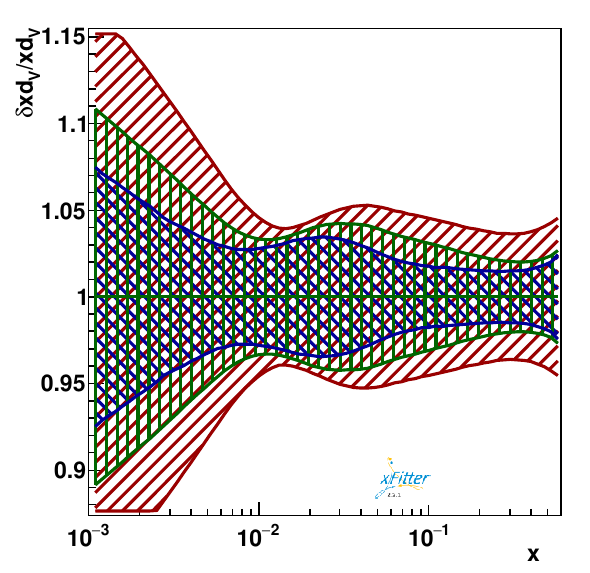}
	\includegraphics[width=0.32\textwidth]{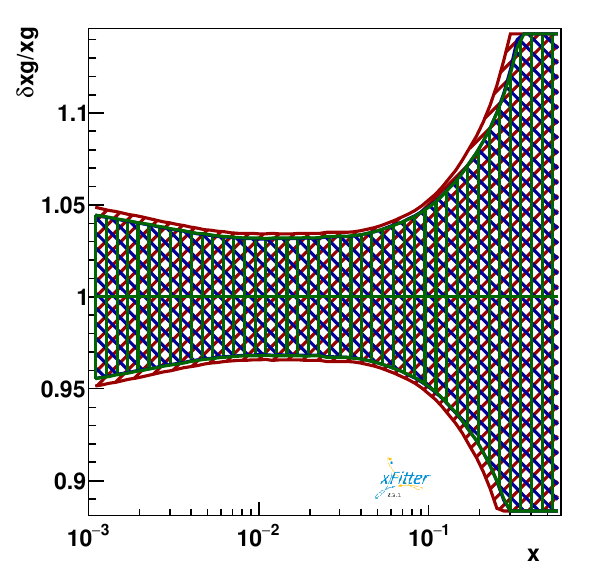}\\
	\includegraphics[width=0.32\textwidth]{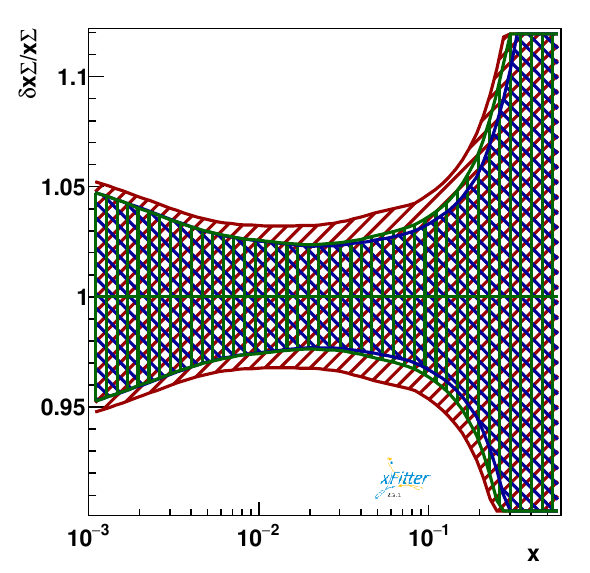}
	\includegraphics[width=0.32\textwidth]{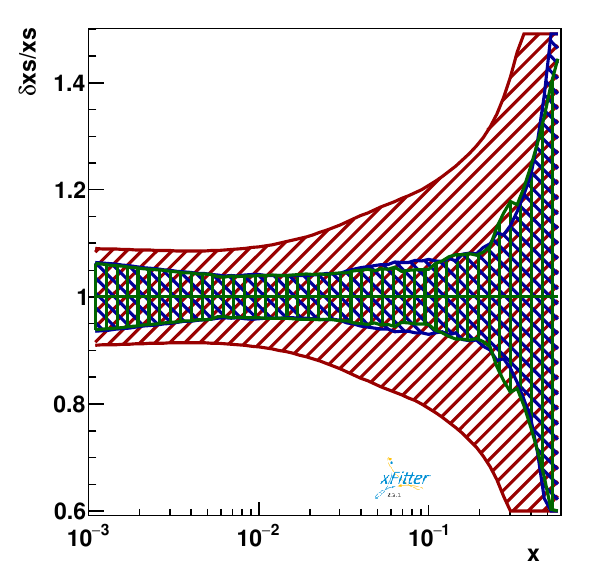}
	\caption{Same as Fig.~\ref{fig:FASERnu2_nocharm} (upper panels)
		and  Fig.~\ref{fig:FASERnu2_nochargeID} (bottom panels) in the case of EPPS21.
	}
	\label{fig:EPPS21_nochargeID}
\end{figure*}

The impact of FPF structure function measurements on the
EPPS21 nuclear PDF determination has been studied in Sect.~\ref{sec:nuclearPDFs}.
Here we provide additional results from this study, and specifically quantify the impact
that removing charm-tagged structure functions and flavour-charge
identification capabilities would have on the projected results.
Fig.~\ref{fig:EPPS21_nochargeID} displays
the analogous comparisons as in Figs.~\ref{fig:FASERnu2_nocharm}
and~\ref{fig:FASERnu2_nochargeID} in the case of EPPS21.
In both cases, results are consistent with the proton PDF profiling analysis.

First of all, charm-tagged events are essential to achieve the best
sensitivity to the strange PDF, while they have a vanishing impact on the
other PDF combinations.
Second, not being able to identify the charge of the outgoing final-state lepton
does not markedly affect  the baseline results, with the possible exception
of the down valence quark PDF.
As discussed in Sect.~\ref{sec:nudis_revisited}, for a non-isoscalar target
such as tungsten, with $Z=74$ and $A-Z=110$, event rates with neutrino projectiles
will differ from those arising from antineutrino scattering, introducing
additional information as compared to an isoscalar target in terms of PDF sensitivity.

\bibliographystyle{utphys}

\begin{thebibliography}{100}

\bibitem{Kling:2021gos}
F.~Kling and L.~J. Nevay, ``{Forward neutrino fluxes at the LHC},''
  \href{http://dx.doi.org/10.1103/PhysRevD.104.113008}{{\em Phys. Rev. D}
  {\bfseries 104} no.~11, (2021) 113008},
  \href{http://arxiv.org/abs/2105.08270}{{\ttfamily arXiv:2105.08270
  [hep-ph]}}.

\bibitem{FASER:2023zcr}
{\bfseries FASER} Collaboration, H.~Abreu {\em et~al.}, ``{First Direct
  Observation of Collider Neutrinos with FASER at the LHC},''
  \href{http://dx.doi.org/10.1103/PhysRevLett.131.031801}{{\em Phys. Rev.
  Lett.} {\bfseries 131} no.~3, (2023) 031801},
  \href{http://arxiv.org/abs/2303.14185}{{\ttfamily arXiv:2303.14185
  [hep-ex]}}.

\bibitem{SNDLHC:2023pun}
{\bfseries SND@LHC} Collaboration, R.~Albanese {\em et~al.}, ``{Observation of
  Collider Muon Neutrinos with the SND@LHC Experiment},''
  \href{http://dx.doi.org/10.1103/PhysRevLett.131.031802}{{\em Phys. Rev.
  Lett.} {\bfseries 131} no.~3, (2023) 031802},
  \href{http://arxiv.org/abs/2305.09383}{{\ttfamily arXiv:2305.09383
  [hep-ex]}}.

\bibitem{CERN-FASER-CONF-2023-002}
{\bfseries FASERcollaboration} Collaboration, ``{Observation of high-energy
  electron neutrino interactions with FASER's emulsion detector at the LHC},''
  tech. rep., CERN, Geneva, 2023.
\newblock \url{https://cds.cern.ch/record/2868284}.

\bibitem{FASER:2019dxq}
{\bfseries FASER} Collaboration, H.~Abreu {\em et~al.}, ``{Detecting and
  Studying High-Energy Collider Neutrinos with FASER at the LHC},''
  \href{http://dx.doi.org/10.1140/epjc/s10052-020-7631-5}{{\em Eur. Phys. J. C}
  {\bfseries 80} no.~1, (2020) 61},
  \href{http://arxiv.org/abs/1908.02310}{{\ttfamily arXiv:1908.02310
  [hep-ex]}}.

\bibitem{FASER:2022hcn}
{\bfseries FASER} Collaboration, H.~Abreu {\em et~al.}, ``{The FASER
  Detector},'' \href{http://arxiv.org/abs/2207.11427}{{\ttfamily
  arXiv:2207.11427 [physics.ins-det]}}.

\bibitem{SHiP:2020sos}
{\bfseries SHiP} Collaboration, C.~Ahdida {\em et~al.}, ``{SND@LHC},''
  \href{http://arxiv.org/abs/2002.08722}{{\ttfamily arXiv:2002.08722
  [physics.ins-det]}}.

\bibitem{SNDLHC:2022ihg}
{\bfseries SND@LHC} Collaboration, G.~Acampora {\em et~al.}, ``{SND@LHC: The
  Scattering and Neutrino Detector at the LHC},''
  \href{http://arxiv.org/abs/2210.02784}{{\ttfamily arXiv:2210.02784
  [hep-ex]}}.

\bibitem{Anchordoqui:2021ghd}
L.~A. Anchordoqui {\em et~al.}, ``{The Forward Physics Facility: Sites,
  experiments, and physics potential},''
  \href{http://dx.doi.org/10.1016/j.physrep.2022.04.004}{{\em Phys. Rept.}
  {\bfseries 968} (2022) 1--50},
  \href{http://arxiv.org/abs/2109.10905}{{\ttfamily arXiv:2109.10905
  [hep-ph]}}.

\bibitem{Feng:2022inv}
J.~L. Feng {\em et~al.}, ``{The Forward Physics Facility at the High-Luminosity
  LHC},'' \href{http://dx.doi.org/10.1088/1361-6471/ac865e}{{\em J. Phys. G}
  {\bfseries 50} no.~3, (2023) 030501},
  \href{http://arxiv.org/abs/2203.05090}{{\ttfamily arXiv:2203.05090
  [hep-ex]}}.

\bibitem{Azzi:2019yne}
P.~Azzi {\em et~al.}, ``{Report from Working Group 1}: {Standard Model Physics
  at the HL-LHC and HE-LHC},''
  \href{http://dx.doi.org/10.23731/CYRM-2019-007.1}{{\em CERN Yellow Rep.
  Monogr.} {\bfseries 7} (2019) 1--220},
  \href{http://arxiv.org/abs/1902.04070}{{\ttfamily arXiv:1902.04070
  [hep-ph]}}.

\bibitem{Cepeda:2019klc}
M.~Cepeda {\em et~al.}, ``{Report from Working Group 2}: {Higgs Physics at the
  HL-LHC and HE-LHC},''
  \href{http://dx.doi.org/10.23731/CYRM-2019-007.221}{{\em CERN Yellow Rep.
  Monogr.} {\bfseries 7} (2019) 221--584},
  \href{http://arxiv.org/abs/1902.00134}{{\ttfamily arXiv:1902.00134
  [hep-ph]}}.

\bibitem{Conrad:1997ne}
J.~M. Conrad, M.~H. Shaevitz, and T.~Bolton, ``{Precision measurements with
  high-energy neutrino beams},''
  \href{http://dx.doi.org/10.1103/RevModPhys.70.1341}{{\em Rev. Mod. Phys.}
  {\bfseries 70} (1998) 1341--1392},
  \href{http://arxiv.org/abs/hep-ex/9707015}{{\ttfamily arXiv:hep-ex/9707015}}.

\bibitem{Mangano:2001mj}
M.~L. Mangano {\em et~al.}, ``{Physics at the front end of a neutrino factory:
  A Quantitative appraisal},''
  \href{http://arxiv.org/abs/hep-ph/0105155}{{\ttfamily arXiv:hep-ph/0105155}}.

\bibitem{Candido:2023utz}
A.~Candido, A.~Garcia, G.~Magni, T.~Rabemananjara, J.~Rojo, and R.~Stegeman,
  ``{Neutrino Structure Functions from GeV to EeV Energies},''
  \href{http://dx.doi.org/10.1007/JHEP05(2023)149}{{\em JHEP} {\bfseries 05}
  (2023) 149}, \href{http://arxiv.org/abs/2302.08527}{{\ttfamily
  arXiv:2302.08527 [hep-ph]}}.

\bibitem{Ethier:2020way}
J.~J. Ethier and E.~R. Nocera, ``{Parton Distributions in Nucleons and
  Nuclei},'' \href{http://dx.doi.org/10.1146/annurev-nucl-011720-042725}{{\em
  Ann. Rev. Nucl. Part. Sci.} {\bfseries 70} (2020) 43--76},
  \href{http://arxiv.org/abs/2001.07722}{{\ttfamily arXiv:2001.07722
  [hep-ph]}}.

\bibitem{Gao:2017yyd}
J.~Gao, L.~Harland-Lang, and J.~Rojo, ``{The Structure of the Proton in the LHC
  Precision Era},'' \href{http://dx.doi.org/10.1016/j.physrep.2018.03.002}{{\em
  Phys. Rept.} {\bfseries 742} (2018) 1--121},
  \href{http://arxiv.org/abs/1709.04922}{{\ttfamily arXiv:1709.04922
  [hep-ph]}}.

\bibitem{Kovarik:2019xvh}
K.~Kova\v{r}\'\i{}k, P.~M. Nadolsky, and D.~E. Soper, ``{Hadronic structure in
  high-energy collisions},''
  \href{http://dx.doi.org/10.1103/RevModPhys.92.045003}{{\em Rev. Mod. Phys.}
  {\bfseries 92} no.~4, (2020) 045003},
  \href{http://arxiv.org/abs/1905.06957}{{\ttfamily arXiv:1905.06957
  [hep-ph]}}.

\bibitem{NuTeV:2007uwm}
{\bfseries NuTeV} Collaboration, D.~Mason {\em et~al.}, ``{Measurement of the
  Nucleon Strange-Antistrange Asymmetry at Next-to-Leading Order in QCD from
  NuTeV Dimuon Data},''
  \href{http://dx.doi.org/10.1103/PhysRevLett.99.192001}{{\em Phys. Rev. Lett.}
  {\bfseries 99} (2007) 192001}.

\bibitem{CCFR:1994ikl}
{\bfseries CCFR} Collaboration, A.~O. Bazarko {\em et~al.}, ``{Determination of
  the strange quark content of the nucleon from a next-to-leading order QCD
  analysis of neutrino charm production},''
  \href{http://dx.doi.org/10.1007/BF01571875}{{\em Z. Phys. C} {\bfseries 65}
  (1995) 189--198}, \href{http://arxiv.org/abs/hep-ex/9406007}{{\ttfamily
  arXiv:hep-ex/9406007}}.

\bibitem{Faura:2020oom}
F.~Faura, S.~Iranipour, E.~R. Nocera, J.~Rojo, and M.~Ubiali, ``{The Strangest
  Proton?},'' \href{http://dx.doi.org/10.1140/epjc/s10052-020-08749-3}{{\em
  Eur. Phys. J. C} {\bfseries 80} no.~12, (2020) 1168},
  \href{http://arxiv.org/abs/2009.00014}{{\ttfamily arXiv:2009.00014
  [hep-ph]}}.

\bibitem{Alekhin:2014sya}
S.~Alekhin, J.~Blumlein, L.~Caminada, K.~Lipka, K.~Lohwasser, S.~Moch,
  R.~Petti, and R.~Placakyte, ``{Determination of Strange Sea Quark
  Distributions from Fixed-target and Collider Data},''
  \href{http://dx.doi.org/10.1103/PhysRevD.91.094002}{{\em Phys. Rev. D}
  {\bfseries 91} no.~9, (2015) 094002},
  \href{http://arxiv.org/abs/1404.6469}{{\ttfamily arXiv:1404.6469 [hep-ph]}}.

\bibitem{CHORUS:2005cpn}
{\bfseries CHORUS} Collaboration, G.~Onengut {\em et~al.}, ``{Measurement of
  nucleon structure functions in neutrino scattering},''
  \href{http://dx.doi.org/10.1016/j.physletb.2005.10.062}{{\em Phys. Lett. B}
  {\bfseries 632} (2006) 65--75}.

\bibitem{NuTeV:2001dfo}
{\bfseries NuTeV} Collaboration, M.~Goncharov {\em et~al.}, ``{Precise
  Measurement of Dimuon Production Cross-Sections in $\nu_{\mu}$ Fe and
  $\bar{\nu}_{\mu}$ Fe Deep Inelastic Scattering at the Tevatron.},''
  \href{http://dx.doi.org/10.1103/PhysRevD.64.112006}{{\em Phys. Rev. D}
  {\bfseries 64} (2001) 112006},
  \href{http://arxiv.org/abs/hep-ex/0102049}{{\ttfamily arXiv:hep-ex/0102049}}.

\bibitem{Yang:2000ju}
{\bfseries CCFR/NuTeV} Collaboration, U.-K. Yang {\em et~al.}, ``{Measurements
  of $F_2$ and $xF^{\nu}_3 - x F^{\bar{\nu}}_3$ from CCFR $\nu_\mu-$Fe and
  $\bar{\nu}_\mu-$Fe data in a physics model independent way},''
  \href{http://dx.doi.org/10.1103/PhysRevLett.86.2742}{{\em Phys. Rev. Lett.}
  {\bfseries 86} (2001) 2742--2745},
\href{http://arxiv.org/abs/hep-ex/0009041}{{\ttfamily arXiv:hep-ex/0009041
  [hep-ex]}}.

\bibitem{NOMAD:2013hbk}
{\bfseries NOMAD} Collaboration, O.~Samoylov {\em et~al.}, ``{A Precision
  Measurement of Charm Dimuon Production in Neutrino Interactions from the
  NOMAD Experiment},''
  \href{http://dx.doi.org/10.1016/j.nuclphysb.2013.08.021}{{\em Nucl. Phys. B}
  {\bfseries 876} (2013) 339--375},
  \href{http://arxiv.org/abs/1308.4750}{{\ttfamily arXiv:1308.4750 [hep-ex]}}.

\bibitem{Berge:1989hr}
J.~P. Berge {\em et~al.}, ``{A Measurement of Differential Cross-Sections and
  Nucleon Structure Functions in Charged Current Neutrino Interactions on
  Iron},''
\href{http://dx.doi.org/10.1007/BF01555493}{{\em Z. Phys.} {\bfseries C49}
  (1991) 187--224}.

\bibitem{NNPDF:2021njg}
{\bfseries NNPDF} Collaboration, R.~D. Ball {\em et~al.}, ``{The path to proton
  structure at 1\% accuracy},''
  \href{http://dx.doi.org/10.1140/epjc/s10052-022-10328-7}{{\em Eur. Phys. J.
  C} {\bfseries 82} no.~5, (2022) 428},
  \href{http://arxiv.org/abs/2109.02653}{{\ttfamily arXiv:2109.02653
  [hep-ph]}}.

\bibitem{Hou:2019efy}
T.-J. Hou {\em et~al.}, ``{New CTEQ global analysis of quantum chromodynamics
  with high-precision data from the LHC},''
  \href{http://dx.doi.org/10.1103/PhysRevD.103.014013}{{\em Phys. Rev. D}
  {\bfseries 103} no.~1, (2021) 014013},
  \href{http://arxiv.org/abs/1912.10053}{{\ttfamily arXiv:1912.10053
  [hep-ph]}}.

\bibitem{Bailey:2020ooq}
S.~Bailey, T.~Cridge, L.~A. Harland-Lang, A.~D. Martin, and R.~S. Thorne,
  ``{Parton distributions from LHC, HERA, Tevatron and fixed target data:
  MSHT20 PDFs},'' \href{http://dx.doi.org/10.1140/epjc/s10052-021-09057-0}{{\em
  Eur. Phys. J. C} {\bfseries 81} no.~4, (2021) 341},
  \href{http://arxiv.org/abs/2012.04684}{{\ttfamily arXiv:2012.04684
  [hep-ph]}}.

\bibitem{Eskola:2021nhw}
K.~J. Eskola, P.~Paakkinen, H.~Paukkunen, and C.~A. Salgado, ``{EPPS21: a
  global QCD analysis of nuclear PDFs},''
  \href{http://dx.doi.org/10.1140/epjc/s10052-022-10359-0}{{\em Eur. Phys. J.
  C} {\bfseries 82} no.~5, (2022) 413},
  \href{http://arxiv.org/abs/2112.12462}{{\ttfamily arXiv:2112.12462
  [hep-ph]}}.

\bibitem{AbdulKhalek:2022fyi}
R.~Abdul~Khalek, R.~Gauld, T.~Giani, E.~R. Nocera, T.~R. Rabemananjara, and
  J.~Rojo, ``{nNNPDF3.0: evidence for a modified partonic structure in heavy
  nuclei},'' \href{http://dx.doi.org/10.1140/epjc/s10052-022-10417-7}{{\em Eur.
  Phys. J. C} {\bfseries 82} no.~6, (2022) 507},
  \href{http://arxiv.org/abs/2201.12363}{{\ttfamily arXiv:2201.12363
  [hep-ph]}}.

\bibitem{Muzakka:2022wey}
K.~F. Muzakka {\em et~al.}, ``{Compatibility of neutrino DIS data and its
  impact on nuclear parton distribution functions},''
  \href{http://dx.doi.org/10.1103/PhysRevD.106.074004}{{\em Phys. Rev. D}
  {\bfseries 106} no.~7, (2022) 074004},
  \href{http://arxiv.org/abs/2204.13157}{{\ttfamily arXiv:2204.13157
  [hep-ph]}}.

\bibitem{AbdulKhalek:2018rok}
R.~Abdul~Khalek, S.~Bailey, J.~Gao, L.~Harland-Lang, and J.~Rojo, ``{Towards
  Ultimate Parton Distributions at the High-Luminosity LHC},''
  \href{http://dx.doi.org/10.1140/epjc/s10052-018-6448-y}{{\em Eur. Phys. J. C}
  {\bfseries 78} no.~11, (2018) 962},
  \href{http://arxiv.org/abs/1810.03639}{{\ttfamily arXiv:1810.03639
  [hep-ph]}}.

\bibitem{AbdulKhalek:2021gbh}
R.~Abdul~Khalek {\em et~al.}, ``{Science Requirements and Detector Concepts for
  the Electron-Ion Collider}: {EIC Yellow Report},''
  \href{http://dx.doi.org/10.1016/j.nuclphysa.2022.122447}{{\em Nucl. Phys. A}
  {\bfseries 1026} (2022) 122447},
  \href{http://arxiv.org/abs/2103.05419}{{\ttfamily arXiv:2103.05419
  [physics.ins-det]}}.

\bibitem{Khalek:2021ulf}
R.~A. Khalek, J.~J. Ethier, E.~R. Nocera, and J.~Rojo, ``{Self-consistent
  determination of proton and nuclear PDFs at the Electron Ion Collider},''
  \href{http://dx.doi.org/10.1103/PhysRevD.103.096005}{{\em Phys. Rev. D}
  {\bfseries 103} no.~9, (2021) 096005},
  \href{http://arxiv.org/abs/2102.00018}{{\ttfamily arXiv:2102.00018
  [hep-ph]}}.

\bibitem{AbdulKhalek:2019mzd}
{\bfseries NNPDF} Collaboration, R.~Abdul~Khalek, J.~J. Ethier, and J.~Rojo,
  ``{Nuclear parton distributions from lepton-nucleus scattering and the impact
  of an electron-ion collider},''
  \href{http://dx.doi.org/10.1140/epjc/s10052-019-6983-1}{{\em Eur. Phys. J. C}
  {\bfseries 79} no.~6, (2019) 471},
  \href{http://arxiv.org/abs/1904.00018}{{\ttfamily arXiv:1904.00018
  [hep-ph]}}.

\bibitem{AbdulKhalek:2019mps}
R.~Abdul~Khalek, S.~Bailey, J.~Gao, L.~Harland-Lang, and J.~Rojo, ``{Probing
  Proton Structure at the Large Hadron electron Collider},''
  \href{http://dx.doi.org/10.21468/SciPostPhys.7.4.051}{{\em SciPost Phys.}
  {\bfseries 7} no.~4, (2019) 051},
  \href{http://arxiv.org/abs/1906.10127}{{\ttfamily arXiv:1906.10127
  [hep-ph]}}.

\bibitem{LHeC:2020van}
{\bfseries LHeC, FCC-he Study Group} Collaboration, P.~Agostini {\em et~al.},
  ``{The Large Hadron\textendash{}Electron Collider at the HL-LHC},''
  \href{http://dx.doi.org/10.1088/1361-6471/abf3ba}{{\em J. Phys. G} {\bfseries
  48} no.~11, (2021) 110501}, \href{http://arxiv.org/abs/2007.14491}{{\ttfamily
  arXiv:2007.14491 [hep-ex]}}.

\bibitem{LHeCStudyGroup:2012zhm}
{\bfseries LHeC Study Group} Collaboration, J.~L. Abelleira~Fernandez {\em
  et~al.}, ``{A Large Hadron Electron Collider at CERN: Report on the Physics
  and Design Concepts for Machine and Detector},''
  \href{http://dx.doi.org/10.1088/0954-3899/39/7/075001}{{\em J. Phys. G}
  {\bfseries 39} (2012) 075001},
  \href{http://arxiv.org/abs/1206.2913}{{\ttfamily arXiv:1206.2913
  [physics.acc-ph]}}.

\bibitem{Batell:2021blf}
B.~Batell, J.~L. Feng, and S.~Trojanowski, ``{Detecting Dark Matter with
  Far-Forward Emulsion and Liquid Argon Detectors at the LHC},''
  \href{http://dx.doi.org/10.1103/PhysRevD.103.075023}{{\em Phys. Rev. D}
  {\bfseries 103} no.~7, (2021) 075023},
  \href{http://arxiv.org/abs/2101.10338}{{\ttfamily arXiv:2101.10338
  [hep-ph]}}.

\bibitem{Batell:2021aja}
B.~Batell, J.~L. Feng, A.~Ismail, F.~Kling, R.~M. Abraham, and S.~Trojanowski,
  ``{Discovering dark matter at the LHC through its nuclear scattering in
  far-forward emulsion and liquid argon detectors},''
  \href{http://dx.doi.org/10.1103/PhysRevD.104.035036}{{\em Phys. Rev. D}
  {\bfseries 104} no.~3, (2021) 035036},
  \href{http://arxiv.org/abs/2107.00666}{{\ttfamily arXiv:2107.00666
  [hep-ph]}}.

\bibitem{Paukkunen:2014zia}
H.~Paukkunen and P.~Zurita, ``{PDF reweighting in the Hessian matrix
  approach},'' \href{http://dx.doi.org/10.1007/JHEP12(2014)100}{{\em JHEP}
  {\bfseries 12} (2014) 100}, \href{http://arxiv.org/abs/1402.6623}{{\ttfamily
  arXiv:1402.6623 [hep-ph]}}.

\bibitem{Schmidt:2018hvu}
C.~Schmidt, J.~Pumplin, C.~P. Yuan, and P.~Yuan, ``{Updating and optimizing
  error parton distribution function sets in the Hessian approach},''
  \href{http://dx.doi.org/10.1103/PhysRevD.98.094005}{{\em Phys. Rev. D}
  {\bfseries 98} no.~9, (2018) 094005},
  \href{http://arxiv.org/abs/1806.07950}{{\ttfamily arXiv:1806.07950
  [hep-ph]}}.

\bibitem{HERAFitterdevelopersTeam:2015cre}
{\bfseries HERAFitter developers' team} Collaboration, S.~Camarda {\em et~al.},
  ``{QCD analysis of $W$- and $Z$-boson production at Tevatron},''
  \href{http://dx.doi.org/10.1140/epjc/s10052-015-3655-7}{{\em Eur. Phys. J. C}
  {\bfseries 75} no.~9, (2015) 458},
  \href{http://arxiv.org/abs/1503.05221}{{\ttfamily arXiv:1503.05221
  [hep-ph]}}.

\bibitem{PDF4LHCWorkingGroup:2022cjn}
{\bfseries PDF4LHC Working Group} Collaboration, R.~D. Ball {\em et~al.},
  ``{The PDF4LHC21 combination of global PDF fits for the LHC Run III},''
  \href{http://dx.doi.org/10.1088/1361-6471/ac7216}{{\em J. Phys. G} {\bfseries
  49} no.~8, (2022) 080501}, \href{http://arxiv.org/abs/2203.05506}{{\ttfamily
  arXiv:2203.05506 [hep-ph]}}.

\bibitem{Alekhin:2014irh}
S.~Alekhin {\em et~al.}, ``{HERAFitter},''
  \href{http://dx.doi.org/10.1140/epjc/s10052-015-3480-z}{{\em Eur. Phys. J. C}
  {\bfseries 75} no.~7, (2015) 304},
  \href{http://arxiv.org/abs/1410.4412}{{\ttfamily arXiv:1410.4412 [hep-ph]}}.

\bibitem{Bertone:2017tig}
V.~Bertone, M.~Botje, D.~Britzger, {\em et~al.}, ``{xFitter} 2.0.0: An open
  source {QCD} fit framework,''
  \href{http://dx.doi.org/10.22323/1.297.0203}{{\em PoS} {\bfseries DIS2017}
  (2018) 203},
\href{http://arxiv.org/abs/1709.01151}{{\ttfamily arXiv:1709.01151 [hep-ph]}}.

\bibitem{xFitter:2022zjb}
{\bfseries xFitter} Collaboration, H.~Abdolmaleki {\em et~al.}, ``{xFitter: An
  Open Source QCD Analysis Framework. A resource and reference document for the
  Snowmass study},''
\newblock 6, 2022.
\newblock \href{http://arxiv.org/abs/2206.12465}{{\ttfamily arXiv:2206.12465
  [hep-ph]}}.

\bibitem{xFitter:web}
xFitter Developers'~Team.
\newblock \url{https://www.xfitter.org/xFitter/}.

\bibitem{NNPDF:2021uiq}
{\bfseries NNPDF} Collaboration, R.~D. Ball {\em et~al.}, ``{An open-source
  machine learning framework for global analyses of parton distributions},''
  {\em Eur. Phys. J. C} {\bfseries 81} no.~10, (2021) 958,
  \href{http://arxiv.org/abs/2109.02671}{{\ttfamily arXiv:2109.02671
  [hep-ph]}}.

\bibitem{Gao:2017kkx}
J.~Gao, ``{Massive charged-current coefficient functions in deep-inelastic
  scattering at NNLO and impact on strange-quark distributions},''
  \href{http://dx.doi.org/10.1007/JHEP02(2018)026}{{\em JHEP} {\bfseries 02}
  (2018) 026}, \href{http://arxiv.org/abs/1710.04258}{{\ttfamily
  arXiv:1710.04258 [hep-ph]}}.

\bibitem{Candido:2022tld}
A.~Candido, F.~Hekhorn, and G.~Magni, ``{EKO: evolution kernel operators},''
  \href{http://dx.doi.org/10.1140/epjc/s10052-022-10878-w}{{\em Eur. Phys. J.
  C} {\bfseries 82} no.~10, (2022) 976},
  \href{http://arxiv.org/abs/2202.02338}{{\ttfamily arXiv:2202.02338
  [hep-ph]}}.

\bibitem{yadism}
A.~Candido and F.~Hekhorn, ``{YADISM: Yet Another DIS Module},'' {\em in
  preparation} (2023) .

\bibitem{Carrazza:2020gss}
S.~Carrazza, E.~R. Nocera, C.~Schwan, and M.~Zaro, ``{PineAPPL: combining EW
  and QCD corrections for fast evaluation of LHC processes},''
  \href{http://dx.doi.org/10.1007/JHEP12(2020)108}{{\em JHEP} {\bfseries 12}
  (2020) 108}, \href{http://arxiv.org/abs/2008.12789}{{\ttfamily
  arXiv:2008.12789 [hep-ph]}}.

\bibitem{Sufian:2018cpj}
R.~S. Sufian, T.~Liu, G.~F. de~T\'eramond, H.~G. Dosch, S.~J. Brodsky, A.~Deur,
  M.~T. Islam, and B.-Q. Ma, ``{Nonperturbative strange-quark sea from lattice
  QCD, light-front holography, and meson-baryon fluctuation models},''
  \href{http://dx.doi.org/10.1103/PhysRevD.98.114004}{{\em Phys. Rev. D}
  {\bfseries 98} no.~11, (2018) 114004},
  \href{http://arxiv.org/abs/1809.04975}{{\ttfamily arXiv:1809.04975
  [hep-ph]}}.

\bibitem{Sufian:2020coz}
R.~S. Sufian, T.~Liu, A.~Alexandru, S.~J. Brodsky, G.~F. de~T\'eramond, H.~G.
  Dosch, T.~Draper, K.-F. Liu, and Y.-B. Yang, ``{Constraints on
  charm-anticharm asymmetry in the nucleon from lattice QCD},''
  \href{http://dx.doi.org/10.1016/j.physletb.2020.135633}{{\em Phys. Lett. B}
  {\bfseries 808} (2020) 135633},
  \href{http://arxiv.org/abs/2003.01078}{{\ttfamily arXiv:2003.01078
  [hep-lat]}}.

\bibitem{Ball:2020xqw}
R.~D. Ball, E.~R. Nocera, and R.~L. Pearson, ``{Deuteron Uncertainties in the
  Determination of Proton PDFs},''
  \href{http://dx.doi.org/10.1140/epjc/s10052-020-08826-7}{{\em Eur. Phys. J.
  C} {\bfseries 81} no.~1, (2021) 37},
  \href{http://arxiv.org/abs/2011.00009}{{\ttfamily arXiv:2011.00009
  [hep-ph]}}.

\bibitem{Ball:2018twp}
{\bfseries NNPDF} Collaboration, R.~D. Ball, E.~R. Nocera, and R.~L. Pearson,
  ``{Nuclear Uncertainties in the Determination of Proton PDFs},''
  \href{http://dx.doi.org/10.1140/epjc/s10052-019-6793-5}{{\em Eur. Phys. J. C}
  {\bfseries 79} no.~3, (2019) 282},
  \href{http://arxiv.org/abs/1812.09074}{{\ttfamily arXiv:1812.09074
  [hep-ph]}}.

\bibitem{NNPDF:2019vjt}
{\bfseries NNPDF} Collaboration, R.~Abdul~Khalek {\em et~al.}, ``{A first
  determination of parton distributions with theoretical uncertainties},''
  \href{http://dx.doi.org/10.1140/epjc/s10052-019-7364-5}{{\em Eur. Phys. J.}
  {\bfseries C} (2019) 79:838},
  \href{http://arxiv.org/abs/1905.04311}{{\ttfamily arXiv:1905.04311
  [hep-ph]}}.

\bibitem{NNPDF:2019ubu}
{\bfseries NNPDF} Collaboration, R.~Abdul~Khalek {\em et~al.}, ``{Parton
  Distributions with Theory Uncertainties: General Formalism and First
  Phenomenological Studies},''
  \href{http://dx.doi.org/10.1140/epjc/s10052-019-7401-4}{{\em Eur. Phys. J. C}
  {\bfseries 79} no.~11, (2019) 931},
  \href{http://arxiv.org/abs/1906.10698}{{\ttfamily arXiv:1906.10698
  [hep-ph]}}.

\bibitem{FASER:2019aik}
{\bfseries FASER} Collaboration, A.~Ariga {\em et~al.}, ``{FASER: ForwArd
  Search ExpeRiment at the LHC},''
  \href{http://arxiv.org/abs/1901.04468}{{\ttfamily arXiv:1901.04468
  [hep-ex]}}.

\bibitem{Aachen-Bonn-CERN-Munich-Oxford:1981lfk}
{\bfseries Aachen-Bonn-CERN-Munich-Oxford} Collaboration, P.~Allen {\em
  et~al.}, ``{Multiplicity Distributions in Neutrino - Hydrogen
  Interactions},'' \href{http://dx.doi.org/10.1016/0550-3213(81)90532-0}{{\em
  Nucl. Phys. B} {\bfseries 181} (1981) 385--402}.

\bibitem{Nason:2004rx}
P.~Nason, ``{A New method for combining NLO QCD with shower Monte Carlo
  algorithms},'' \href{http://dx.doi.org/10.1088/1126-6708/2004/11/040}{{\em
  JHEP} {\bfseries 11} (2004) 040},
  \href{http://arxiv.org/abs/hep-ph/0409146}{{\ttfamily arXiv:hep-ph/0409146}}.

\bibitem{Frixione:2007vw}
S.~Frixione, P.~Nason, and C.~Oleari, ``{Matching NLO QCD computations with
  Parton Shower simulations: the POWHEG method},''
  \href{http://dx.doi.org/10.1088/1126-6708/2007/11/070}{{\em JHEP} {\bfseries
  11} (2007) 070}, \href{http://arxiv.org/abs/0709.2092}{{\ttfamily
  arXiv:0709.2092 [hep-ph]}}.

\bibitem{Alioli:2010xd}
S.~Alioli, P.~Nason, C.~Oleari, and E.~Re, ``{A general framework for
  implementing NLO calculations in shower Monte Carlo programs: the POWHEG
  BOX},'' \href{http://dx.doi.org/10.1007/JHEP06(2010)043}{{\em JHEP}
  {\bfseries 06} (2010) 043}, \href{http://arxiv.org/abs/1002.2581}{{\ttfamily
  arXiv:1002.2581 [hep-ph]}}.

\bibitem{Sjostrand:2014zea}
T.~Sj\"ostrand, S.~Ask, J.~R. Christiansen, R.~Corke, N.~Desai, P.~Ilten,
  S.~Mrenna, S.~Prestel, C.~O. Rasmussen, and P.~Z. Skands, ``{An introduction
  to PYTHIA 8.2}'' \href{http://dx.doi.org/10.1016/j.cpc.2015.01.024}{{\em
  Comput. Phys. Commun.} {\bfseries 191} (2015) 159--177},
  \href{http://arxiv.org/abs/1410.3012}{{\ttfamily arXiv:1410.3012 [hep-ph]}}.

\bibitem{Bierlich:2022pfr}
C.~Bierlich {\em et~al.}, ``{A comprehensive guide to the physics and usage of
  PYTHIA 8.3}'' \href{http://arxiv.org/abs/2203.11601}{{\ttfamily
  arXiv:2203.11601 [hep-ph]}}.

\bibitem{Pierog:2013ria}
T.~Pierog, I.~Karpenko, J.~M. Katzy, E.~Yatsenko, and K.~Werner, ``{EPOS LHC:
  Test of collective hadronization with data measured at the CERN Large Hadron
  Collider},'' \href{http://dx.doi.org/10.1103/PhysRevC.92.034906}{{\em Phys.
  Rev. C} {\bfseries 92} no.~3, (2015) 034906},
  \href{http://arxiv.org/abs/1306.0121}{{\ttfamily arXiv:1306.0121 [hep-ph]}}.

\bibitem{Sominka:2023}
L.~Buonocore, F.~Kling, L.~Rottoli, and J.~Sominka, ``{Predictions for
  Neutrinos and New Physics from Forward Heavy Hadron Production at the LHC}.''
  In preparation, 2023.

\bibitem{Kling:2023tgr}
F.~Kling, T.~M\"akel\"a, and S.~Trojanowski, ``{Investigating the fluxes and
  physics potential of LHC neutrino experiments},''
  \href{http://dx.doi.org/10.1103/PhysRevD.108.095020}{{\em Phys. Rev. D}
  {\bfseries 108} no.~9, (2023) 095020},
  \href{http://arxiv.org/abs/2309.10417}{{\ttfamily arXiv:2309.10417
  [hep-ph]}}.

\bibitem{Fieg:2023kld}
M.~Fieg, F.~Kling, H.~Schulz, and T.~Sj\"ostrand, ``{Tuning Pythia for Forward
  Physics Experiments},'' \href{http://arxiv.org/abs/2309.08604}{{\ttfamily
  arXiv:2309.08604 [hep-ph]}}.

\bibitem{Amoroso:2023pey}
S.~Amoroso {\em et~al.}, ``{Compatibility and combination of world W-boson mass
  measurements},'' \href{http://arxiv.org/abs/2308.09417}{{\ttfamily
  arXiv:2308.09417 [hep-ex]}}.

\bibitem{Ball:2022qtp}
R.~D. Ball, A.~Candido, S.~Forte, F.~Hekhorn, E.~R. Nocera, J.~Rojo, and
  C.~Schwan, ``{Parton distributions and new physics searches: the
  Drell\textendash{}Yan forward\textendash{}backward asymmetry as a case
  study},'' \href{http://dx.doi.org/10.1140/epjc/s10052-022-11133-y}{{\em Eur.
  Phys. J. C} {\bfseries 82} no.~12, (2022) 1160},
  \href{http://arxiv.org/abs/2209.08115}{{\ttfamily arXiv:2209.08115
  [hep-ph]}}.

\bibitem{Ethier:2021ydt}
J.~J. Ethier, R.~Gomez-Ambrosio, G.~Magni, and J.~Rojo, ``{SMEFT analysis of
  vector boson scattering and diboson data from the LHC Run II},''
  \href{http://dx.doi.org/10.1140/epjc/s10052-021-09347-7}{{\em Eur. Phys. J.
  C} {\bfseries 81} no.~6, (2021) 560},
  \href{http://arxiv.org/abs/2101.03180}{{\ttfamily arXiv:2101.03180
  [hep-ph]}}.

\bibitem{Greljo:2021kvv}
A.~Greljo, S.~Iranipour, Z.~Kassabov, M.~Madigan, J.~Moore, J.~Rojo, M.~Ubiali,
  and C.~Voisey, ``{Parton distributions in the SMEFT from high-energy
  Drell-Yan tails},'' \href{http://dx.doi.org/10.1007/JHEP07(2021)122}{{\em
  JHEP} {\bfseries 07} (2021) 122},
  \href{http://arxiv.org/abs/2104.02723}{{\ttfamily arXiv:2104.02723
  [hep-ph]}}.

\bibitem{christopher_schwan_2023_7995675}
C.~Schwan, A.~Candido, F.~Hekhorn, S.~Carrazza, and A.~Barontini,
  \href{http://dx.doi.org/10.5281/zenodo.7995675}{``Nnpdf/pineappl: v0.6.0,''}
\newblock Zenodo, June, 2023.
\newblock \url{https://doi.org/10.5281/zenodo.7995675}.

\bibitem{Forte:2010ta}
S.~Forte, E.~Laenen, P.~Nason, and J.~Rojo, ``{Heavy quarks in deep-inelastic
  scattering},'' \href{http://dx.doi.org/10.1016/j.nuclphysb.2010.03.014}{{\em
  Nucl. Phys. B} {\bfseries 834} (2010) 116--162},
  \href{http://arxiv.org/abs/1001.2312}{{\ttfamily arXiv:1001.2312 [hep-ph]}}.

\bibitem{Ball:2011mu}
R.~D. Ball, V.~Bertone, F.~Cerutti, L.~Del~Debbio, S.~Forte, A.~Guffanti, J.~I.
  Latorre, J.~Rojo, and M.~Ubiali, ``{Impact of Heavy Quark Masses on Parton
  Distributions and LHC Phenomenology},''
  \href{http://dx.doi.org/10.1016/j.nuclphysb.2011.03.021}{{\em Nucl. Phys. B}
  {\bfseries 849} (2011) 296--363},
  \href{http://arxiv.org/abs/1101.1300}{{\ttfamily arXiv:1101.1300 [hep-ph]}}.

\bibitem{Ball:2010de}
R.~D. Ball, L.~Del~Debbio, S.~Forte, A.~Guffanti, J.~I. Latorre, J.~Rojo, and
  M.~Ubiali, ``{A first unbiased global NLO determination of parton
  distributions and their uncertainties},''
  \href{http://dx.doi.org/10.1016/j.nuclphysb.2010.05.008}{{\em Nucl. Phys. B}
  {\bfseries 838} (2010) 136--206},
  \href{http://arxiv.org/abs/1002.4407}{{\ttfamily arXiv:1002.4407 [hep-ph]}}.

\bibitem{Barontini:2023vmr}
A.~Barontini, A.~Candido, J.~M. Cruz-Martinez, F.~Hekhorn, and C.~Schwan,
  ``{Pineline: Industrialization of High-Energy Theory Predictions},''
  \href{http://arxiv.org/abs/2302.12124}{{\ttfamily arXiv:2302.12124
  [hep-ph]}}.

\bibitem{Watt:2012tq}
G.~Watt and R.~S. Thorne, ``{Study of Monte Carlo approach to experimental
  uncertainty propagation with MSTW 2008 PDFs},''
  \href{http://dx.doi.org/10.1007/JHEP08(2012)052}{{\em JHEP} {\bfseries 08}
  (2012) 052}, \href{http://arxiv.org/abs/1205.4024}{{\ttfamily arXiv:1205.4024
  [hep-ph]}}.

\bibitem{Carrazza:2015hva}
S.~Carrazza, J.~I. Latorre, J.~Rojo, and G.~Watt, ``{A compression algorithm
  for the combination of PDF sets},''
  \href{http://dx.doi.org/10.1140/epjc/s10052-015-3703-3}{{\em Eur. Phys. J. C}
  {\bfseries 75} (2015) 474}, \href{http://arxiv.org/abs/1504.06469}{{\ttfamily
  arXiv:1504.06469 [hep-ph]}}.

\bibitem{NNPDF:2017mvq}
{\bfseries NNPDF} Collaboration, R.~D. Ball {\em et~al.}, ``{Parton
  distributions from high-precision collider data},''
  \href{http://dx.doi.org/10.1140/epjc/s10052-017-5199-5}{{\em Eur. Phys. J. C}
  {\bfseries 77} no.~10, (2017) 663},
  \href{http://arxiv.org/abs/1706.00428}{{\ttfamily arXiv:1706.00428
  [hep-ph]}}.

\bibitem{Gao:2013bia}
J.~Gao and P.~Nadolsky, ``{A meta-analysis of parton distribution functions},''
  \href{http://dx.doi.org/10.1007/JHEP07(2014)035}{{\em JHEP} {\bfseries 07}
  (2014) 035}, \href{http://arxiv.org/abs/1401.0013}{{\ttfamily arXiv:1401.0013
  [hep-ph]}}.

\bibitem{Carrazza:2015aoa}
S.~Carrazza, S.~Forte, Z.~Kassabov, J.~I. Latorre, and J.~Rojo, ``{An Unbiased
  Hessian Representation for Monte Carlo PDFs},''
  \href{http://dx.doi.org/10.1140/epjc/s10052-015-3590-7}{{\em Eur. Phys. J. C}
  {\bfseries 75} no.~8, (2015) 369},
  \href{http://arxiv.org/abs/1505.06736}{{\ttfamily arXiv:1505.06736
  [hep-ph]}}.

\bibitem{Carrazza:2016htc}
S.~Carrazza, S.~Forte, Z.~Kassabov, and J.~Rojo, ``{Specialized minimal PDFs
  for optimized LHC calculations},''
  \href{http://dx.doi.org/10.1140/epjc/s10052-016-4042-8}{{\em Eur. Phys. J. C}
  {\bfseries 76} no.~4, (2016) 205},
  \href{http://arxiv.org/abs/1602.00005}{{\ttfamily arXiv:1602.00005
  [hep-ph]}}.

\bibitem{Ahn:2009wx}
E.-J. Ahn, R.~Engel, T.~K. Gaisser, P.~Lipari, and T.~Stanev, ``{Cosmic ray
  interaction event generator SIBYLL 2.1}''
  \href{http://dx.doi.org/10.1103/PhysRevD.80.094003}{{\em Phys. Rev. D}
  {\bfseries 80} (2009) 094003},
  \href{http://arxiv.org/abs/0906.4113}{{\ttfamily arXiv:0906.4113 [hep-ph]}}.

\bibitem{Ahn:2011wt}
E.-J. Ahn, R.~Engel, T.~K. Gaisser, P.~Lipari, and T.~Stanev, ``{Sibyll with
  charm},'' in {\em 16th International Symposium on Very High Energy Cosmic Ray
  Interactions}.
\newblock 2, 2011.
\newblock \href{http://arxiv.org/abs/1102.5705}{{\ttfamily arXiv:1102.5705
  [astro-ph.HE]}}.

\bibitem{Riehn:2015oba}
F.~Riehn, R.~Engel, A.~Fedynitch, T.~K. Gaisser, and T.~Stanev, ``{A new
  version of the event generator Sibyll},''
  \href{http://dx.doi.org/10.22323/1.236.0558}{{\em PoS} {\bfseries ICRC2015}
  (2016) 558}, \href{http://arxiv.org/abs/1510.00568}{{\ttfamily
  arXiv:1510.00568 [hep-ph]}}.

\bibitem{Fedynitch:2018cbl}
A.~Fedynitch, F.~Riehn, R.~Engel, T.~K. Gaisser, and T.~Stanev, ``{Hadronic
  interaction model sibyll 2.3c and inclusive lepton fluxes},''
  \href{http://dx.doi.org/10.1103/PhysRevD.100.103018}{{\em Phys. Rev. D}
  {\bfseries 100} no.~10, (2019) 103018},
  \href{http://arxiv.org/abs/1806.04140}{{\ttfamily arXiv:1806.04140
  [hep-ph]}}.

\bibitem{Buonocore:2023kna}
L.~Buonocore, F.~Kling, L.~Rottoli, and J.~Sominka, ``{Predictions for
  Neutrinos and New Physics from Forward Heavy Hadron Production at the LHC},''
  \href{http://arxiv.org/abs/2309.12793}{{\ttfamily arXiv:2309.12793
  [hep-ph]}}.

\bibitem{Roesler:2000he}
S.~Roesler, R.~Engel, and J.~Ranft,
  \href{http://dx.doi.org/10.1007/978-3-642-18211-2_166}{``{The Monte Carlo
  event generator DPMJET-III},''} in {\em International Conference on Advanced
  Monte Carlo for Radiation Physics, Particle Transport Simulation and
  Applications (MC 2000)}, pp.~1033--1038.
\newblock 12, 2000.
\newblock \href{http://arxiv.org/abs/hep-ph/0012252}{{\ttfamily
  arXiv:hep-ph/0012252}}.

\bibitem{Fedynitch:2015kcn}
A.~Fedynitch, \href{http://dx.doi.org/10.5445/IR/1000055433}{{\em Cascade
  equations and hadronic interactions at very high energies}}.
\newblock PhD thesis, KIT, Karlsruhe, Dept. Phys., 11, 2015.

\bibitem{Bai:2020ukz}
W.~Bai, M.~Diwan, M.~V. Garzelli, Y.~S. Jeong, and M.~H. Reno, ``{Far-forward
  neutrinos at the Large Hadron Collider},''
  \href{http://dx.doi.org/10.1007/JHEP06(2020)032}{{\em JHEP} {\bfseries 06}
  (2020) 032}, \href{http://arxiv.org/abs/2002.03012}{{\ttfamily
  arXiv:2002.03012 [hep-ph]}}.

\bibitem{Bai:2021ira}
W.~Bai, M.~Diwan, M.~V. Garzelli, Y.~S. Jeong, F.~K. Kumar, and M.~H. Reno,
  ``{Parton distribution function uncertainties in theoretical predictions for
  far-forward tau neutrinos at the Large Hadron Collider},''
  \href{http://dx.doi.org/10.1007/JHEP06(2022)148}{{\em JHEP} {\bfseries 06}
  (2022) 148}, \href{http://arxiv.org/abs/2112.11605}{{\ttfamily
  arXiv:2112.11605 [hep-ph]}}.

\bibitem{Bai:2022xad}
W.~Bai, M.~Diwan, M.~V. Garzelli, Y.~S. Jeong, K.~Kumar, and M.~H. Reno,
  ``{Forward production of prompt neutrinos from charm in the atmosphere and at
  high energy colliders},'' \href{http://arxiv.org/abs/2212.07865}{{\ttfamily
  arXiv:2212.07865 [hep-ph]}}.

\bibitem{Ostapchenko:2010vb}
S.~Ostapchenko, ``{Monte Carlo treatment of hadronic interactions in enhanced
  Pomeron scheme: I. QGSJET-II model},''
  \href{http://dx.doi.org/10.1103/PhysRevD.83.014018}{{\em Phys. Rev. D}
  {\bfseries 83} (2011) 014018},
  \href{http://arxiv.org/abs/1010.1869}{{\ttfamily arXiv:1010.1869 [hep-ph]}}.

\bibitem{Bhattacharya:2023zei}
A.~Bhattacharya, F.~Kling, I.~Sarcevic, and A.~M. Stasto, ``{Forward Neutrinos
  from Charm at Large Hadron Collider},''
  \href{http://arxiv.org/abs/2306.01578}{{\ttfamily arXiv:2306.01578
  [hep-ph]}}.

\bibitem{Maciula:2022lzk}
R.~Maciula and A.~Szczurek, ``{Far-forward production of charm mesons and
  neutrinos at forward physics facilities at the LHC and the intrinsic charm in
  the proton},'' \href{http://dx.doi.org/10.1103/PhysRevD.107.034002}{{\em
  Phys. Rev. D} {\bfseries 107} no.~3, (2023) 034002},
  \href{http://arxiv.org/abs/2210.08890}{{\ttfamily arXiv:2210.08890
  [hep-ph]}}.

\bibitem{Ball:2016neh}
{\bfseries NNPDF} Collaboration, R.~D. Ball, V.~Bertone, M.~Bonvini,
  S.~Carrazza, S.~Forte, A.~Guffanti, N.~P. Hartland, J.~Rojo, and L.~Rottoli,
  ``{A Determination of the Charm Content of the Proton},''
  \href{http://dx.doi.org/10.1140/epjc/s10052-016-4469-y}{{\em Eur. Phys. J. C}
  {\bfseries 76} no.~11, (2016) 647},
  \href{http://arxiv.org/abs/1605.06515}{{\ttfamily arXiv:1605.06515
  [hep-ph]}}.

\bibitem{Ball:2022qks}
{\bfseries NNPDF} Collaboration, R.~D. Ball, A.~Candido, J.~Cruz-Martinez,
  S.~Forte, T.~Giani, F.~Hekhorn, K.~Kudashkin, G.~Magni, and J.~Rojo,
  ``{Evidence for intrinsic charm quarks in the proton},''
  \href{http://dx.doi.org/10.1038/s41586-022-04998-2}{{\em Nature} {\bfseries
  608} no.~7923, (2022) 483--487},
  \href{http://arxiv.org/abs/2208.08372}{{\ttfamily arXiv:2208.08372
  [hep-ph]}}.

\bibitem{Duwentaster:2022kpv}
P.~Duwent\"aster, T.~Je\v{z}o, M.~Klasen, K.~Kova\v{r}\'\i{}k, A.~Kusina, K.~F.
  Muzakka, F.~I. Olness, R.~Ruiz, I.~Schienbein, and J.~Y. Yu, ``{Impact of
  heavy quark and quarkonium data on nuclear gluon PDFs},''
  \href{http://dx.doi.org/10.1103/PhysRevD.105.114043}{{\em Phys. Rev. D}
  {\bfseries 105} no.~11, (2022) 114043},
  \href{http://arxiv.org/abs/2204.09982}{{\ttfamily arXiv:2204.09982
  [hep-ph]}}.

\bibitem{Frederix:2018nkq}
R.~Frederix, S.~Frixione, V.~Hirschi, D.~Pagani, H.~S. Shao, and M.~Zaro,
  ``{The automation of next-to-leading order electroweak calculations},''
  \href{http://dx.doi.org/10.1007/JHEP07(2018)185}{{\em JHEP} {\bfseries 07}
  (2018) 185}, \href{http://arxiv.org/abs/1804.10017}{{\ttfamily
  arXiv:1804.10017 [hep-ph]}}.

\bibitem{Jeong:2023hwe}
Y.~S. Jeong and M.~H. Reno, ``{Neutrino Cross Sections: Interface of shallow-
  and deep-inelastic scattering for collider neutrinos},''
  \href{http://arxiv.org/abs/2307.09241}{{\ttfamily arXiv:2307.09241
  [hep-ph]}}.

\bibitem{cruz_martinez_juan_2023_8355209}
J.~Cruz-Martinez, M.~Fieg, T.~Giani, P.~Krack, T.~Makela, T.~Rabemananjara, and
  J.~Rojo, ``Lhcfitnikhef/fpf-wg1,'' Sept., 2023.
\newblock \url{https://doi.org/10.5281/zenodo.8355209}.

\bibitem{Buckley:2014ana}
A.~Buckley, J.~Ferrando, S.~Lloyd, K.~Nordstr\"om, B.~Page, M.~R\"ufenacht,
  M.~Sch\"onherr, and G.~Watt, ``{LHAPDF6: parton density access in the LHC
  precision era},''
  \href{http://dx.doi.org/10.1140/epjc/s10052-015-3318-8}{{\em Eur. Phys. J. C}
  {\bfseries 75} (2015) 132}, \href{http://arxiv.org/abs/1412.7420}{{\ttfamily
  arXiv:1412.7420 [hep-ph]}}.

\end{thebibliography}
\providecommand{\href}[2]{#2}\begingroup\raggedright\endgroup

\end{document}